\documentclass[11pt]{article}
\usepackage{amsmath,amssymb,graphicx,epsfig,color,float,cancel,cite,soul}
\usepackage[titletoc,title]{appendix}
\renewcommand\baselinestretch{1.15}
\numberwithin{equation}{section}

\begin{document}
\thispagestyle{empty}

\hfill TIFR-TH/16-02

\begin{center}
{\LARGE\bf A Detailed Analysis of Flavour-changing Decays of \\ [2mm]
Top Quarks as a Probe of New Physics at the LHC}

\bigskip

{\large\sl Debjyoti Bardhan}\,$^{a,1}$,
{\large\sl Gautam Bhattacharyya}\,$^{b,2}$,
{\large\sl Diptimoy Ghosh}\,$^{c,3}$, \\
{\large\sl Monalisa Patra}\,$^{d,4}$
and
{\large\sl Sreerup Raychaudhuri}\,$^{a,5}$
 
\bigskip 
 
{\small
$^a$ Department of Theoretical Physics, Tata Institute of Fundamental 
Research, \\ 1 Homi Bhabha Road, Mumbai 400005, India. 

$^b$ Saha Institute of Nuclear Physics, \\
1/AF Bidhan Nagar, Kolkata 700064, India.

$^c$ Department of Particle Physics and Astrophysics, Weizmann Institute 
of Science, \\ Rehovot 76100, Israel.

$^d$ Rudjer Boskovic Institute, Division of Theoretical Physics, \\
Bijeni cka 54, HR-10000 Zagreb, Croatia. 
}
\end{center}

\begin{center} 
{\Large\bf Abstract} 
\end{center}
\vspace*{-0.35in}
\begin{quotation}
\noindent If the LHC should fail to observe direct signals for new 
physics, it may become necessary to look for new physics effects in rare 
events such as flavour-changing decays of the top quark, which, in the 
Standard Model, are predicted to be too small to be observed. We set up 
the theoretical framework in which experimentally accessible results can 
be expected in models of new physics, and go on to discuss two models of 
supersymmetry -- one with conserved $R$-parity, and one without 
$R$-parity -- to illustrate how the flavour-changing signals are 
predicted in these models. In the latter case, there is a distinct 
possibility of detecting the rare decay $t \to c + Z^0$ at the LHC. We 
also present a detailed set of very general formulae which can be used 
to make similar calculations in diverse models of new physics.
\end{quotation}

\bigskip

\centerline{\sf Pacs Nos: 11.30.Pb, 12,20.Ds, 12.60.-i, 14.65.Ha}

\vfill


\bigskip

\hrule
\vspace*{-0.1in}
$^1$ debjyoti@theory.tifr.res.in \hfill
$^2$ gautam.bhattacharyya@saha.ac.in \\
$^3$ diptimoy.ghosh@weizmann.ac.il \hfill
$^4$ mpatra@irb.hr \\
\centerline{$^5$ sreerup@theory.tifr.res.in}

\newpage
\section{Introduction : FCNC portal to new physics}

The Run-I of the CERN Large Hadron Collider (LHC) has already led to the 
discovery of the long-sought Higgs boson \cite{Higgs}, and, probably, 
the elusive pentaquark \cite{Aaij:2015tga} as well. As the LHC has now 
commenced its crucial Run-II, the eyes of the whole world are focussed 
on CERN with the hope that there will be startling discoveries at this 
machine, which is designed to probe an energy regime hitherto 
inaccessible to terrestrial experiments. Indeed, some hints of this 
kind~\cite{750GeV} have already created considerable 
excitement~\cite{nature}.

It is natural, at this stage, to inquire into the different 
possibilities, and ask how sure we are that any such discovery will be 
made. Unfortunately, it turns out that there is no really {\it 
compelling} reason to expect a new discovery at the LHC Run-2 -- though 
it is certainly possible. This is because the whole range of experiments 
done at low, intermediate and the highest available energies are 
beautifully explained by the Standard Model (SM), a portmanteau theory 
which incorporates three or four disparate ideas and holds them together 
with a set of phenomenological parameters. Ad hoc as it may seem, this 
clumsy model has been remarkably successful -- perhaps too successful -- 
in explaining every known measurement, sometimes to four or five decimal 
places. Ironically, it is the LHC, in its Run-I, which has put the 
strongest stamp of authenticity on the SM by discovering the missing 
Higgs boson, measuring its properties to be consistent with the SM 
predictions and, at the same time, failing to find any significant 
deviations from the SM in a host of highly precise measurements. The 
discovery of the pentaquark is as consistent with the SM as any of the 
other results.

When we extend our consideration beyond purely terrestrial experiments 
to the cosmos at large, we immediately realise that the SM fails to 
explain several outstanding problems. These include the problems of dark 
matter\cite{dark_matter} , dark energy \cite{dark_energy} and ultra-high 
energy cosmic rays above the Greisen-Zatsepin-Kuzmin (GZK) bound 
\cite{Zatsepin:1966jv}. In particular, if the Earth is immersed in a 
distribution of dark matter, as appears to be the case, there must be 
some way to detect this fact. This is a subject of intense experimental 
investigation around the world 
\cite{DMCryo,DMLiquid,DMSpace,Bertone:2004pz}. It is also hoped that 
discoveries at the LHC could shed light on the problem of dark matter, 
which, if particulate, would appear in a collision as missing energy and 
momentum. Some of the theoretical deficiencies of the SM are addressed 
in theories which extend or go beyond it to postulate new structures and 
symmetries at higher energy scales -- these are generically referred to 
as `new physics'. A few of these models also have dark matter 
candidates. The great hope of the present moment is that unambiguous 
signals for such new physics will be discovered in Run-II of the LHC.

There are two ways in which new physics can be discovered at the LHC. 
The first -- and simplest -- way is to `directly' discover evidence for 
new particles, which could appear either as resonances or pairs, or be 
produced in association with SM particles. Denoting a `new' particle by 
$P$, the simplest tree-level processes are:
\begin{equation}
pp \to P ~{\rm or}~ P^\ast \to X + Y  \qquad\qquad 
pp \to P + \bar{P} \qquad\qquad 
pp \to P + X 
\end{equation}
where $X$ and $Y$ stand for SM particles. Taking into account the fact 
that a `new' particle will either decay into SM particles, or, if it is 
a component of dark matter, lead to missing energy and momentum signals, 
one can enumerate the possible final states and then analyse the LHC 
data to see if there is any evidence for such signals. An answer in the 
affirmative would, of course, be very exciting, and hopefully this is 
what will occur in the near future.

While we have no wish to pour cold water on optimistic predictions of 
the above nature, one cannot ignore the possibility that the mass of the 
`new' particle(s) may very well lie outside the kinematic reach of the 
LHC. Curiously, the last undiscovered particle for whose mass we had a 
theoretical {\it upper} bound was the Higgs boson, and, in fact, the LHC 
was designed to find it within the entire range of possibilities 
\footnote{As it happens, the Higgs boson was found rather soon, and that 
too, near its lower mass bound rather than the upper.}. For `new' 
particles, however, all that we have are experimental lower bounds 
\cite{Aad:2015iea, Aad:2014wea, Aad:2014vma, Aad:2014nua, Aad:2014iza, 
Khachatryan:2015vra, Khachatryan:2014qwa} -- which are more a measure of 
the failure of experimental searches than a reflection of any physical 
principle. Thus, future failures to find any signals of new physics can 
always be explained away as due to higher and higher masses of the `new' 
particle(s). In such a case, there would arise a serious problem in 
falsifying the theories in question.

There does, however, exist an escape route, and this happens when we 
consider the quantum effects of the `new' physics. When we consider, 
say, tree-level decays of a SM particle which have been mediated by a 
heavy `new' particle $P$, e.g. a decay of the form
$$
Q \to X + P^\ast \to X + Y + Z
$$ 
where the $Q, X, Y, Z$ are all SM particles, then these are generally 
subject to a propagator suppression by a factor $M_Q^2/M_P^2$ --- which 
can be quite severe if $M_Q \ll M_P$, which is usually the case. 
However, if, instead of a decaying particle, we have a scattering 
experiment
$$ 
Q + \bar{X} \to P^\ast \to Y + Z 
$$ 
conducted at an energy $\sqrt{s} < M_P$, the corresponding `suppression' 
factor will be $s/M_P^2$ --- which may be orders of magnitude larger 
than the earlier factor since it is possible to make $\sqrt{s} \gg M_Q$. 
Even then, it could very well be that $M_P$ is so large that even with 
the effective values $\sqrt{s} \sim 1 - 2$~TeV available at the LHC, the 
propagator suppression will still make the process unobservable at the 
LHC, especially if there are large backgrounds arising from purely SM 
production of $Y + Z$ final states.

What we need to find, therefore, is a process which, for some reason, is 
severely suppressed in the SM, but, for some equally valid reason, is 
not so severely suppressed in the new physics sector. Here we are lucky, 
for there exists a whole class of SM processes which are severely 
suppressed by the unitarity constraints of the Cabibbo-Kobayashi-Maskawa 
(CKM) matrix. These are the so-called flavour-changing neutral current 
(FCNC) processes involving at least two generations of fermions in the 
initial and final states, and all the generations in the loop. Though 
this suppression, commonly called the Glashow-Iliopoulos-Maiani (GIM) 
mechanism \cite{Glashow:1970gm}, is described in any textbook on the SM 
\cite{Cheng:1985bj}, it is worthwhile to take a quick look at the main 
argument, since it will form the crux of some of the discussions in this 
article. The idea is that if we have an initial quark flavour $q$ and a 
final quark flavour $q'$ of the same charge, and the only 
flavour-changing couplings we have are due to the charged currents 
coupling to the $W$-boson, then the transition amplitude must have the 
form
\begin{equation}
M_{qq'} = \sum_{i=1}^3 V_{qi}^\ast V_{q'i} A(x_i, M_W) 
= \sum_{i=1}^3 \lambda_{i} A(x_i, M_W) 
\end{equation}
where $x_i \equiv m_i^2/M_W^2$ carries the generation dependence and $M_W$ 
sets the mass scale for charged-current interactions. Moreover, 
$\lambda_{i} = V_{qi}^\ast V_{q'i}$, and the unitarity of the CKM matrix 
ensures that if $q \neq q'$, then $\sum_i \lambda_i = 0$. Obviously, we 
can expand the $A(x_i, M_W)$ in a Maclaurin series
\begin{equation}
A(x_i, M_W) = A_0(M_W)  + x_i A'_i(M_W) + \frac{1}{2} x_i^2 A''_i(M_W)  + \dots
\end{equation}
where
\begin{equation*}
A_0(M_W) = A(0,M_W) \ ,\qquad
A'_i(M_W) = \left[\frac{\partial A}{\partial x_i}\right]_{x_i = 0} \ , \qquad
A''_i(M_W) = \left[\frac{\partial^2 A}{\partial x_i^2}\right]_{x_i = 0}
\end{equation*}
and so on, where we make the assumption that $x_i \ll 1$. The leading 
term in $M_{qq'}$ cancels out and what is left is therefore suppressed by 
$x_i$. Obviously, this will work nicely if we take the quarks $q,q'$ to 
have charge $+2/3$, for then we automatically get a suppression in the 
probability by $x_b = (m_b/M_W)^2 \sim 10^{-3}$, or by even smaller 
factors for the other generations\footnote{For FCNC decays of the $b$ 
quark, we need to expand about $x_t$ rather than $x_i = 0$, since $x_t > 
1$. However, this article focusses only on decays of the $t$ quark.}.

If we now assume that the `new' particle(s) $P$ make(s) contributions of 
the form
\begin{equation}
M_{qq'}^{\rm new} = \sum_{i=1}^3 \lambda_{i} \eta_i \tilde{A}(y_i, M_P)
\label{eqn:GIMviolation}
\end{equation}
where the $y_i \equiv m_i^2/M_P^2$ are similar to the $x_i$ and the 
$\eta_i$ are arbitrary flavour-dependent factors, then we immediately see 
that the leading order contribution stays, for $\sum_i \lambda_i \eta_i 
\neq 0$. Such contributions are unaffected by the GIM suppression, and, 
therefore, could, in principle, be three orders of magnitude larger than 
the SM contributions.

The beauty of the above argument lies in the fact that in the above 
process, all that we need to observe is the transition of a $t$ quark to 
a quark of a different flavour but the same charge, i.e. a $u$ or a $c$. 
There is no requirement to produce heavy `new' particles on-shell. Thus, 
in the disappointing situation that all direct searches for `new' physics 
at the LHC fail, one can fall back upon GIM-suppressed processes as a 
portal through which we can still peer into that otherwise-inaccessible 
new world.

The major loop-induced FCNC processes involving the top quark which have 
been studied in the literature are:
\begin{enumerate}
\item the decays $t \to q + S$, where $q = u,c$ and $S$ is a scalar -- 
either the Higgs boson $H^0$ or its counterpart(s) in new physics models; 
and
\item the decays $t \to q + V$, where $q = u,c$ and $V$ is a vector gauge 
boson -- which can be a photon or a gluon or a $Z^0$-boson or any 
counterpart(s) in new physics models;
\end{enumerate}
\vspace*{-0.1in}
In the SM, we have well known results for the branching ratio
\begin{equation}
B(t \to c + H^0) \sim 10^{-15}  \qquad\qquad  B(t \to c + Z^0) \sim 10^{-13}
\label{eqn:SMBR}
\end{equation}
These are many, many orders of magnitude too small to be measured at 
Run-2 of the LHC, where estimates are that at best branching ratios at 
the level of $10^{-5}$ may become accessible when enough data are 
eventually collected (see Figure 8). There have been several predictions 
in the literature that new physics processes could provide the necessary 
enhancement and predict branching ratios at this level. The purpose of 
this article is to investigate these claims critically and try to 
determine the model requirements which could lead to an actual discovery 
of new physics at the LHC through the top quark FCNC portal.

Before proceeding further, we address the question of the rare decay $t 
\to q + \gamma$, which is bound to happen if its counterpart $t \to q + 
Z$ is possible. Electromagnetic gauge invariance demands that $t \to q + 
\gamma$ be mediated only by the magnetic dipole moment operator 
\cite{AguilarSaavedra:2002ns}. This process, however, turns out to be 
less interesting for two reasons. In the first place, one loop 
contributions to $t \to q + \gamma$ are suppressed by about an order of 
magnitude compared to the corresponding process with a final-state $Z$. 
This turns out to be essentially because the coupling of a photon to 
$d_i$-quark pairs is suppressed by their fractional charge of $-1/3$. A 
more serious hurdle is that experimental measurement of the rare decay $t 
\to q + \gamma$ is plagued with much larger backgrounds because of the 
ease with which photons can be radiated at tree-level. For this reason, 
experiments\cite{Khachatryan:2015att} can only achieve an accuracy for $t 
\to c + \gamma$ which is an order of magnitude poorer than that for $t 
\to c + Z$. Taken together, these two factors ensure that the search for 
$t \to q + Z$ should clearly take precedence\footnote{As we will see in 
the final section, the process $t \to c + Z$ is somewhat marginal at the 
LHC. This makes the case hopeless for $t \to c + \gamma$. Replacing $c$ 
by $u$ leads to even smaller decay widths.} over that for $t \to q + 
\gamma$. Hence, we do not discuss the latter process further. For similar 
reasons, we do not consider the process $t \to q + g$ either.

This article is organised as follows. In the following section, we 
consider generic FCNC decays of the top quark \cite{Agashe:2013hma}, 
taking a toy model, and determine the conditions required to have maximal 
contributions to an FCNC process like $t \to c + B$, where $B$ is a 
scalar or a vector boson. As an example we take up, in the next section, 
a supersymmetric model which is quite likely to evade direct searches at 
the LHC. The following section extends this to the case of a 
supersymmetric model with $R$-party violation, which relies on non-CKM 
sources of FCNC. Finally we present a summary of our results and a 
conclusion. In the interests of smooth reading, most of the more 
cumbersome formulae are relegated to the Appendix.

\section{Generic FCNC Decays of the top quark in a toy model}

\begin{figure}[H]
\begin{minipage}{0.4\textwidth}
In this section, we investigate a toy model which could be taken as a 
prototype for FCNC decays for the top quark. Let us assume there are a 
pair of charged scalars $\omega^\pm$ with couplings of the form
\begin{eqnarray}
{\cal L}_{\rm int} & = & \xi \, \omega^+\omega^- H \\
& + &
\sum_{i,j=1}^3 \left( \eta \, V_{ij}\, \bar{u}_{iL} \, d_{jR} \, \omega^+ 
+ {\rm H.c.} \right) \nonumber 
\label{eqn:toyLagS}
\end{eqnarray}
where $H$ is the SM Higgs boson and $\xi, \eta$ are unknown couplings. 
These $\omega^\pm$'s are rather like scalar versions of the 
$W^\pm$-bosons. The choice of scalars
\vspace*{-0.2in}
\end{minipage}
\hskip 0.05\textwidth
\begin{minipage}{0.55\textwidth}
\begin{center}
\includegraphics[width=\textwidth,angle=0]{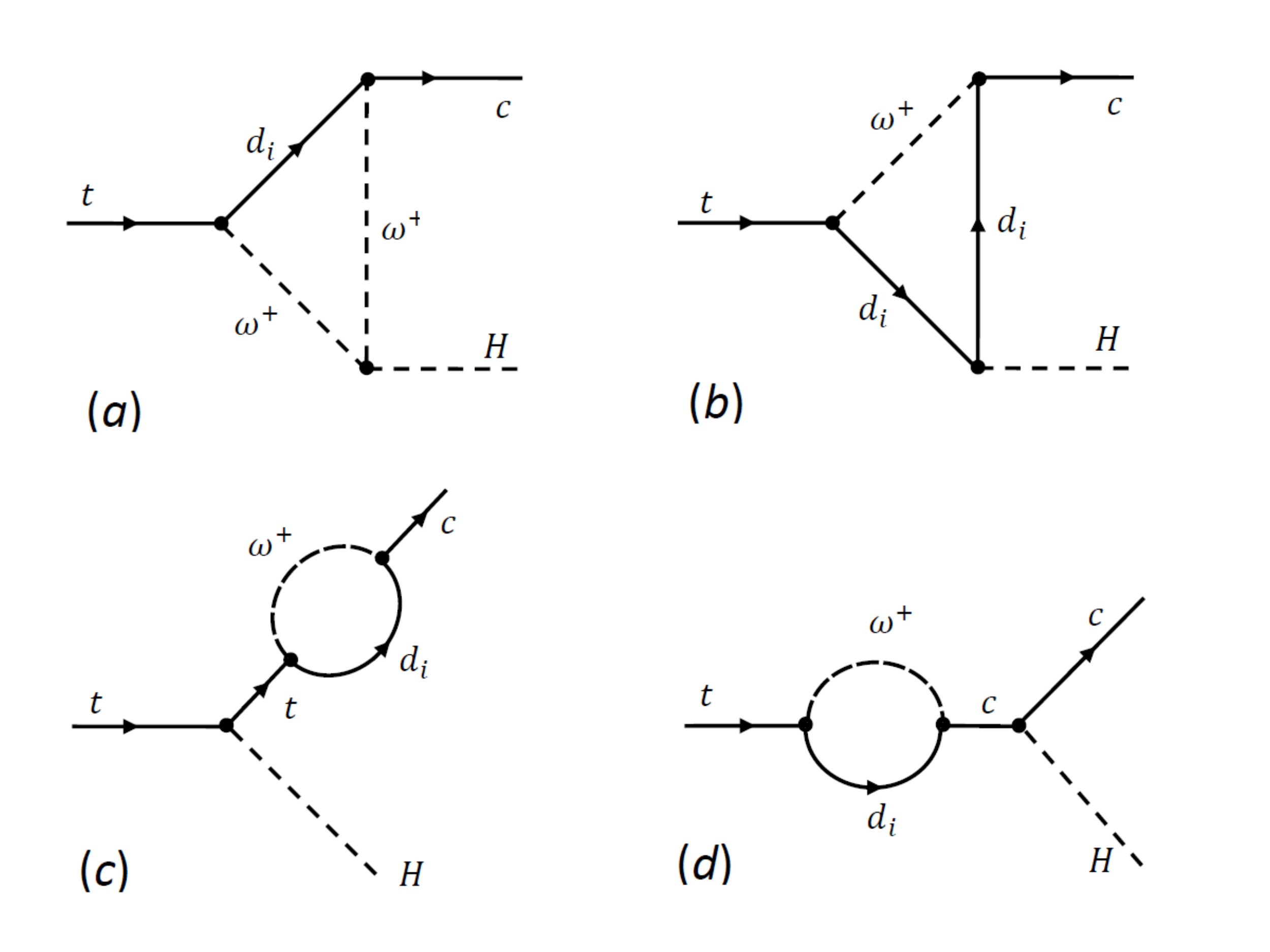}
\caption{\footnotesize\rm Set of Feynman Diagrams leading to the decay $t 
\to c + H$ in our toy model.}
\label{fig:Feyntoy}
\end{center}
\end{minipage}
\end{figure}   

makes the calculation simple and sidesteps complications due to gauge 
choice which arise with the $W^\pm$. For this part we stay within the 
minimal flavour violation (MFV) paradigm (see for example, Ref. 
\cite{Buras:2000dm}) insofar as the only flavour-changing effects happen 
through the `CKM' matrix elements $V_{ij}$.

Let us now consider the decay $t \to c + H$ as predicted in this model. 
Using the SM Yukawa couplings for the $H$-boson and Feynman rules for 
$\omega^\pm$ (which can quite easily be read off from the above 
Lagrangian), we obtain four diagrams, shown in Figure~1. It is then a 
straightforward matter to calculate the helicity amplitudes for the decay 
$t \to c + H$. In terms of the $\lambda_{i} = V_{ti}^\ast V_{ci}$, these 
can be written in the generic form
\begin{equation}
{\cal M}_{h_c h_t} = \sum_{i=1}^3  \lambda_i {\cal A}_i(h_c,h_t) 
\end{equation}
where $h_c$ and $h_t$ are the helicities of the $c$ and the $t$ quarks 
respectively, and $\lambda_1 + \lambda_2 + \lambda_3 = 0$ by unitarity of 
the CKM-like matrix $V$. Explicit expressions for these in terms of 
Passarino-'tHooft-Veltman functions \cite{Passarino:1978jh} are given in 
Appendix~A. We require to calculate only two non-vanishing amplitudes
\begin{equation}
(a) \ \ {\cal M}_{++} = \sum_{i=1}^3 \lambda_i {\cal A}_i(+1,+1)
\qquad\qquad
(b) \ \ {\cal M}_{--} = \sum_{i=1}^3 \lambda_i {\cal A}_i(-1,-1)
\label{eqn:Samplitudes}
\end{equation}
which become analogues of the SM amplitudes if we put $\xi = gM_W$ and 
$\eta = g/\sqrt{2}$. To calculate the branching ratio, we note that the 
squared and spin-summed/averaged matrix element, in terms of the helicity 
amplitudes of Eqn.~(\ref{eqn:Samplitudes}), is
\begin{equation}
\overline{|{\cal M}|^2} = \frac{1}{2} \left[ 
\left|{\cal M}_{++}\right|^2 + \left|{\cal M}_{--}\right|^2 
\right]	
\label{eqn:squaredS}
\end{equation}
The partial width can now be written as
\begin{equation}
\Gamma(t \to c + H) = \frac{1}{16\pi m_t} \left( 1 - \frac{M_H^2}{m_t^2}\right)
\overline{|{\cal M}|^2}
\label{eqn:widthS}
\end{equation}
and (if necessary) the branching ratio is easily obtained by dividing by 
the total decay width $\Gamma_t \simeq 1.29$~GeV.

At this point we pause to make a rough numerical estimate of the above 
quantities. As may be seen from Eqn.~(\ref{eqn:widthS}), the helicity 
amplitudes must have a mass dimension $+1$. Since these arise from 
one-loop computations, and if $M_\omega$ is close to $M_W$, a crude 
approximation for the amplitude factor will be
\begin{equation}
\overline{|{\cal M}|^2} \approx \left(\frac{m_t}{16\pi^2} \right)^2
\end{equation}
Substituting this into Eqn.~(\ref{eqn:widthS}), leads to a numerical 
estimate
\begin{equation}
\Gamma(t \to c + H) \approx 5.9 \times 10^{-5} \ {\rm GeV}
\label{eqn:estimate0}
\end{equation}
which is ten orders of magnitude larger than the SM prediction. 

It is natural to ask why the SM prediction is so much smaller than what 
one would naively have expected. The answer is that the SM amplitude is 
suppressed by a combination of three different effects, each reducing the 
amplitude by a few orders of magnitude. These are explained below.
\vspace*{-0.2in}
\begin{enumerate}
\item The first of these suppression effects is, of course, the GIM 
cancellation, which we have already shown to lead to suppression by a 
factor
$$
\left[ \frac{m_b(m_t)}{M_W} \right]^2 = \left[ \frac{2.6~{\rm GeV}}{80.4 
{\rm GeV}} \right]^2 \simeq 1.0 \times 10^{-3}
$$
in the decay amplitude. 

\item In this toy model, we have taken the flavour-violating coupling to 
be $\eta V_{ij}$ (or $\eta_i V_{ij}$), where the flavour-violation arises 
exactly as in the SM -- from the off-diagonal terms in the `CKM' matrix. 
This makes it a model with {\it minimal} flavour violation (MFV). Since 
the CKM matrix exhibits a strong hierarchy as we move away from the 
diagonal, this results in a further suppression in all MFV models -- 
which may not hold in a new physics model which deviates from the MFV 
paradigm. To make matters explicit, we have $\lambda_i = 
V_{2i}V^\ast_{3i}$ for $i = 1,2,3$. If we choose the $\eta_i$ as in 
Eqn.~(\ref{eqn:fccc}), the only relevant one is $\lambda_3 = V_{23} 
V^\ast_{33} \simeq V_{23}$ since $V_{33} \simeq 1$. Now, $|V_{23}| 
\approx 0.04$ \cite{Agashe:2014kda}. This gives us a suppression by two 
orders of magnitude.

There is a subtle issue, however. If we consider the flavour mixing in a 
model of new physics to be arbitrary and of unknown origin, it is 
perfectly fine to set $\lambda_3 = 1$ and thereby obtain an enhancement 
factor of $1/0.04 = 25$. In fact, this is what we shall assume in 
Section~5 of this paper. However, in a large class of non-MFV models, 
flavour mixing does arise from mixing effects of the quarks, and there 
exists some {\it unitary} matrix $V'_{ij}$ which is not the measured CKM 
matrix. To get a maximal value of $V'_{23}$, we take
\begin{equation}
V' = \left( \begin{array}{ccc}  
1 &  0                   & 0 \\ 
0 & \cos\theta  & \sin\theta \\
0 & -\sin\theta  & \cos\theta \end{array} \right)
\end{equation}  
so that $\lambda'_3 = \sin\theta \cos\theta = \frac{1}{2} \sin 2\theta$. 
Obviously, the maximum occurs for $\theta = \pi/4$ and the corresponding 
value of $\lambda_3$ is 0.5 --- an enhancement by a factor of 12.5 instead 
of 25. Thus, what we can achieve by abandoning the MFV paradigm is an 
enhancement by half of what we would get by discarding the CKM-type 
mechanism altogether.

\item Finally, in a model of new physics, there is always the possibility 
that the actual couplings may be enhanced over the SM ones. To see this, 
we put\footnote{Strictly speaking, the couplings can be taken up to 
$\sqrt{4\pi} \approx 3.5$, but then we will have to worry about 
higher-order effects.}  $\xi = M_\omega$ instead of $gM_W$ and $\eta_3 = 
1$ instead of $g/\sqrt{2}$, and recalculate the amplitudes, thereby 
achieving a modest enhancement by a factor of $2/g^3 \simeq 7.3$, 
assuming that $M_\omega \simeq M_W$. This means that the `SM' amplitude 
is suppressed by a factor $1/7.3 \simeq 0.14$ .
\end{enumerate}
\vspace*{-0.2in}
If we now combine the three effects, then the amplitude will have an
overall suppression factor
\begin{equation}
\left(1.0 \times 10^{-3}\right) \times 0.04 \times 0.14 \simeq 5.6\times 
10^{-6}
\label{eqn:suppression}
\end{equation}
Multiplying the amplitude by this factor and squaring leads to a 
suppression of the estimated partial decay width in 
Eqn.~(\ref{eqn:estimate0}) by ten orders of magnitude to $1.85\times 
10^{-15}$ --- which is in the right ballpark.

\begin{figure}[ht]
\def\baselinestretch{0.95}
\begin{center}
\includegraphics[height=2.6in,angle=0]{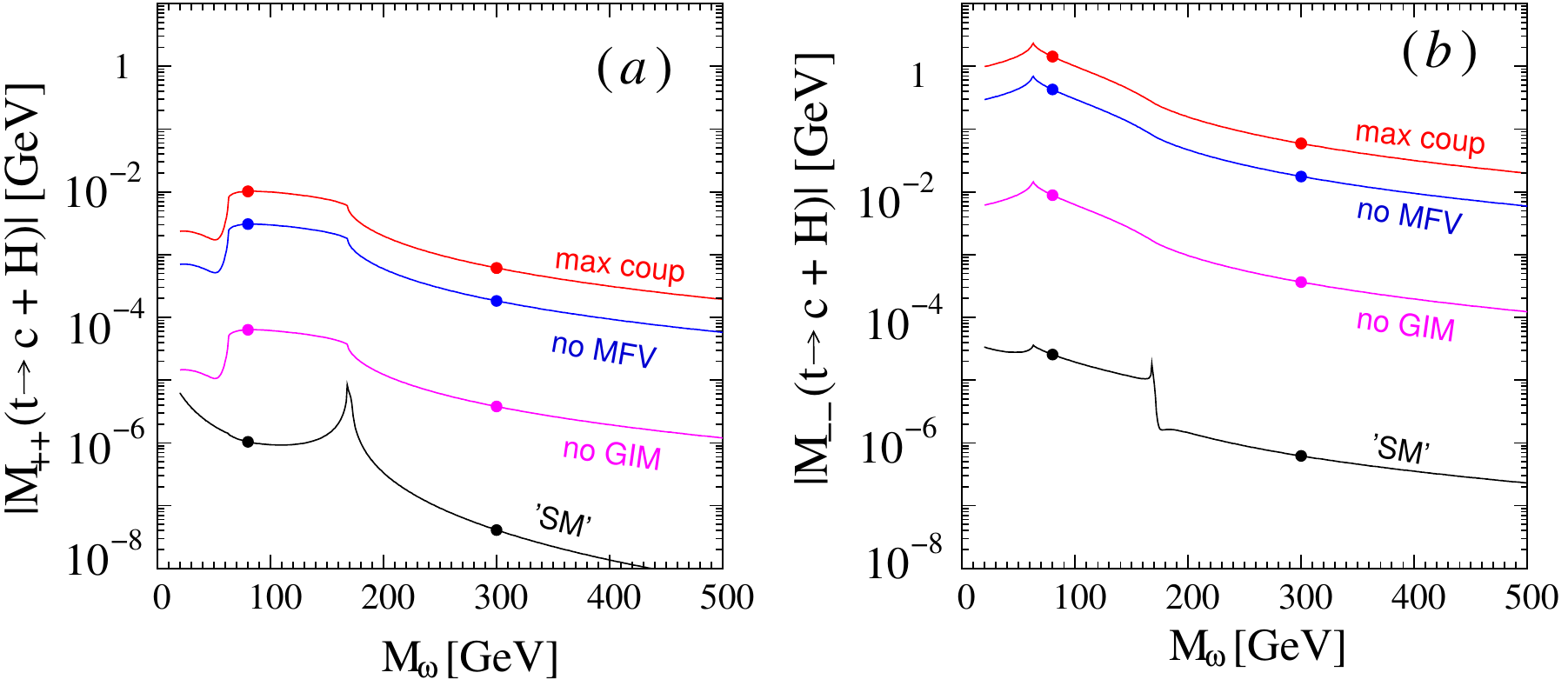}
\caption{\footnotesize\rm The two non-vanishing helicity amplitudes for 
the decay $t \to c + H$, as calculated in our toy model as a function of 
the mass $M_\omega$ of the scalar field $\omega$. The legends next to 
each curve are explained in the text. The small solid circles indicate 
the values $M_\omega = 80, 300$~GeV used in Table~1.}
\label{fig:toyS}
\end{center}
\def\baselinestretch{1.15}
\end{figure}

Now that we have a clear understanding of the nature of the FCNC 
suppression in the SM (or a SM-like model), we can remove these effects 
one by one to see how much the amplitude can be enhanced in a new physics 
model. In order to predict really significant deviations from the SM 
branching ratio any new physics model requires to meet the following 
conditions:
\vspace*{-0.2in}
\begin{enumerate}
\item[A.] Frustration of the GIM cancellation.
\item[B.] Non-MFV pattern of flavour mixing.
\item[C.] Enhanced couplings.
\end{enumerate}  
\vspace*{-0.2in}
To illustrate these in a concrete manner, we perform detailed numerical 
computations of the helicity amplitudes of Eqn.~(\ref{eqn:Samplitudes}) 
using the formulae of Appendix~A.1. The loop integrals in these formulae 
are evaluated using the well-known package FF 
\cite{vanOldenborgh:1990yc}, and our numerical results are given in 
Figure~\ref{fig:toyS}.

\vspace*{-0.2in}

The `normal case', when the couplings in Eqn.~(\ref{eqn:toyLagS}) are 
exactly like those in the SM corresponds to the black curves marked `SM' 
in Figure~2. The dots correspond to the values $M_\omega = 80, 300$~GeV 
(see Table~1). These amplitudes are suppressed due to a combination of 
all the three effects described above\footnote{It may be seen in 
Appendix~A.1 that the form factors $F_{1i}^{(b)}$ and $F_{2i}^{(b)}$ 
would violate the GIM cancellation. This is indeed true, and arises from 
the helicity-flipping nature of the scalar $\omega$ interaction. However, 
the contributions of $F_{1i}^{(b)}$ and $F_{2i}^{(b)}$ are very small, 
and hence, for all practical purposes, may be ignored in the numerical 
evaluation.} (see below).

We can disrupt the GIM cancellation partially or wholly by replacing the 
coupling constant $\eta$ in Eqn.~(\ref{eqn:toyLagS}) by a 
generation-dependent factor $\eta_i$. The maximal effect will be obtained 
if, for example, we consider
\begin{equation}
\eta_1 = \eta_2 = 0 \qquad \eta_3 = \frac{g}{\sqrt{2}} 
\label{eqn:fccc}
\end{equation}
The corresponding numerical curves are shown in Figure~2 in magenta, and 
labelled `no GIM'. It is immediately obvious that the amplitude increases 
by $2 - 3$ orders of magnitude, exactly as expected.

Next, we eschew MFV and consider the case $\lambda_3 = 1$. This gives an 
enhancement by a factor of 25. The blue lines marked `no MFV' in Figure~2 
represent the case in question. Finally, we set the couplings to the 
maximal values $\xi = M_\omega$ and $\eta_3 = 1$ and obtain a further 
enhancement illustrated by the curves shown in red in Figure~2 and marked 
`max coup'. This, as predicted, is enhanced by one order of magnitude.

If we consider the combination of all these effects, as we have done in 
Figure~2, we get an enhancement factor around $2.04\times 10^4$ 
($5.43\times 10^4$) for $|{\cal M}_{++}|$ ($|{\cal M}_{--}|$) taking 
$M_\omega = 80$~GeV. This is a more modest enhancement than estimated in 
Eqn.~(\ref{eqn:suppression}), but that is not surprising, given the fact 
that the earlier estimate was made under a very crude approximation to 
the decay amplitude. The actual enhancements available are made explicit 
in Table~1, where we list the partial widths for $t \to c + H$ in the toy 
model for $M_\omega = 80, \, 300$~GeV, for the SM-like case as well as 
with the three suppression mechanisms successively disabled.

\begin{table}[h!]
\begin{center}
\begin{tabular}{lllll}
\hline
$M_\omega$ &  `SM'  &  $\oplus$ no GIM   &  $\oplus$ no MFV   &  $\oplus$ max coup  \\ \hline\hline
 $80$ & $1.81\times 10^{-14}$ & $2.04\times 10^{-9}$ & $4.74\times 10^{-6}$ & $5.31\times 10^{-5}$ \\
$300$ & $4.31\times 10^{-18}$ & $5.12\times 10^{-11}$ & $1.19\times 10^{-7}$ & $1.33\times 10^{-6}$  \\ \hline
\end{tabular}
\end{center}
\caption{\footnotesize Partial decay widths for the decay $t \to c + H$ 
in the toy model, with successive application of the three enhancement 
conditions. All numerical values are in units of GeV.}
\label{tab:toyS}
\end{table}

\begin{figure}[htb]
\def\baselinestretch{0.95}
\begin{center}
\includegraphics[height=5.4in,angle=0]{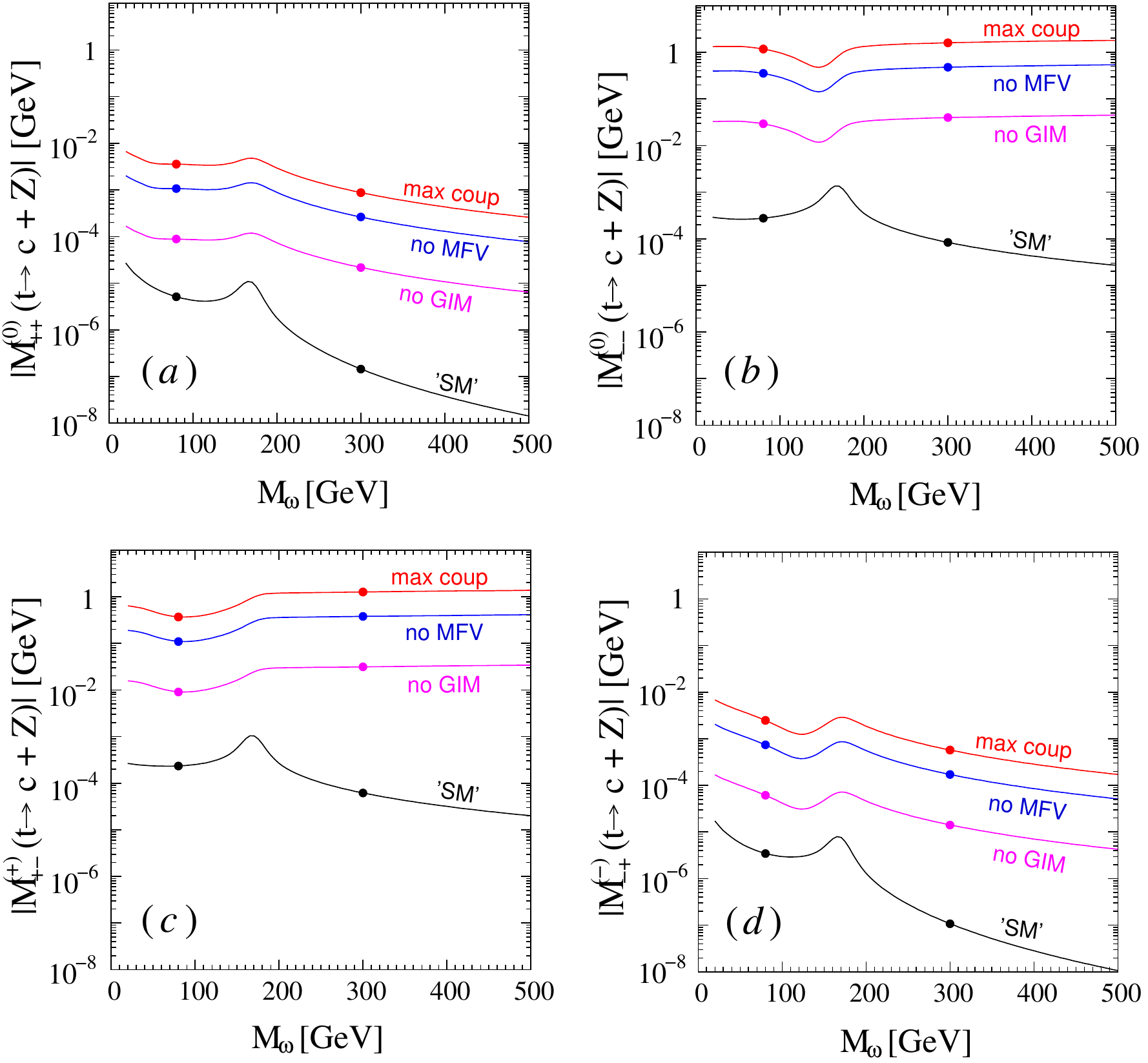}
\end{center}
\vspace*{-0.2in}
\caption{\footnotesize\rm Helicity amplitudes for the decay $t \to c + Z$ 
in our toy model. The notations and conventions follow those of 
Figure~\ref{fig:toyS}.}
\label{fig:toyZ}
\def\baselinestretch{1.15}
\end{figure}   

Another process of interest at the LHC is the decay $t \to c + Z$. The 
diagrams for this are identical to those in Figure~1, except that the 
scalar $H$ line must be replaced by a wiggly $Z$ line. We do not exhibit 
these diagrams in the interest of brevity, though we keep the same 
configuration and numbering. In this case, the computation is rendered a 
little more complicated because of the vector nature of the $Z$ boson. 
The toy Lagrangian will be
\begin{equation}
{\cal L}_{\rm int} = i\xi \omega^+ \overleftrightarrow{\partial_\mu} \omega^-
Z^\mu + \sum_{i,j=1}^3 \left( \eta \, V_{ij}\, \bar{u}_{iL} \, d_{jR} \, 
\omega^+ + {\rm H.c.} \right)
\label{eqn:toyLagZ}
\end{equation}
where $\xi, \eta$ are unknown couplings, as before. We can now compute 
the partial width for the decay $t \to c + Z$. The Feynman amplitude 
will assume the form
\begin{equation}
{\cal M}^{(h_Z)}_{h_c h_t} = \sum_{i=1}^3 \lambda_i {\cal A}_i(h_Z; h_c,h_t)
\end{equation}
where the sum over $h_Z$ runs over the longitudinal polarisation 
$\varepsilon_L = \varepsilon(h_Z)|_{h_Z = 0}$ and the transverse 
polarisations $\varepsilon_T^\pm = \varepsilon(h_Z)|_{h_Z=\pm 1}$. The 
only non-vanishing amplitudes are
\begin{equation}
\begin{array}{lcrll}
(a) & {\cal M}^{(+)}_{-+} = \sum_{i=1}^3 \lambda_i {\cal A}_i(+1;-1,+1)
& \qquad
(b) & {\cal M}^{(-)}_{+-} = \sum_{i=1}^3 \lambda_i {\cal A}_i(-1;+1,-1) 
\\ [4mm]
(c) & {\cal M}^{(0)}_{++} = \sum_{i=1}^3 \lambda_i {\cal A}_i(0;+1,+1)
& \qquad
(d) & {\cal M}^{(0)}_{--} = \sum_{i=1}^3 \lambda_i {\cal A}_i(0;-1,-1)
\end{array}
\label{eqn:Zamplitudes}
\end{equation}
and these may be regarded as `SM' amplitudes, if we take $\xi = 
gM_\omega$ and $\eta = g/\sqrt{2}$ as before. Once again, we plot these 
amplitudes in Figure~\ref{fig:toyZ} as a function of $M_\omega$ and 
relegate the detailed formulae to Appendix~A.
 
In Figure~\ref{fig:toyZ}, the four panels marked ($a$)--($d$) correspond 
to the four amplitudes ($a$)--($d$) indicated in 
Eqn.~(\ref{eqn:Zamplitudes}). The colour coding and conventions for this 
figure are identical to those in Figure~\ref{fig:toyS}. It is not 
difficult to see that once again, we get enhancement factors for these 
amplitudes which are very similar to those for the $t \to c + H$ case, 
when we successively (a) relax the GIM cancellation, (b) abandon the 
minimal flavour-violation paradigm and (c) enhance the couplings. This 
enables us to predict much larger partial widths, as shown in Table~2.

For this calculation, we require the squared and spin-summed/averaged 
matrix element, which is
\begin{eqnarray}
\overline{|{\cal M}|^2} & = & 
\frac{1}{2}\left[ 
\left|{\cal M}^{(+)}_{-+} \right|^2 
+ \left|{\cal M}^{(-)}_{+-} \right|^2
+ \left|{\cal M}^{(0)}_{++}\right|^2 
+ \left|{\cal M}^{(0)}_{--}\right|^2 \right]
\label{eqn:squaredZ}
\end{eqnarray}
in terms of the helicity amplitudes of Eqn.~(\ref{eqn:Zamplitudes}). 
The partial width can now be written
\begin{equation}
\Gamma(t \to c + Z) = \frac{1}{16\pi m_t} \left( 1 - \frac{M_Z^2}{m_t^2}\right)
\overline{|{\cal M}|^2}
\label{eqn:widthZ}
\end{equation}
as before, with $M_Z$ replacing $M_H$. In this case, of course, the 
partial width in more enhanced cases far exceeds the measured top quark 
width of 1.29~GeV, but this is not a serious matter, since this is, after 
all, a toy model. The enhancement in this case due to, successively, 
frustration of the GIM mechanism, saturation of the flavour off-diagonal 
terms and saturation of the coupling constant, have the same magnitudes 
as in the case of the top decaying through a scalar $H$ boson. We may, 
therefore, apply the same insights to both cases.

In general, the summed amplitudes for the decay $t \to c + Z^0$ are about 
an order of magnitude larger than the similar summed amplitudes for the 
decay $t \to c + H^0$. This is principally because a major contribution 
comes from the diagram with a $\omega^+\omega^-Z$ or $\omega^+\omega^-Z$ 
vertex, which are proportional to $g\cos\theta_W$ and $\lambda$ 
respectively, other factors being equal or similar. Since the measurement 
of the Higgs boson mass tells us that $\lambda \simeq 0.12$ it follows 
that $g\cos\theta_W/\lambda \simeq 5$. A further factor of around 2 is 
obtained because of the four non-vanishing helicity amplitudes for $t \to 
c + Z^0$ as opposed to the two obtained for $t \to c + H^0$. Thus, we get 
an enhancement of around 10, which becomes around $10^2$ when we consider 
the partial decay width. As this is a generic feature of the SM and most 
new physics models, it is obvious that the decay mode $t \to c + Z^0$ is 
more promising for discovery than the $t \to c + H^0$ mode.
   
\begin{table}[h!]
\begin{center}
\begin{tabular}{rllll}
$M_\omega$ 
&  `SM'    &  no GIM   &  no MFV            &  max coup  
\\ \hline\hline
$ 80$ & $4.23\times 10^{-11}$ & $3.55\times 10^{-4}$ & $5.15\times 10^{-2}$ & $0.58$ \\
$300$ & $8.16\times 10^{-12}$ & $8.32\times 10^{-3}$ & $1.21$               & $13.5$
\\ \hline
\end{tabular}
\caption{\footnotesize Partial widths for the decay $t \to c + Z$ in the 
toy model, with successive application (L to R) of the three enhancement 
conditions. All numerical values are in units of GeV. }
\end{center}
\label{tab:toyZ}
\end{table}
\vspace*{-0.2in}

\section{FCNC decays of the top quark in the SM}

\vspace*{-0.1in}
\begin{figure}[H]
\begin{minipage}{0.45\textwidth}
\begin{center}
\includegraphics[width=0.9\textwidth,angle=0]{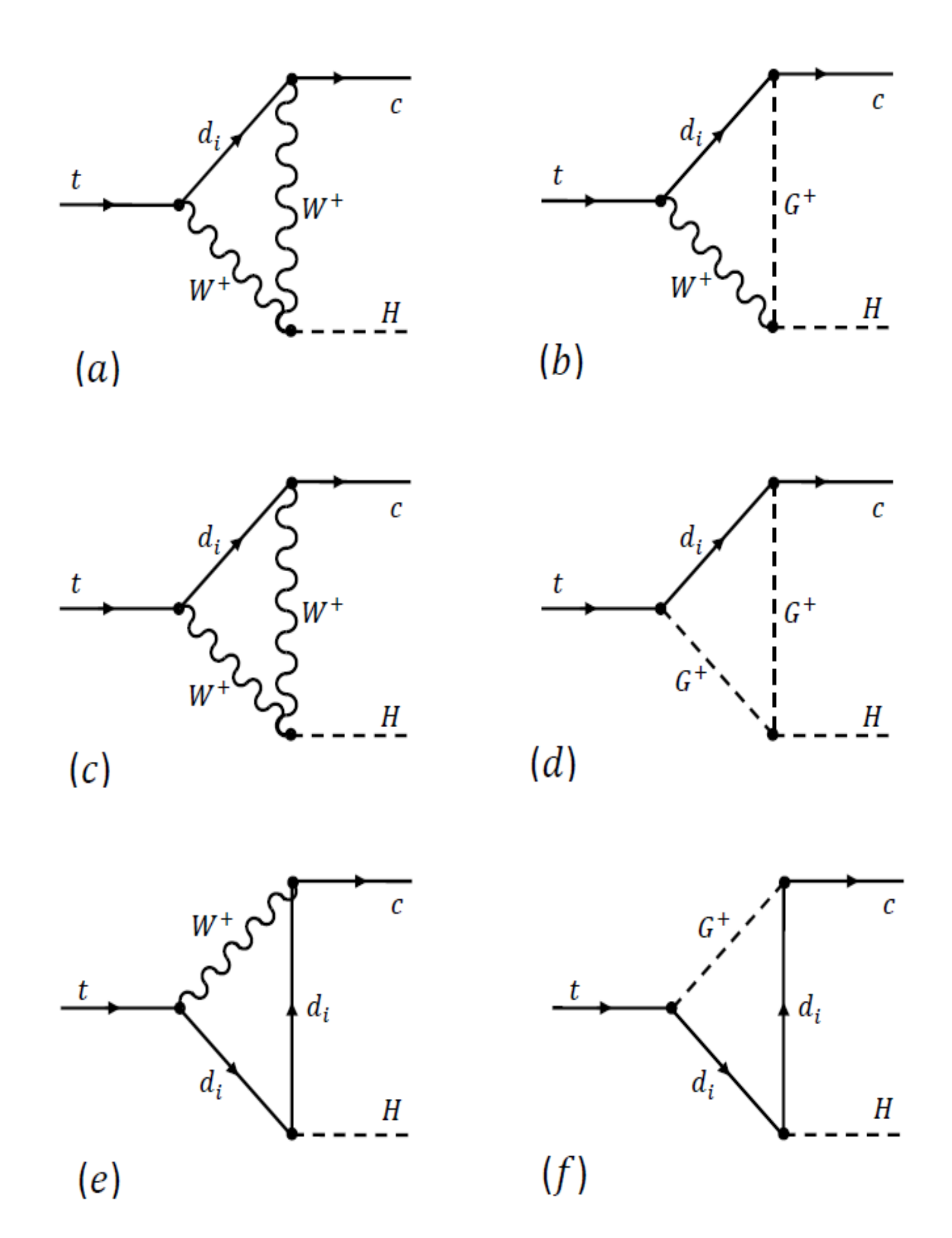}
\vspace*{-0.2in}
\includegraphics[width=0.9\textwidth,angle=0]{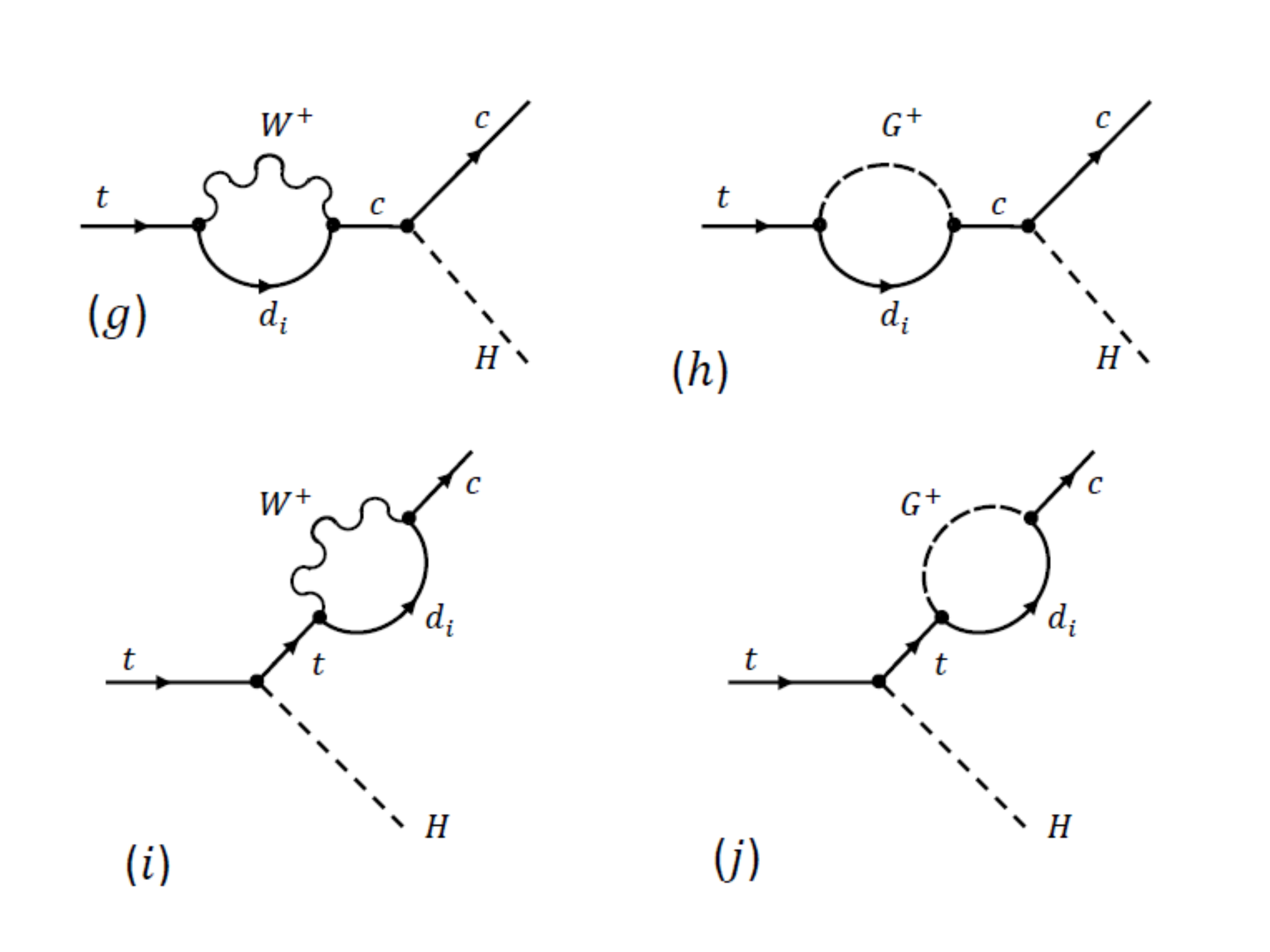}
\caption{\footnotesize\rm Feynman Diagrams leading to the decay $t \to c + H$ 
in the SM.}
\label{fig:FeynSM}
\end{center}
\end{minipage}
\hskip 0.02\textwidth
\begin{minipage}{0.53\textwidth}
We are now in a position to explore the decays $t \to c + H$ and $t \to c 
+ Z$ in the Standard Model, using insights from the toy model in the 
previous section. We start with $t \to c + H$. This time, of course, we 
have to take into account the exchange of the weak gauge bosons $W^\pm$ 
in the loops, and this requires a choice of gauge in which to work. For 
loop diagrams, it is convenient to choose the 'tHooft-Feynman gauge, 
since that keeps the ultraviolet divergences at a manageable level. Of 
course, this comes at the cost of having extra diagrams with unphysical 
Higgs bosons, and hence, in the SM, the four diagram topologies of 
Figure~\ref{fig:Feyntoy} become the ten diagrams in Figure~4.

\medskip

There is a small catch in using the 'tHooft-Feynman gauge, however, and 
that lies in the appearance of the unphysical Higgs bosons. The couplings 
of these to quarks depend on the $d$-quark masses $m_i$, and hence, would 
apparently lead to frustration of the GIM mechanism. However, these 
contributions cancel out when all the diagrams are added, as may be 
expected, since after all, they constitute a gauge artefact. The largest 
contributions to the amplitudes from individual diagrams (once the 
singularities are isolated) are of the order of $10^{-3}$ -- this already
\end{minipage} 
\end{figure} 
\vspace*{-0.35in} 

contains the suppression of one order due to the electroweak couplings 
and the factor $1/16\pi^2$ which appears in all loop diagrams. When all 
the contributions are summed-up, the GIM cancellation becomes manifest, 
and there is a reduction by ${\cal O}(m_b^2/m_t^2) \approx 6\times 
10^{-4}$. This brings down the amplitude to ${\cal O}(10^{-7})$ and 
hence, its square to ${\cal O}(10^{-14})$. Another order is lost in 
kinematics, and thus we get the final result $5.8 \times 10^{-15}$, as 
quoted in Eq.~(\ref{eqn:SMBR}).

When we turn to the decay $t \to c + Z$, we have a situation similar to 
the toy model in the previous section. The Feynman diagrams for this can 
be obtained from those of Fig.~\ref{fig:FeynSM} by replacing the dashed 
lines for $H$ by wiggly lines for $Z$ and changing the labels 
accordingly.

We then go on the calculate the helicity amplitudes of 
Eqn.~(\ref{eqn:Zamplitudes}) in terms of four form factors, which are 
given in Appendix~B. Most of the arguments given in the case of $t \to c 
+ H$ above hold for this case as well, except that the presence of four 
separate helicity amplitudes leads to a somewhat larger branching ratio, 
${\cal O}(10^{-13})$ as quoted in Eq.~(\ref{eqn:SMBR}).

The most important thing we learn from this exercise has already been 
stated in the Introduction -- the branching ratios for flavour-changing 
$t$-quark decays in the SM are severely suppressed, being far too small 
to be detected at the LHC, or even the most ambitious futuristic machine 
that can be conceived. This has the effect of making these decays a very 
sensitive probe of new physics, for any enhancement to measurable levels 
must arise from new physics beyond the SM.

\section{FCNC decays of the top quark in the cMSSM}

When we turn to new physics beyond the SM, the very first option must be 
the one which has captivated the imagination of high energy physicists 
for the last few decades, viz., supersymmetry (SUSY). The merits and 
demerits of SUSY have been exhaustively discussed in the literature 
\cite{SUSY} and do not require to be repeated here. Instead, we focus on 
the effects of SUSY on the flavour-changing processes $t \to c + H$ and 
$t \to c + Z$ which are the subject of this work.

Apart from the fact that every SM field has a supersymmetric partner 
differing from it in spin by one half, one of the most significant new 
features of SUSY models is the fact that they all require the existence 
of two scalar Higgs doublets. Thus, after the electroweak 
symmetry-breaking, these models contain five physical scalar fields, viz. 
a pair of charged Higgs bosons $H^\pm$ and a triplet of neutral Higgs 
bosons, of which two ($h^0, H^0$) are even under $CP$ and one ($A^0$) is 
odd under $CP$. The lighter one $h^0$ of the $CP$-even pair can be 
identified with the near-125~GeV scalar state found at the LHC in recent 
times. All the other states, $H^\pm$, $H^0$ and $A^0$, are presumed to be 
heavier, and, in fact, too heavy to have been detected in any experiments 
so far, including the LHC. As we shall see, it is likely that these 
states are all heavier than the $t$-quark, and hence, the only 
kinematically-permitted decay will be $t \to c + h^0$, which is analogous 
to the SM decay.
 
\begin{figure}[H]
\begin{minipage}{0.52\textwidth}
\begin{center}
\includegraphics[width=0.9\textwidth,angle=0]{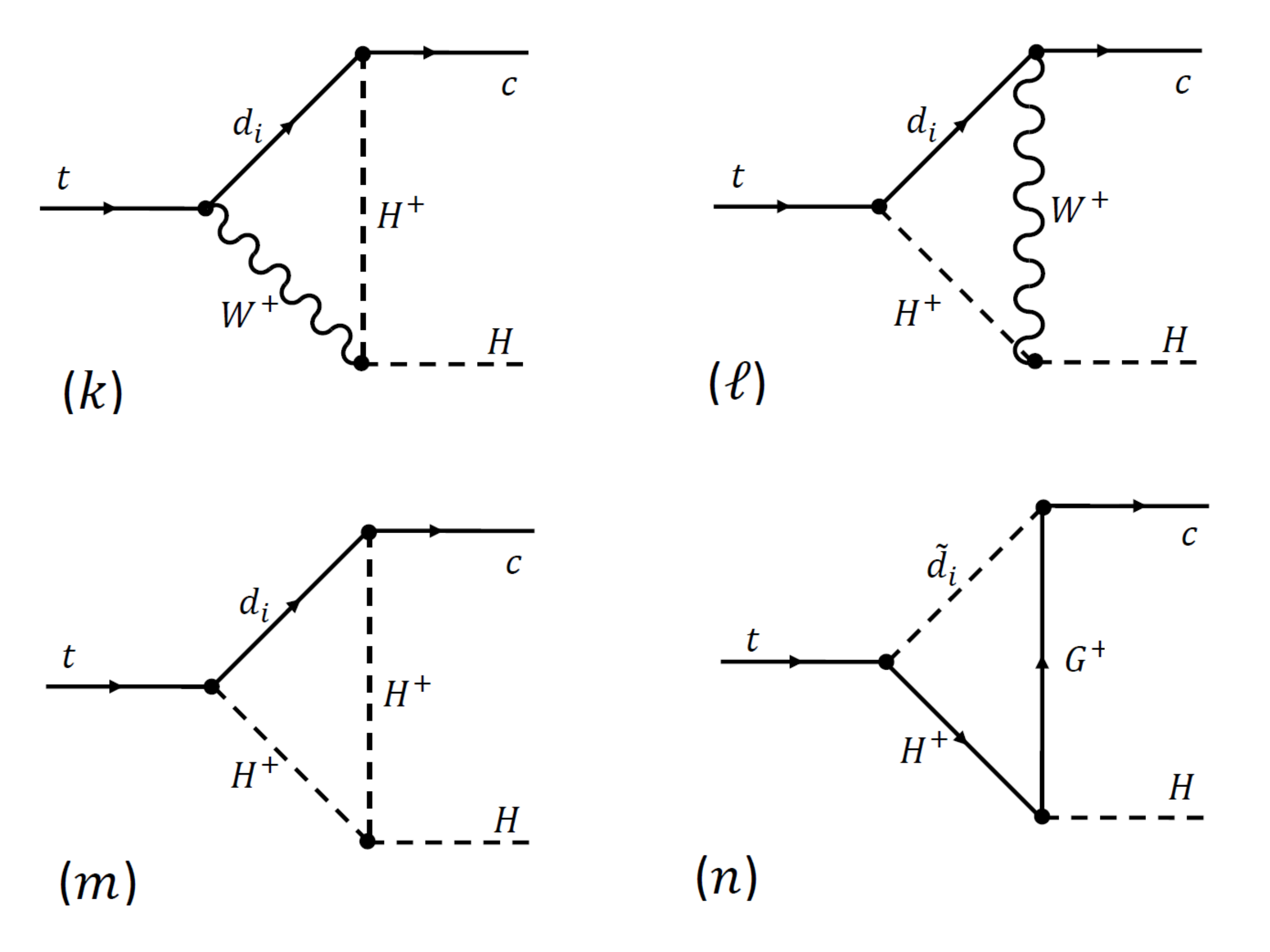}
\vspace*{-0.2in}
\includegraphics[width=0.9\textwidth,angle=0]{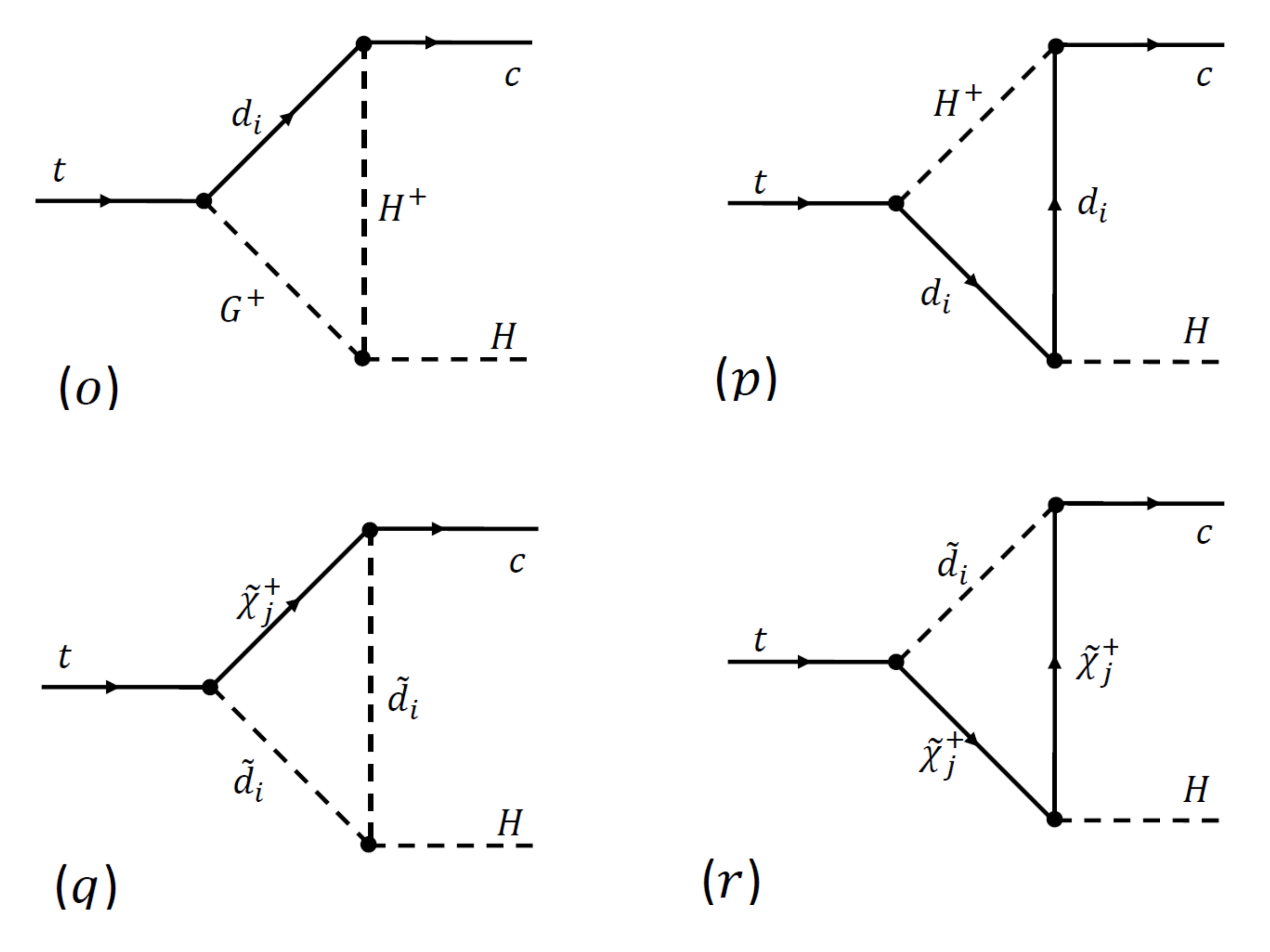}
\vspace*{-0.2in}
\includegraphics[width=0.9\textwidth,angle=0]{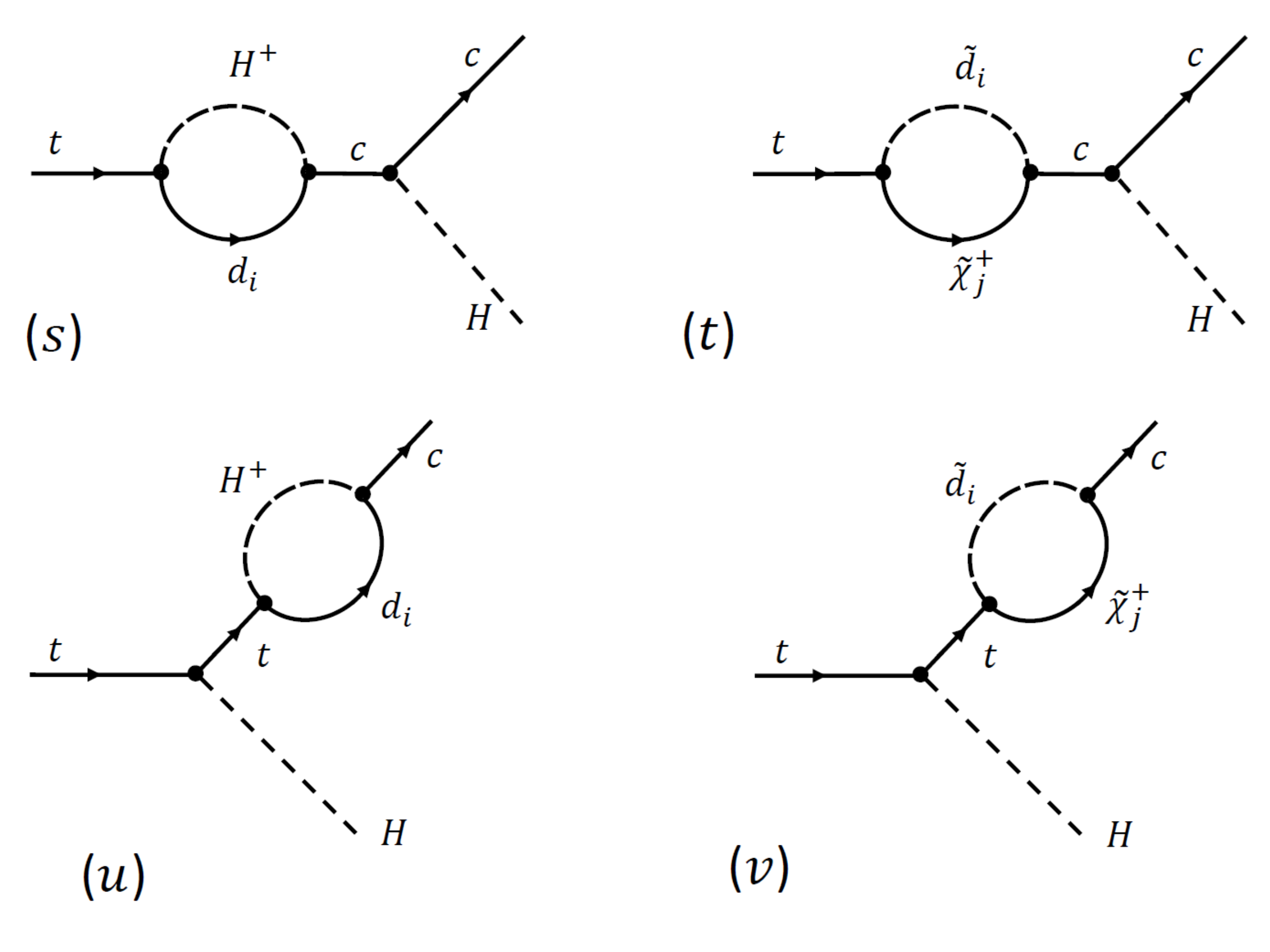}
\caption{\footnotesize\rm Additional Feynman diagrams leading to the 
decay $t \to c + H$ in the cMSSM.}
\label{fig:FeynSUSY}
\end{center}
\end{minipage}
\hskip 0.02\textwidth
\begin{minipage}{0.46\textwidth}
The more important difference from the SM in SUSY models arises because 
of the contributions of new particles in the loops. The most important of 
these are the contributions due to the charged Higgs bosons $H^\pm$, 
which have flavour-changing coupling like the $W$-boson. However, since 
these couplings originate from the Yukawa sector, they are proportional 
to the quark masses and hence will frustrate the GIM mechanism. Then 
there are contributions where the SM particles are replaced by their SUSY 
partners, viz. squarks and charginos. Here the flavour-changing effects 
will arise from the mixing matrices for squarks. In the so-called minimal 
flavour violation (MFV) models, the squark mixing matrices are aligned 
with the quark mixing matrix, i.e. the CKM matrix. This is the paradigm 
we shall adopt in the present study. Non-MFV models have been studied in 
the literature and we shall have occasion to discuss them in the final 
section.

\medskip

Though there are many SUSY versions of the SM and its extensions, the 
minimal version of this is the so-called constrained minimal 
supersymmetric SM, or cMSSM \cite{SUSY}. This is the SUSY model which has 
the minimum number of extra parameters (four parameters and a sign), when 
compared with all the others.  Not surprisingly, it is also
\end{minipage}
\end{figure} 
\vspace*{-0.35in} 
the SUSY model which is most constrained by experiment. However, since a 
light Higgs boson $h^0$ is a common feature of all SUSY models, including 
the cMSSM, the only features which will be affected will be the couplings 
and the super-partner masses. As we have seen, this is not too serious a 
constraint on loop-induced processes, so it is sensible to use the cMSSM 
as a paradigm case for FCNC processes in SUSY. This is adopted in our 
work and it fixes the particle content and the vertex factors, though 
there will be large variations in the latter as the model parameters 
change.

In the cMSSM, the process $t \to c + h^0$ will be mediated by the 10 
diagrams of the SM listed earlier in Fig.~\ref{fig:FeynSM} as well as the 
12 additional one-loop diagrams listed in Fig.~\ref{fig:FeynSUSY}. These 
diagrams have not only charged Higgs bosons but also charginos and 
squarks in the loops. The details for calculating all these 22 diagrams 
are given in Appendix~B, in terms of the usual form factors.  Numerical 
evaluation of these form factors, and hence the branching ratio, is not, 
however, very simple.

\begin{table}[h!]
\begin{center}
\begin{tabular}{rcccccccc}
gauginos : & $\tilde{\chi}_1^\pm$ & $\tilde{\chi}_2^\pm$ & $\tilde{\chi}_1^0$ 
& $\tilde{\chi}_2^0$ & $\tilde{\chi}_3^0$ & $\tilde{\chi}_4^0$ & $\tilde{g}$ & \\
\hline 
mass bound (GeV) : & 94 & 94 & 46 & 63 & 100 & 116 & 520 & \\ [3mm]
squarks : & $\tilde{u}_1$ & $\tilde{u}_2$  & $\tilde{d}_1$ & $\tilde{d}_2$
& $\tilde{t}_1$ & $\tilde{t}_2$ & $\tilde{b}_1$ & $\tilde{b}_2$ \\ 
\hline 
mass bound (GeV) : & 1100 & 1100 &  1100 & 1100 & 96 & 96 & 89 &  247 \\ [3mm] 
gauginos : & $\tilde{e}_1$ & $\tilde{e}_2$ & $\tilde{\tau}_1$ & $\tilde{\tau}_2$
& $\tilde{\nu_e}$ & $\tilde{\nu_\tau}_1$  &  & \\
\hline 
mass bound (GeV) : &  82 &   82 &   73 &   94 &  94 &  94 & & \\  [3mm] 
Higgs bosons : & $H^0$ & $A^0$ & $H^\pm$ & & & & &  \\
\hline 
mass bound (GeV) : & 500 & 0 & 80 & & & & & \\
\end{tabular}
\caption{\footnotesize Experimental lower bounds on new particle masses 
relevant to SUSY models. The results for the second generation of quarks 
and leptons are the same as those shown for the first generation. The 
most conservative bounds have been taken. The numbers shown in this Table 
correspond to the case when $R$-parity is conserved, but they do not 
change very much when $R$-parity is violated.}
\end{center}
\label{tab:sparticles}
\end{table}
\vspace*{-0.2in}
The problem here is that we cannot make {\it any} random choice of the 
four parameters and one sign in the cMSSM, for large ranges of these have 
been ruled out by experimental data on a variety of measured processes. 
We, therefore, must evaluate the branching ratio for $t \to c + h^0$ {\it 
only} for points in the parameter space which are permitted by all the 
experimental constraints \cite{Cao:2014udj}. At a first glance, this is a 
daunting prospect, given the wide range and diverse nature of 
experimental data which impact the cMSSM, but the task is made much 
easier by the presence of public domain software which do most of the 
computation automatically. We have, therefore, made free use of these 
software to constrain the cMSSM parameter space. The exact procedure 
followed is described below.
\begin{enumerate}
\item A set of random choices is made of the four parameters of the 
cMSSM, viz. the universal scalar mass $m_0$, the universal fermion mass 
$m_{1/2}$, the universal trilinear coupling $A_0$ and the ratio of Higgs 
boson vevs $\tan\beta$, within the ranges
\begin{eqnarray*}
100~{\rm GeV} \leq & m_0 & \leq 10~{\rm TeV} \quad\qquad\qquad
100~{\rm GeV} \leq  m_{1/2}  \leq 10~{\rm TeV} \\
-10 ~{\rm TeV} \leq & A_0 & \leq 10 ~{\rm TeV} \qquad\qquad\qquad\qquad
2 \leq  \tan\beta  \leq 50
\end{eqnarray*}

The sign of the $\mu$ parameter is chosen positive, since it is known 
that the negative sign is disfavoured by measurements of the muon 
anomalous magnetic moment.

\item Given a choice of the above parameters, we find the low-energy 
cMSSM mass spectrum by using the software {\sc SuSpect} 
\cite{Djouadi:2002ze}, which takes these values at the scale of grand 
unification and uses the renormalisation group equations to evolve them 
down to the electroweak scale, and also calculates mixing induced by the 
electroweak symmetry-breaking.

\item We eliminate parameter sets which are inconsistent with the 
observed $h^0$ mass $125 \pm 2$~GeV. This turns out to be a very severe 
constraint for low values of $m_0$, $m_{1/2}$ and $A_0$. \item Of the 
surviving parameter sets, we eliminate those that are inconsistent with 
the results of direct searches for SUSY, i.e. which yield masses for the 
SUSY particles which are smaller than the experimental lower bounds given 
in Table~3 below \cite{Agashe:2014kda, Aad:2015iea, Aad:2014wea, 
Aad:2014vma, Aad:2014nua, Aad:2014iza, Khachatryan:2015vra, 
Khachatryan:2014qwa, Aad:2015lea}.

\item With the remaining parameter sets, we calculate a clutch of 
low-energy variables measured in $K$ and $B$ decays, using the software 
{\sc SuperISO} \cite{Mahmoudi:2008tp}. We then eliminate parameter sets 
which are inconsistent with the 95\% C.L. experimental data on these 
variables, as given in Table~4. 
\end{enumerate}

\vspace*{-0.2in} 

\begin{figure}[htb!]
\def\baselinestretch{0.95}
\begin{center}
\includegraphics[height=2.8in,angle=0]{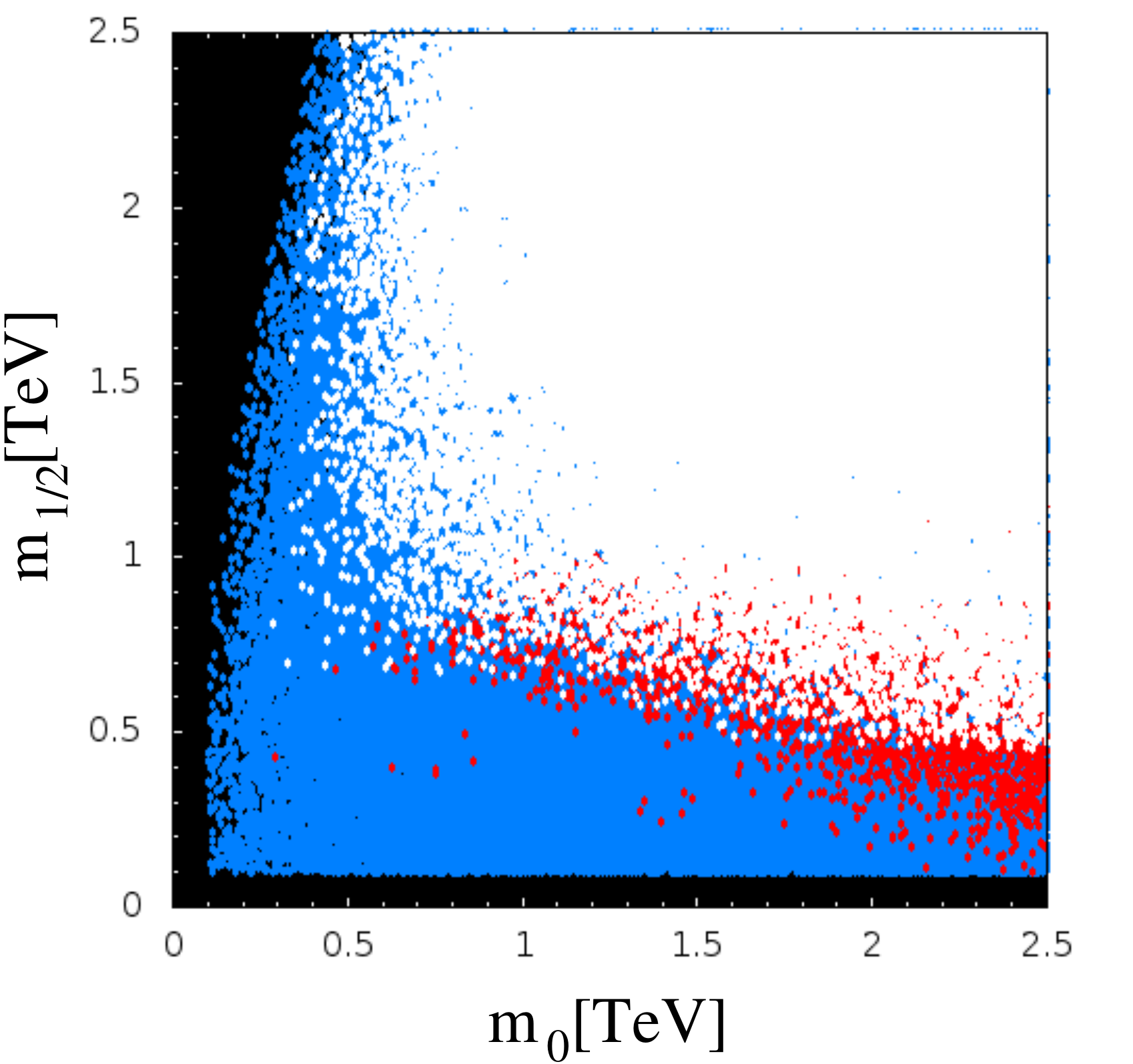}
\includegraphics[height=2.8in,angle=0]{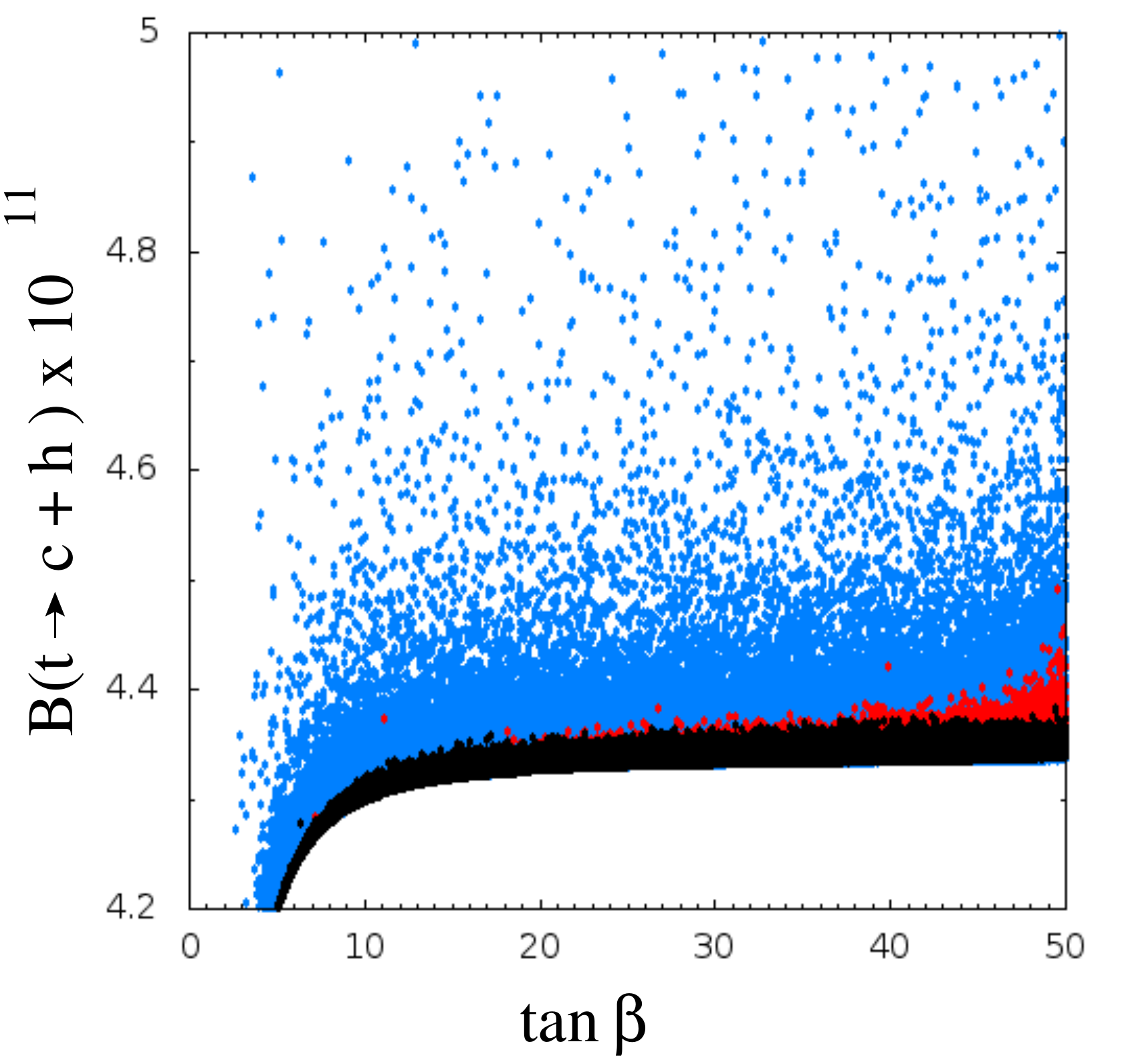}
\caption{\footnotesize\rm The panel on the left shows the parts of the 
$m_0$-$m_{1/2}$ plane in the cMSSM which are ruled out for all chosen 
values of $A_0$ and $\tan\beta$. In the left panel, the black region is 
ruled out by theory constraints, the blue dots by the Higgs boson mass 
constraints, and the red dots by all low-energy constraints. In the right 
panel, blue and red dots follow the same convention as in the left panel, 
while the black dots are {\it allowed} by all constraints.  }
\label{fig:cMSSMplots}
\end{center}
\def\baselinestretch{1.15}
\end{figure}

The most restrictive of these are the branching ratios ${\cal B}(B \to 
X_s \gamma)$ and ${\cal B}(B_s \to \mu^+\mu^-)$. The former is known to 
be highly sensitive to low values of the charged Higgs boson mass and the 
latter is important for precluding very large values of $\tan\beta$. Once 
a parameter set survives all the above filters, we consider it acceptable 
and use it to evaluate the $t \to c + h^0$ branching ratio. Our results 
are then set out in Figure~\ref{fig:cMSSMplots}.

\begin{table}[h!]
\small
\begin{center}
\begin{tabular}{lll}
Variable & Lower Bound & Upper Bound \\
\hline\hline
${\cal B}(B \to X_s \gamma)$        &   $2.766\times 10^{-4}$  &   $4.094\times 10^{-4}$ \\
$\Delta_0(B \to K^\ast \gamma)$                          &   $-3.8\times 10^{-2}$   &   $1.0\times 10^{-1}$ \\
${\cal B}(B_s \to \mu^+\mu^-)$      &   $7.261\times 10^{-10}$ &   $6.173\times 10^{-9}$ \\
${\cal B}(B_d \to \mu^+\mu^-)$      &   $4.0\times 10^{-11}$   &   $6.8\times 10^{-10}$ \\
${\cal B}(B \to X_s \mu^+\mu^-)$ \ \ {\rm (low} \ $Q^2$)    
                                  &   $2.4\times 10^{-7}$    &   $2.96\times 10^{-6}$  \\
${\cal B}(B \to X_s \mu^+\mu^-)$  \ \ {\rm (high} $Q^2$)   
                                  &   $1.48\times 10^{-7}$   &   $6.88\times 10^{-7}$  \\
${\cal B}(B \to \tau^+\nu_\tau)$    &   $7.388\times 10^{-5}$  &   $2.993\times 10^{-4}$ \\
$R[{\cal B}(B \to \tau^+\nu_\tau)]$ &   $5.5\times 10^{-1}$    &   $2.71$ \\
${\cal B}(B \to D \tau\nu)]$        &   $5.2\times 10^{-3}$    &   $1.02\times 10^{-2}$ \\
${\cal B}(D_s \to \tau\nu)$         &   $5.06\times 10^{-2}$   &   $5.7\times 10^{-2}$ \\
${\cal B}(D_s \to \mu\nu)$          &   $4.95\times 10^{-3}$   &   $6.67\times 10^{-3}$ \\
${\cal B}(D \to \mu^+\mu^-)$        &   $3.49\times 10^{-4}$   &   $4.15\times 10^{-4}$ \\
$R[{\cal B}(K \to \mu\nu)]$         &   $6.325\times 10^{-1}$  &   $6.391\times 10^{-1}$ \\
$R_\mu^{23}$                        &   $9.92\times 10^{-1}$   &   $1.006$ \\
$\delta(a_\mu)$                     &   $-6.5\times 10^{-10}$  &   $5.75\times 10^{-9}$
\\ \hline
\end{tabular}
\caption{\footnotesize Experimental bounds \cite{Mahmoudi:2008tp,Aaij:2012nna,CMS:2014xfa,Lees:2012wg,Lees:2012ju,Ambrosino:2005fw,Mahmoudi:2010jb,Ghosh:2012dh, Dighe:2013wfa,Bhattacherjee:2010ju} at 95\% C.L. on low energy 
parameters calculable in the software {\sc SuperISO}. For detailed definitions, see 
\cite{Mahmoudi:2008tp}.}
\end{center}
\label{tab:lowenergy}
\normalsize
\end{table}
\vspace*{-0.2in}

The left panel in Figure~\ref{fig:cMSSMplots} shows a scatter plot 
indicating the allowed regions in the $m_0$--$m_{1/2}$ plane, which is 
probably the best way to indicate constraints on the cMSSM. We note that 
every point on this plane corresponds to all possible random choices of 
the other parameters in the model, which accounts for the fuzziness in 
shapes. The black regions are disallowed by `theory' constraints, which 
include the proper shape of the electroweak potential 
\cite{vacuum_stability,top_mess} and the requirement that the lightest 
supersymmetric particle -- a prime dark matter candidate -- should be 
electrically neutral and have no colour quantum numbers. The extensive 
region in blue is ruled out by a combination of the $h^0$ mass constraint 
and the direct searches for supersymmetry, while the comparatively 
limited red regions are ruled out by constraints from low-energy 
measurements. Points falling in the white region are all allowed, and it 
is for these that we can legitimately try to evaluate top FCNC processes. 
It is important to note that almost the entire region for $m_0$ and 
$m_{1/2}$ within a TeV is ruled out -- this is another way of stating 
that there are no light squarks, unless we consider the third generation, 
where a seesaw-type mechanism can give us one lighter squark state.

The panel on the right in Figure~\ref{fig:cMSSMplots} contains our actual 
results. The scale on the $y$-axis,where we have plotted the branching 
ratio of $t \to c + h^0$ immediately tells that this always comes of the 
order of $10^{-11}$, which is just two-orders of magnitude above the SM 
prediction. On the $x$-axis we have plotted the $\tan\beta$ variable, 
even though the actual branching ratio is not a very sensitive function 
of this, except when $\tan\beta$ is around 5. As before, the blue points 
are ruled out by Higgs mass constraints and direct constraints, and the 
red points are ruled out by low-energy measurements. Unlike the left 
panel, however, the {\it black} points are the ones which represent the 
allowed parameter sets. It is immediately obvious, therefore, that the 
cMSSM prediction for $B(t \to c + h)$ is around $4.3\times 10^{-11}$, and 
this holds for almost all the points in the allowed parameter space.

Why is this branching ratio so small in the cMSSM, when there exist 
charged Higgs bosons to frustrate the GIM mechanism, as well as a wide 
range of possible couplings? The reason is quite simple. We do indeed 
have contributions which frustrate the GIM mechanism. This raises the 
branching ratio from the SM value of ${\cal O}(10^{-15})$ to ${\cal 
O}(10^{-11})$. However, if the factor had been as large as $m_W^2/m_b^2 
\simeq 5 \times 10^5$, we should have expected the prediction to be one 
order larger. That this does not happen is a phenomenon rather peculiar 
to the cMSSM, which is more constrained than other SUSY models. The 
requirement of a light Higgs boson with a mass as high as 125~GeV above 
the tree-level value, which is $M_Z$, requires most of the SUSY partners 
in this model to be very heavy, and this, being essentially a logarithmic 
effect, leads to the additional suppression of one order of magnitude in 
the $t \to c + Z$ branching ratio. Once this is understood, we cannot get 
the other enhancements, since ($a$) we have adopted the MFV paradigm, and 
($b$) the couplings in SUSY closely resemble the gauge couplings. The 
Yukawa couplings of the charged Higgs boson are, indeed, dependent on 
$\tan\beta$, but they are proportional to $$ \frac{m_t}{M_W}\cot\beta + 
\frac{m_b}{M_W}\tan\beta $$ and hence do not grow very large in the range 
$3 \leq \tan\beta \leq 50$.

As shown in the right panel in Figure~\ref{fig:cMSSMplots}, the 
application of the Higgs mass and direct search constraints pushes the 
branching ratio down by a factor around 3, which is expected since these 
are known to push up the SUSY partner masses from the 100~GeV to the TeV 
range. The application of low-energy constraints (especially $B_s \to 
\mu^+\mu^-$) further kills the feeble enhancement due to large 
$\tan\beta$, leading to the somewhat disappointing prediction of 
$4.3\times 10^{-11}$.

When we come to the process $t \to c + Z^0$, this will be mediated by the 
whole set of diagrams in Figures~\ref{fig:FeynSM} and \ref{fig:FeynSUSY} 
where, as in the previous case, the $h^0$ is replaced by the $Z^0$ and 
the corresponding broken line by a wiggly line. As in the previous 
section, we can calculate the four helicity amplitudes in terms of 
$F_1$--$F_4$ form factors which are listed in Appendix B.2 and make a 
numerical evaluation. As in the case of the toy model, we predict 
branching ratios which are about two orders of magnitude greater than the 
branching ratios for $t \to c + h^0$, i.e. we get ${\cal B}(t \to c + 
Z^0) \sim 10^{-9}$, which is still far too small to be accessed by 
experiment. The reason is, of course, the same -- breakdown of the GIM 
mechanism leads to a value about four orders of magnitude greater than 
the SM prediction, but so long as we stay within the MFV paradigm and 
have couplings which are not significantly greater than gauge coupling, 
no further enhancements will be obtained.

We see, therefore, that not only does the cMSSM fail to produce enough 
enhancement of the top FCNC decays for observation, but this will be a 
generic feature of any MSSM variant which follows the MFV paradigm. Not 
much can be gained, therefore, by relaxing the universality constraints 
on the SUSY-breaking parameters, as is done in, for example, the 
so-called phenomenological MSSM or pMSSM models. However, it is possible 
to break the MFV paradigm by choosing squark mixing matrices which are 
not aligned with the CKM matrix \cite{Dedes:2014asa}. This provides some 
enhancement of the branching ratios for top FCNC decay, but only to the 
level of about $10^{-7}$, partly because the squarks are already 
constrained to be rather heavy.
 
\section{Beyond the MFV paradigm : R-parity violation}

In the preceding section we have discussed how the cMSSM and its variants 
fail to produce top FCNC effects at a measurable level. Within SUSY, 
however, there exists another scenario which can provide the necessary 
enhancements, and that is the scenario when $R$-parity is violated. It is 
well-known that the conservation of the $Z_2$ quantum number $R = 
(-1)^{L+2S+3B}$, where $L$, $S$ and $B$ stand for lepton number, spin and 
baryon number of a particle, is a condition which must be imposed by hand 
on all SUSY models if we want the lightest SUSY particle, or LSP, to be a 
candidate for cold dark matter. Thus, when we consider a scenario in 
which the $R$-parity is not conserved, we abandon the idea of explaining 
dark matter in a SUSY model -- a feature which has contributed to making 
such models far less popular than the opposite variant. It is important 
to note, however, that $R$-parity conservation is not demanded by SUSY at 
all -- it is an add-on which was originally believed to be necessary to 
explain the long lifetime of the proton \cite{Senjanovic:2009kr}. 
However, ever since it was pointed out that this can be done be 
separately conserving either lepton number $L$ {\it or} baryon number 
$B$, it has been known that one can easily have $R$-parity violating 
models which are consistent with both exact and broken SUSY. In that 
case, $R$-parity loses its special position, for the way in which 
$R$-parity produces a dark matter candidate is no different from any 
other $Z_2$ symmetry imposed by hand on a new physics model, such as, for 
example, the KK-parity imposed in models with a universal extra dimension 
\cite{Macesanu:2005jx} and the $T$-parity imposed in the littlest Higgs 
models \cite{littlest_higgs}. Thus, at the cost of decoupling SUSY from 
the search for an explanation of dark matter in terms of new particles, 
it is legitimate to consider models where $R$-parity is violated.

Once we allow $R$-parity violation, it is straightforward to write down 
the extra interactions allowed. These will arise from a superpotential 
term \cite{Barbier:2004ez}
\begin{equation}
\widehat{W}_{\not{\!R}} = \sum_{i,j,k=1}^3 \left( 
\frac{1}{2}\lambda_{ijk}  \widehat{L}_i \widehat{L}_j \widehat{E}_k^c + 
\lambda'_{ijk}  \widehat{L}_i \widehat{Q}_j \widehat{D}_k^c +
\frac{1}{2}\lambda''_{ijk}  \widehat{U}_i^c \widehat{D}_j^c \widehat{D}_k^c
\right)
\end{equation}  
where the $\widehat{L}$ and $\widehat{Q}$ superfields are SU(2) doublets 
(suitably combined) and the $\widehat{E}^c$, $\widehat{U}^c$ and 
$\widehat{D}^c$ are SU(2) singlets. The indices $i$, $j$ and $k$ run over 
the three matter generations. It is immediately clear that the 
$\lambda_{ijk}$ are antisymmetric in $i$ and $j$, i.e. there are 9 
independent $\lambda_{ijk}$'s and the $\lambda''_{ijk}$ are antisymmetric 
in $j$ and $k$, i.e. there are 9 independent $\lambda''_{ijk}$'s. The 
$\lambda'_{ijk}$ have no such symmetry properties and hence there will be 
27 independent $\lambda'_{ijk}$'s, bringing the total number of 
independent parameters to 45. However, to avoid fast proton decay, we 
must either conserve lepton number and set all the $\lambda_{ijk}$'s and 
$\lambda'_{ijk}$'s to zero, or conserve baryon number and set all the 
$\lambda''_{ijk}$'s to zero. Either alternative leads to FCNC processes, 
including, when the third generation is considered, the top quark. In 
this work, all RPV couplings will be considered real.

Constraints on the $R$-parity violating couplings from various low-energy 
FCNC processes have been industriously studied in the literature 
\cite{Barbier:2004ez, Dreiner:1997uz, Bhattacharyya:1996nj, 
Bhattacharya_review, Csaki:2011ge,Kao:2009fg,Davidson:2015zza} and a 
first look would lead to the conclusion that the $\lambda$, $\lambda'$ 
and $\lambda''$ couplings must be rather small. Such constraints depend, 
however, on two crucial assumptions, viz.,
\vspace*{-0.2in}
\begin{table}[b!]
\def\baselinestretch{1.2}
\centering
\small
\begin{tabular}{|llccrrr|}
\hline
          &              &  Scales         &       & Upper    &  Sfermion    & Current  \\
         & Strongest Constraint   &  as mass &  Scaling     & bound    & mass & upper \\ 
 & arises from &  of & Exponent & (100~GeV)& ${\rm (GeV)}$ & bound \\ 
\hline\hline
\\ [-12pt]
$\lambda'_{121}$ & Atomic Parity Violation \cite{Dreiner:1997uz} & $\tilde{q}_L$ & 1 & 0.035 & 1350 \cite{Aad:2014pda} & 0.473 \\ \hline
$\lambda'_{122}$ & $\nu_e$ mass bound \cite{Bhattacharyya:1999tv}& $\tilde{d}_R$ & $^1\!/_2$ & 0.004 & 1100 \cite{Agashe:2014kda}& 0.013 \\ \hline 
$\lambda'_{123}$ & CC Universality \cite{Dreiner:1997uz} & $\tilde{b}_1$ & $^1\!/_2$ & 0.02 & 620 \cite{Aad:2013ija} & 0.05  \\ \hline
$\lambda'_{131}$ & Atomic parity violation \cite{Eilam:2001dh} & $\tilde{t}_L$ & 1 & 0.019  & 300 \cite{ATLAS-CONF-2015-026} & 0.057 \\ \hline
$\lambda'_{132}$ & FB asymmetry ($e^+ e^-$) \cite{Eilam:2001dh} \cite{Barbier:2004ez} & $\tilde{t}_L$ & 1 & 0.28 & 300 \cite{ATLAS-CONF-2015-026} & 0.84  \\ \hline 
$\lambda'_{133}$ & $\nu_e$ mass bound \cite{Bhattacharyya:1999tv} & $\tilde{b}_1$ & $^1\!/_2$ & 0.0002 & 620 \cite{Aad:2013ija} & 0.0005 \\ \hline
$\lambda'_{221}$ & Bounds on $R_{\mu e}$ \cite{Bhattacharyya:1995pq} & $\tilde{d}_R$ & 1 & 0.18 & 1100 \cite{Agashe:2014kda}& 1.98 \\ \hline
$\lambda'_{222}$ & $\nu_\mu$ mass bound \cite{Bhattacharyya:1999tv} & $\tilde{d}_R$ & $^1\!/_2$ & 0.015 & 1100 \cite{Agashe:2014kda}& 0.05 \\ \hline
$\lambda'_{223}$ & $D_s$ meson decay \cite{Bhattacharyya:1995pq} & $\tilde{b}_1$ & 1 & 0.18 & 620 \cite{Aad:2013ija} & 1.1 \\ \hline
$\lambda'_{231}$ & $\nu_\mu$ DIS \cite{Dreiner:1997uz,Barbier:2004ez} & $\tilde{\nu}_{\tau}$  & 1 & 0.22 & 1700 \cite{Aad:2015pfa} & 2.00 \\ \hline
$\lambda'_{232}$  & Bounds on $R_\mu (Z)$ \cite{Yang:1999ms,Bhattacharyya:1995pr} & $\tilde{s}$ & 1 & 0.39 & 1000 \cite{Agashe:2014kda} & 2.00  \\
 &  & $\tilde{\mu}$ & -1 & & 100 \cite{Agashe:2014kda} &   \\ \hline
$\lambda'_{233}$ & $\nu_\mu$ mass bound \cite{Bhattacharyya:1999tv} & $\tilde{d}_R$ & $^1\!/_2$ & 0.001 & 1100 \cite{Agashe:2014kda}& 0.003 \\ \hline
$\lambda'_{321}$  & $D_s$ decays \cite{Barbier:2004ez} & $\tilde{d}_R$ & $1$ & 0.52 & 1100 \cite{Agashe:2014kda}& 0.66  \\ \hline
$\lambda'_{322}$ & $\nu_\tau$ mass bound \cite{Bhattacharyya:1999tv} & $\tilde{d}_R$ & $^1\!/_2$ & 0.02 & 1100 \cite{Agashe:2014kda}& 0.07 \\ \hline
$\lambda'_{323}$ & $D_s$ decay \cite{Barbier:2004ez} & $\tilde{b}_1$ & $1$ & 0.52 & 620 \cite{Aad:2013ija} & 2.00 \\ \hline
$\lambda'_{331}$  & Bounds on $R_\tau (Z)$ \cite{Yang:1999ms} & $\tilde{d}$ & 1, & 0.22 & 1000 \cite{Agashe:2014kda}& 2.00  \\ 
$\lambda'_{332}$  &  &$\tilde{\tau}$  & -1 & 0.22 & 100 \cite{Agashe:2014kda}& 2.00  \\ \hline
$\lambda'_{333}$ & $\nu_\tau$ mass bound \cite{Bhattacharyya:1999tv} & $\tilde{b}_1$ & $^1\!/_2$ & 0.001 & 620 \cite{Aad:2013ija} & 0.003 \\ [2mm] 
\hline
\end{tabular}
\def\baselinestretch{0.95}
\caption{\footnotesize Showing the experimental constraints on the 
$R$-parity-violating couplings $\lambda'_{i2j}$ and $\lambda'_{i3j}$ 
relevant for FCNC decays of the top quark. The abbreviations used in the 
second column are as follows: charged current (CC), forward-backward 
(FB), deep inelastic scattering (DIS), branching ratio (BR). The upper 
bounds on the $\lambda'$ and $\lambda''$ couplings scale as the masses of 
the sfermions listed in the third column, raised to the powers given in 
the fourth column. The fifth column records the upper bounds when these 
masses are uniformly set to 100~GeV (except for the gluino, whose mass is 
set to 1000~GeV). The sixth column gives the current lower bound on the 
relevant sparticle masses and the last column gives the corresponding 
(scaled) upper bound on the $R$-parity-violating couplings. }
\label{tab:lpcoup}
\def\baselinestretch{1.15}
\end{table}
\normalsize

\begin{itemize}
\item Only one (or at most two) of the $R$-parity couplings are 
substantial and all the others are zero or of negligible value. This 
makes a phenomenological analysis simple, but its virtue ends there. The 
oft-repeated analogy with a similar pattern observed in the SM Yukawa 
couplings is not a very convincing argument.
\item Most of the bounds used to be presented with scaling factors 
depending on the mass of the exchanged squark, which was assumed to be 
around 100~GeV. Today, most of the lower bounds on the squark masses (at 
least in the first two generations) are an order of magnitude higher, 
leading to considerable relaxation in the constraints on the $R$-parity 
violating couplings.
\end{itemize}
\vspace*{-0.2in}  

Once we realise that the $R$-parity violating couplings can, in fact, be 
very large, we also note that they have no need to be aligned with the 
CKM matrix or even satisfy unitarity constraints, for these are 
parameters of the Lagrangian, and do not arise from the mixing of fields. 
The $R$-parity violating scenario, therefore, can satisfy all the 
conditions required for FCNC enhancement, viz. frustration of the GIM 
mechanism, non-MFV mixing terms and almost unconstrained coupling 
constants. We therefore choose, in this section, the $R$-parity violating 
model (RPV-MSSM) as a paradigm to illustrate how large top FCNC effects 
can be obtained.

As a first step to this study, we note that the $\lambda_{ijk}$, while 
interesting enough in their own right, are not relevant for the processes 
of interest in this article, since they do not appear with operators 
involving quark fields. We do not discuss them further in this article. 
The couplings of interest are the $\lambda'_{ijk}$ or the 
$\lambda''_{ijk}$ -- but obviously not both. We therefore list, in 
Table~5 below, the constraints on the $R$-parity violating couplings 
relevant for the processes under consideration, taking into account the 
current constraints on the masses of the sleptons and squarks. These, of 
course, still assume that one (or at most two) coupling(s) at a time is 
dominant. 

\begin{table}[h!] \centering \def\baselinestretch{1.2} \small 
\begin{tabular}{|llccrrr|} \hline
          &              &  Scales         &       & Upper    &  Sfermion    & Current  \\
         & Strongest Constraint   &  as mass &  Scaling     & bound    & mass & upper \\ 
 & arises from &  of & Exponent & (100~GeV)& ${\rm (GeV)}$ & bound \\ 
\hline\hline
\\ [-12pt]
$\lambda''_{212}$  &  &  &  &  &  &  \\ 
$\lambda''_{213}$  & Perturbativity \cite{Brahmachari:1994wd} & -- & -- & 1.24 & -- & 1.24 \\ 
$\lambda''_{223}$  &  &  &  &  &  &  \\ \hline
$\lambda''_{312}$ & $n-\bar{n}$ oscillation \cite{Sher:1994sp, Chemtob:2004xr} & $\tilde{d}_R$ & 2 & $10^{-3}$ & 1100 \cite{Agashe:2014kda} &  $0.1$ \\ 
$ \lambda''_{313}$ & & $\tilde{g}$ & $^1\!/_2$ &  & 1000 \cite{Aad:2015lea} &  $0.1$  \\ \hline
$ \lambda''_{323}$ & Bounds on $R_b (Z)$ \cite{Bhattacharyya:1995bw} & $\tilde{b}$ & 1 & 1.89 & 500 \cite{Aad:2013ija}  &  1.89  \\  
 &  & $\tilde{\tau}$ & -1 & 1.89 & 80 \cite{Agashe:2014kda}  &   \\ \hline
\end{tabular}
\def\baselinestretch{0.95}
\caption{\footnotesize Showing the experimental constraints on the 
$R$-parity-violating couplings $\lambda''_{2jk}$ and $\lambda''_{3jk}$ 
relevant for FCNC decays of the top quark. The notations and 
abbreviations follow the conventions of Table~\ref{tab:lpcoup}. }
\label{tab:lppcoup}
\def\baselinestretch{1.15}
\end{table}
\normalsize

A glance at the last column of Tables~\ref{tab:lpcoup} and 
\ref{tab:lppcoup} will make it clear that with the current values of 
sfermion masses, the constraints on most of the $R$-parity-violating 
couplings are very weak. These couplings can be as large as gauge 
couplings, or, is specific cases, much larger. Top FCNC processes will 
typically involve
\vspace*{-0.2in}
\begin{enumerate}
\item the products $\lambda'_{i2k}\lambda'_{i3k}$ for the decays $t \to 
c + h^0/Z$, where $i$ denotes the leptonic flavour in the loop and $k$ 
denotes the $d$-type quark flavour in the loop. For decays to $t \to u + 
h^0/Z$, we would get the products $\lambda'_{i1k}\lambda'_{i3k}$, but 
these have not been considered in this work.
\item the products $\lambda''_{2jk}\lambda''_{3jk}$ for the decays $t 
\to c + h^0/Z$, where $j$ denotes a quark flavour of the $u$-type and 
$k$ denotes a $d$-type quark flavour. As in the previous case, for the 
decays $t \to u + h^0/Z$, we would get products like 
$\lambda''_{1jk}\lambda''_{3jk}$, which are not considered in this work.
\end{enumerate}
\vspace*{-0.2in}
\begin{table} [H]
\centering
\begin{tabular}{cccccc}
$\lambda'_{121}\lambda'_{131}$ &
$\lambda'_{122}\lambda'_{132}$ &
$\lambda'_{123}\lambda'_{133}$ &
$\lambda'_{221}\lambda'_{231}$ &
$\lambda'_{222}\lambda'_{232}$ &
$\lambda'_{223}\lambda'_{233}$  \\ \hline
0.0269 & 0.0109 & $2.5\times 10^{-5}$ & 3.96 & 0.1 & 0.0033 \\ \hline
$\widetilde{e}_L$, $\widetilde{d}_R$ &  $\widetilde{e}_L$, $\widetilde{s}_R$ &
$\widetilde{e}_L$, $\widetilde{b}_R$ & $\widetilde{\mu}_L$, $\widetilde{d}_R$ &
$\widetilde{\mu}_L$, $\widetilde{s}_R$ & $\widetilde{\mu}_L$, $\widetilde{b}_R$ 
\\ [3mm]
$\lambda'_{321}\lambda'_{331}$ &
$\lambda'_{322}\lambda'_{332}$ &
$\lambda'_{323}\lambda'_{333}$ &
$\lambda''_{212}\lambda''_{312}$ &
$\lambda''_{213}\lambda''_{313}$ &
$\lambda''_{223}\lambda''_{323}$ \\ \hline
1.32 & 0.14 & 0.006 & 0.124 & 0.124 & 2.3436 \\ \hline
$\widetilde{\tau}_L$, $\widetilde{d}_R$ &  $\widetilde{\tau}_L$, $\widetilde{s}_R$ &
$\widetilde{\tau}_L$, $\widetilde{b}_R$ & $\widetilde{s}_R$ &
$\widetilde{b}_R$ & $\widetilde{b}_R$
\end{tabular}
\caption{\footnotesize Showing upper limits on the products of pairs of 
$R$-parity-violating couplings relevant for the decays $t \to c + 
h^0/Z$, as well as the sparticles exchanged in the loops for each 
combination. }
\label{tab:RPVprod}
\end{table}
\vspace*{-0.2in}

\begin{figure}[H]
\begin{minipage}{0.5\textwidth}
\begin{center}
\includegraphics[width=0.95\textwidth,angle=0]{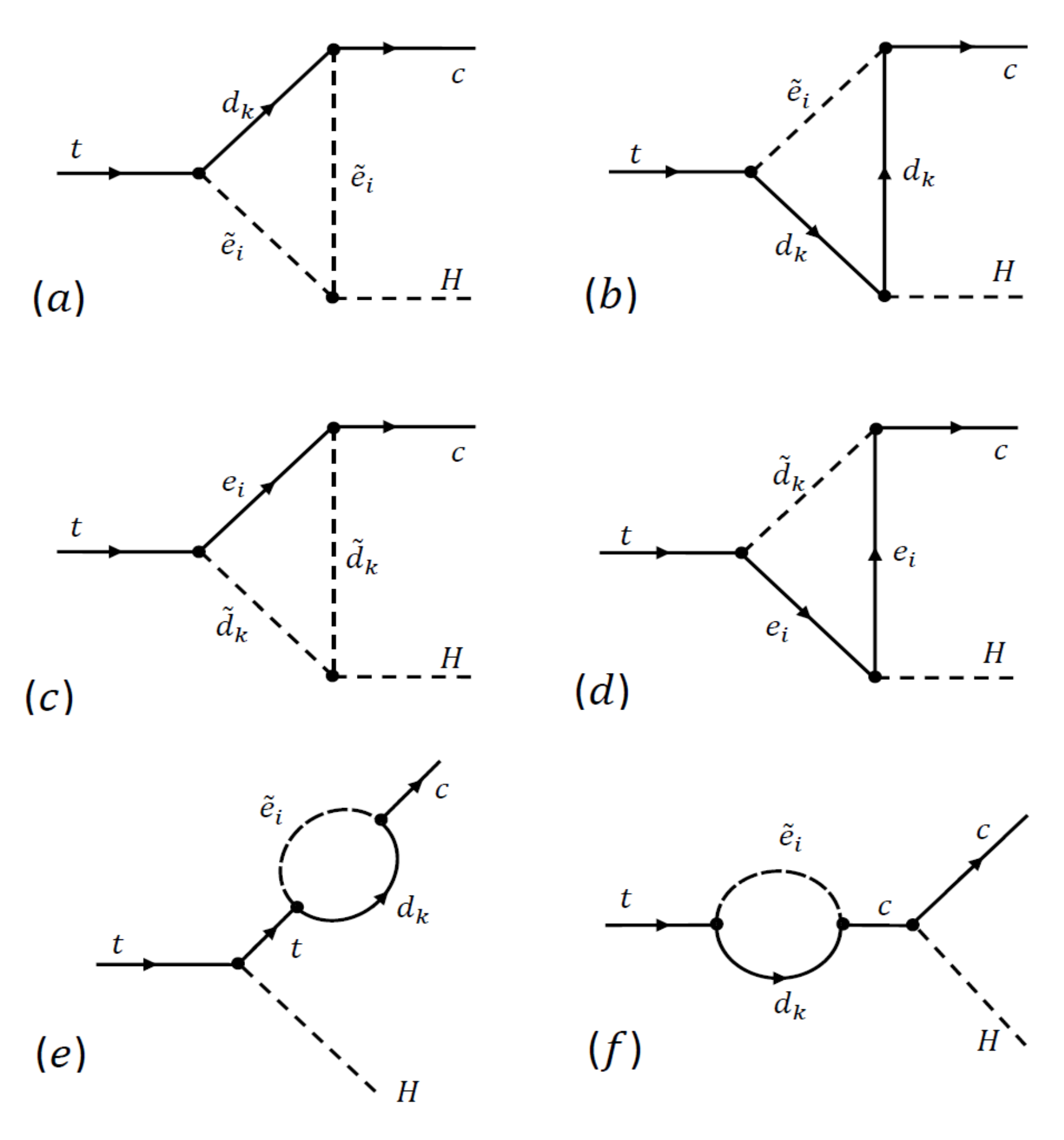}
\vspace*{-0.2in}
\includegraphics[width=0.95\textwidth,angle=0]{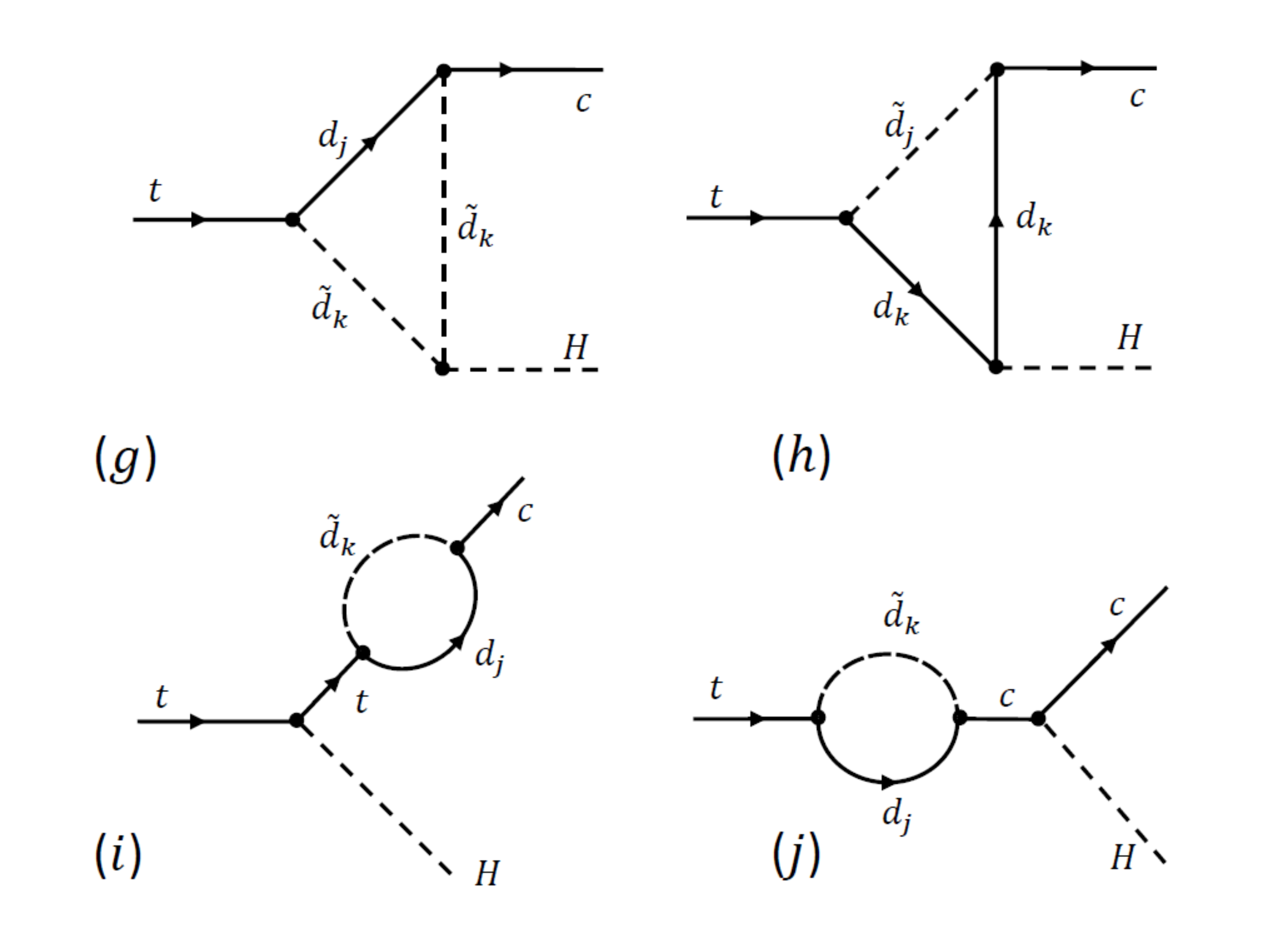}
\caption{\footnotesize\rm Further Feynman Diagrams leading to the decay 
$t \to c + H$ in the RPV MSSM.}
\label{fig:FeynRPV}
\end{center}
\end{minipage}
\begin{minipage}{0.5\textwidth}
In Table \ref{tab:RPVprod}, we list the pairs of $R$-parity-violating 
couplings which can lead to top FCNC processes, together with their 
maximum values corresponding to the last column of 
Tables~\ref{tab:lpcoup} and \ref{tab:lppcoup}. Some of the products are 
rather large, though staying well within the perturbative limit of 
$4\pi$.

\medskip

The Feynman diagrams which contribute to the FCNC decay $t \to c + h^0$ 
in the RPV-MSSM have been listed in Fig.~\ref{fig:FeynRPV}.  Of course, 
since the $R$-parity violating superpotential is added to the MSSM terms, 
we will also have contributions from all the diagrams in 
Figs.~\ref{fig:FeynSM} and \ref{fig:FeynSUSY}. However, these are always 
small -- as we have seen -- and hence the dominant contribution will 
arise from $R$-parity-violating terms alone.

\medskip
 
As before, the details of the calculation are given in Appendix~C. It is 
important to note that we have presented the diagrams mediated by 
$\lambda'$ couplings and the diagrams mediated by $\lambda''$ couplings 
in the same framework. The former include diagrams labelled ($a$)--($f$), 
while the latter are labelled ($g$)--($j$). The corresponding amplitudes 
will be added, as described in Appendix~C. However, there is no harm 
done, so long as we keep all the $\lambda''$ zero when the $\lambda'$
\end{minipage}
\end{figure}
\begin{figure}[H]
\def\baselinestretch{0.95}
\begin{center}
\includegraphics[height=5.4in,angle=0]{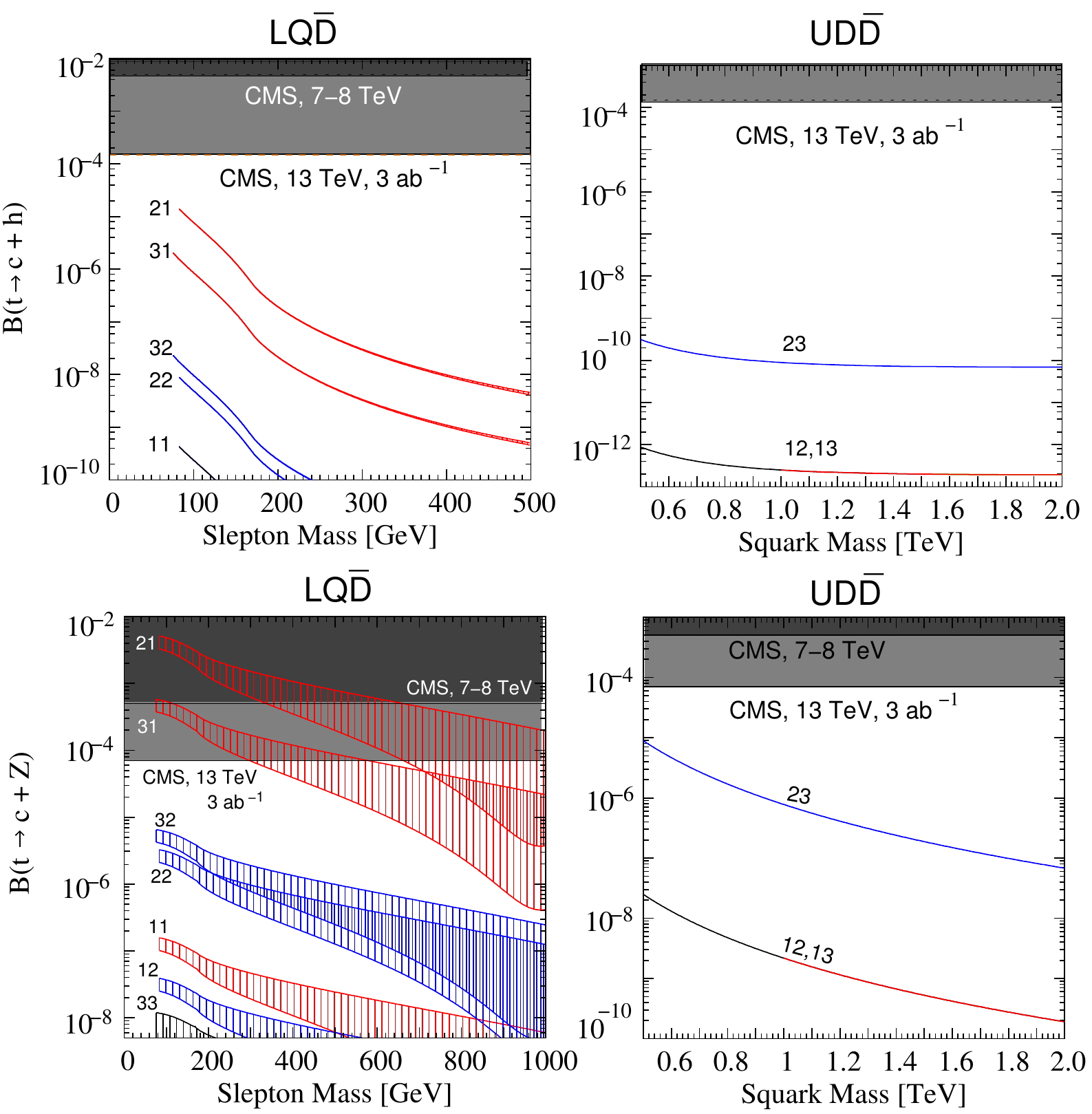}
\caption{\footnotesize\rm Illustrating the variation in the branching 
ratios ${\cal B}(t \to c + h^0)$ (upper panels) and ${\cal B}(t \to c + 
Z^0)$ (lower panels) with increase in the sfermion masses. For the panels 
on the left, which show branching ratios proportional to $(\lambda'_{i2k} 
\lambda'_{i3k})^2$ with the values of $ik$ marked next to each curve, the 
mass of the slepton $\widetilde{e}_{Li}$ is plotted along the abscissa, 
and the mass of the squark $\widetilde{d}_{Rk}$ is responsible for the 
thickness of the lines in the upper panel and the hatched region in the 
lower panel. For the panels on the right, which show branching ratios 
proportional to $(\lambda''_{2jk} \lambda''_{3jk})^2$ with the values of 
$jk$ marked next to each curve, the mass of the squark 
$\widetilde{d}_{Rk}$ is plotted along the abscissa. The dark (light) grey 
shaded regions represent the experimental bounds (discovery limits) from 
the LHC, operating at 7--8~TeV (13~TeV, projected).}
\label{fig:RPVBR}
\end{center}
\def\baselinestretch{1.15}
\end{figure}

   
\vspace*{-0.32in}
are non-zero, and vice versa. The variation of the branching ratios for 
$t \to c + h^0$ and $t \to c + Z$ as a function of the sfermion mass are 
given in Figure~\ref{fig:RPVBR}. The panels on the left, carrying the 
header ${\rm LQ}\bar{\rm D}$, correspond to the $\lambda'$ couplings and 
show values proportional to $(\lambda'_{i2k} \lambda'_{i3k})^2$. The 
relevant values of $ik$ are marked alongside each curve. To illustrate 
the variation with the sfermion masses, we have set these couplings to 
the experimental upper bounds in the last column of Table~5, and 
consequently, the products to the values in Table~7. These, of course, 
will be relaxed further if the concerned sfermion masses are taken 
higher, and would lead to even greater branching ratios, as may be 
imagined. However, we have chosen to keep the couplings fixed to the 
values given in Table~7. In a similar way, the panels on the right, 
carrying the header ${\rm UD}\bar{\rm D}$, correspond to the $\lambda''$ 
couplings, and show values proportional to the products $(\lambda''_{2jk} 
\lambda''_{3jk})^2$. Here, too, we have marked the values of $jk$ next to 
the relevant curves.

\vspace*{-0.4in} 

In Figure~\ref{fig:RPVBR}, the left panels illustrate the behaviour of 
the respective branching ratios with respect to variations in the mass of 
the slepton $\widetilde{e}_{Li}$. Each curve starts on the left from the 
current lower bound on the mass of this slepton and goes up to a TeV. The 
variation of the branching ratio as the mass of the squark 
$\tilde{d}_{Rk}$ varies from $1 - 2$~TeV is represented by the thickness 
of the lines in the upper panel, and by the hatched regions on the lower 
panel (with the upper boundary indicating a squark mass of 1~TeV). Quite 
obviously, the branching ratio ${\cal B}(t \to c + h^0)$ is hardly 
affected by changes in the squark mass, whereas the branching ratio 
${\cal B}(t \to c + Z^0)$ can vary by as much as an order of magnitude as 
the squark grows heavier.

The panels on the right in Figure~\ref{fig:RPVBR} illustrate the 
variation in the respective branching ratios with change in the mass of 
the squark $\tilde{d}_{Rk}$, which is the $b$-squark for $jk = 13, 23$ 
and the $c$-squark for $jk = 12$. The black and blue curves correspond to 
the former two cases and the red curves to the latter. In all the panels, 
the upper region shaded dark grey corresponds to bounds on the relevant 
branching ratios as set by the CMS Collaboration \cite{CMSfcnc}, while 
the regions shaded light grey corresponds to the projected discovery 
limits at the 13~TeV LHC, assuming an integrated luminosity of 
3000~fb$^{-1}$. It is immediately obvious, that even with all the 
enhancements available to us in a model with $R$-parity violation, the 
FCNC branching ratios of the $t$-quark are rather small. For $\lambda''$ 
couplings, in fact, these are hopelessly small -- in fact, so small, that 
even if we take the couplings to their perturbative limits, detection at 
the LHC will become a touch-and-go affair. The situation is better for 
$\lambda'$ couplings, largely because the sleptons can still be quite 
light. However, as the sleptons become heavier, the FCNC branching ratios 
fall rather fast and become unobservable. The best case arises for ${\cal 
B}(t \to c + Z^0)$ when we have the couplings $\lambda'_{221} 
\lambda'_{231}$ and $\lambda'_{321} \lambda'_{331}$, with exchange of 
$\widetilde{\mu}_L$ or $\widetilde{\tau}_L$ in the loops. In the former 
case, the data already available from the LHC constrains the slepton mass 
to be greater than about 350~GeV. In either case, a discovery at the 
13~TeV run is possible for a wide rage of slepton and squark masses. For 
other combinations of the $\lambda'$ couplings, the branching ratios are 
too small to be accessible at the LHC, even at the end of its run.

Before concluding this section, we may take up the issue mentioned 
before, that if the experimental bounds on the sfermion masses increase, 
the upper bounds on the $R$-parity-violating couplings can be relaxed 
still further. This may lead to higher values of the branching ratios is 
question, if the sfermion in the FCNC loop is not the same one which 
leads to relaxation of the bound. However, if we consider the only 
products which lead to sizable results as shown in 
Figure~\ref{fig:RPVBR}, viz. $\lambda'_{221} \lambda'_{231}$, 
$\lambda'_{321} \lambda'_{331}$ and $\lambda''_{223} \lambda''_{323}$, we 
can see from Table~7 that the values are, respectively, 3.96, 1.32 and 
2.34. The maximum value that we can push these to is, of course, $4\pi$, 
and that would provide enhancements in the branching ratios at the level 
of one or two orders of magnitude. This might just make it possible to 
observe the decay $t \to c + Z$ if it is mediated by $\lambda''_{223} 
\lambda''_{323}$, with more optimistic results for the $\lambda'$ 
couplings. However, only if some sign of $R$-parity-violating SUSY is 
found at the LHC will it be worthwhile to investigate further details in 
this regard.

\section{Summary and Conclusions}

This work was undertaken with a definite view, viz. to investigate FCNC 
decays of the $t$ quark which involve heavy particles that cannot be 
discovered directly at the LHC. Several such claims exist in the 
literature, but the results obtained are not always mutually consistent 
(see Table~8 below). By starting with a toy model which closely resembles 
the SM, we have shown that the extremely low values of FCNC branching 
ratios of the $t$-quark in the SM arise from three different sources. 
These are ($i$) the GIM cancellation between one-loop diagrams with 
different $d$-type quarks in the loop, ($ii$) the MFV paradigm, i.e. the 
choice of the hierarchical CKM matrix as the only source of flavour 
violation, and ($iii$) the choice of gauge couplings or their equivalent 
for the new particles. These result in suppression factors of the order 
$10^{-5}$, $10^{-4}$ and $10^{-1}$ respectively, driving the loop-induced 
branching ratios from their naive values around $10^{-4}$ to tiny values 
in the neighbourhood of $10^{-14}$. It follows, therefore, that a new 
physics model will be able to predict enhanced rates of these FCNC decays 
only to the extent that one or more of these conditions is violated. We 
then illustrate this set of conditions by considering ($a$) the cMSSM -- 
a model where GIM cancellation is frustrated, but MFV holds and the 
couplings can be modestly enhanced, and ($b$) the $R$-parity-violating 
extension of the cMSSM, where all three conditions can be broken. In 
vindication of the general principles enunciated above, the branching 
ratios in the cMSSM do not exceed $10^{-10}$ for $t \to c + h^0$ and 
$10^{-8}$ for $t \to c + Z^0$, whereas, for the case when $R$-parity is 
violated, we can predict them to be as large as $10^{-5}$ and $10^{-3}$ 
respectively. The last-mentioned values are well within the range of 
accessibility at the LHC, as illustrated in Figure~\ref{fig:RPVBR} above.

\begin{table}[htb!]
\def\baselinestretch{1.5}
 \centering
 \begin{tabular}{|llccccc|}
 \hline
Reference & Model & {GIM} & {MFV} & g & $t \to ch^0$ & $t \to cZ^0$\\
 \hline\hline
T.-J.~Gao {\it et al.}\cite{Gao:2014lva}& $\not{\!\!B}$, $\not{\!\!L}$ & $\times$ &$\times$ & $\times$ & $~10^{-4(5)}$ & --\\
J.-J.~Cao {\it et al.}\cite{Cao:2007dk} & MSSM & $\times$ & \checkmark & $\times$ & $10^{-5(9)}$ & $10^{-6(7)}$\\
B.~Mele \cite{Mele:1999zx} & MSSM & $\times$ & \checkmark & $\times$ & $10^{-5(9)}$ & $10^{-8(7)}$\\ 
S.~Bejar {\it et al.}\cite{Bejar:2001sj}& 2HDM Type-II & $\times$ & \checkmark  & $\times$ &$~10^{-4(9)}$ & --  \\ 
G.~Eilam {\it et al.}\cite{Eilam:2001dh}  & $\not{\!\!R}$ SUSY &$\times$ &$\times$ & \checkmark & $10^{-5(5)}$ & --   \\
C.~Yue {\it et al.}\cite{Yue:2003wd} & Non-universal $Z'$ &$\times$ & $\times$& $\times$ & -- &$10^{-6(4)}$   \\       
I.~Baum {\it et al.}\cite{Baum:2008qm} & $t$-quark 2HDM & $\times$ & \checkmark & $\times$ &$10^{-6(6)}$ & --   \\
A.~Dedes {\it et al.}\cite{Dedes:2014asa} & SUSY & $\times$& $\times$  & $\times$ & $10^{-7(7)}$& --\\ \hline
\end{tabular}
\def\baselinestretch{0.95}
\caption{\footnotesize\rm A few of the earlier calculations of FCNC 
decays of the top quark. Some of the results are in agreement with our 
predictions, given in parentheses. Those which are not are generally due 
to choice of vastly different parameters, which were allowed when these 
calculations were performed.}
\label{tab:refs}
\def\baselinestretch{1.15}
\end{table}

The utility of identifying the three suppression principles is well 
illustrated in Table~8, where some of the different models considered in 
the literature are classified according to the conditions which hold 
($\checkmark$) or are violated ($\times$). It is, then, easy to utilise 
the suppression levels quoted above to understand/criticise the branching 
ratios predicted by these authors. Moreover, we now have a quick rule of 
thumb to predict the branching ratios for FCNC decays of the top quark 
for {\it any} new physics model, for all that we need is to ask ourselves 
is which of these three conditions are applicable.

The appendices of this article present a collection of the formulae 
needed to perform the computations given in the text, in an explicit and 
user-friendly form, using the 'tHooft-Veltman and Passarino-Veltman 
formalism for one-loop integrals. The formulae are given in terms of 
certain generic couplings, so as to be easily usable to carry out similar 
computations in almost any new model of physics beyond the SM.
     
Finally, a word about the phenomenological implications of this work. It 
turns out that the use of the FCNC decays of the top quark is not such a 
ready handle to new physics at the LHC (and other high energy machines) 
as one might naively think, since the corresponding branching ratios are 
generally rather small. Even when we deviate almost completely from the 
SM, as exemplified in the $R$-parity-violating couplings, we require to 
be lucky to have just the right masses and pairing(s) of couplings in 
order to predict an observable effect. This is something which only the 
future can tell, and it is certain that the eyes of the entire high 
energy community will be turned to the results of the LHC, as they slowly 
unfold over the years to come.

\bigskip
{\small {\sl Acknowledgments}: The authors are grateful to A.~Dighe and 
T.S.~Roy for discussions and to D.~Bhatia and T.~Samui for help in 
computation. Thanks are also due to P.S.~Bhupal~Dev and D.K.~Ghosh for 
pointing out an error in Table~5. The work of SR is partially supported 
under project no. 2013/37C/37/BRNS by the Board of Research in Nuclear 
Studies, Government of India. }

\newpage

\def\baselinestretch{1.15}
\begin{appendices}

\section{Toy model amplitudes}

\subsection{The decay $t \to c + H$}

We consider the decay $t(k) \to c(p) + H(q)$. In the rest frame of the 
$t$ quark, we have $k = \left(m_t, \vec{0}\right)$ and
\begin{equation}
u(k,h_t)= \sqrt{\frac{m_t}{2}} 
\left( \begin{array}{cccc} 1 + h_t & 1 - h_t  &  0   &  0  \end{array} \right)^T 
\label{eqn:tspinor}
\end{equation}
where $h_t = \pm 1$ is the helicity of the $t$ quark. 
Now, the three-momenta $\vec{p}$ and $\vec{q}$ will be back-to-back, and 
we can choose this as the $z$-axis. In this case, we can write
\begin{equation}
p = \left(\begin{array}{cccc} E_c & 0  &  0   &  |\vec{p}|  \end{array} \right)
\qquad\qquad
q = \left(\begin{array}{cccc} E_H & 0  &  0   &  -|\vec{p}|  \end{array} \right) 
\end{equation}
where
\begin{equation}
|\vec{p}| \simeq E_c \simeq \frac{m_t^2 - M_H^2}{2m_t}
\qquad\qquad
E_H \simeq \frac{m_t^2 + M_H^2}{2m_t}
\label{eqn:fourmom}
\end{equation}
taking $m_c \ll m_t, M_H$. In the approximation, the $c$-quark wave function is
\begin{equation}
u(p,h_c) \simeq \sqrt{\frac{m_t^2 - M_H^2}{8m_t}}
\left( \begin{array}{cccc} 1 + h_c & 1 - h_c  &  1 + h_c   &  -1 + h_c  
\end{array} \right)^T 
\end{equation}
The helicity amplitudes ${\cal M}(h_c,h_t)$ now have the explicit form
\begin{equation}
{\cal M}(h_c,h_t) = \sum_{i=1}^3 \lambda _i \, {\cal A}_i(h_c,h_t)
\end{equation}
where $i$ runs over the three $d$-type quarks in the loop, 
$\lambda_i = V_{2i}V_{3i}^\ast$, 
and we parametrise
\begin{equation}
{\cal A}_i(h_c,h_t) = \bar{u}(p,h_c) i\left(F_{1i} P_L + F_{2i} 
P_R\right) u(k,h_t)
\end{equation}
where $P_L, P_R$ are the chiral projection operators
\begin{equation}
P_L = \frac{1}{2}\left(1 - \gamma_5\right)
\qquad\qquad
P_R = \frac{1}{2}\left(1 + \gamma_5\right)
\end{equation}
and $F_{1i}$ and $F_{2i}$ are form factors given below. Four helicity
amplitudes are possible, but the only non-vanishing ones are
\begin{eqnarray}
{\cal A}_i(+1,+1) &\simeq& \sqrt{m_t^2 - M_H^2} \ F_{1i} \nonumber \\
{\cal A}_i(-1,-1) &\simeq& \sqrt{m_t^2 - M_H^2} \ F_{2i} 
\label{eqn:helampS}
\end{eqnarray}
Each of the form factors $F_{1i}$ and $F_{2i}$ can be written
\begin{equation}
F_{ni} = F_{ni}^{(a)} + F_{ni}^{(b)} + F_{ni}^{(c)} + F_{ni}^{(d)}
\label{eqn:formfacS}
\end{equation}
where $n = 1,2$ and the superscripts refer to the graphs ($a$)--($d$) 
shown in Figure~1. These can be written in terms of the 
Passarino-'tHooft-Veltman functions, defined as Euclidean space integrals
\begin{eqnarray}
B_0(m_1,m_2; M) & = &  \int \frac{d^4k}{\pi^2} \, 
\frac{1}{(k^2 + m_1^2)\{ (k+p)^2 + m_2^2\}} \nonumber \\
p_\mu B_1(m_1,m_2; M) & = &  \int \frac{d^4k}{\pi^2} \, 
\frac{k_\mu}{(k^2 + m_1^2)\{ (k+p)^2 + m_2^2\}}
\end{eqnarray}
where $p^2 = - M^2$. In the $\overline{\rm MS}$ scheme, we can write
\begin{eqnarray}
B_0(m_1,m_2; M) & = & \Delta + {\widehat B}_0(m_1,m_2; M) \nonumber \\
B_1(m_1,m_2; M) & = & -\frac{1}{2} \Delta + {\widehat B}_1(m_1,m_2; M)
\end{eqnarray}
where the $\widehat{B}_{0,1}$ are finite. The divergent quantity is 
$\Delta = 2/ \varepsilon - \gamma + \ln 4\pi$ where $\varepsilon \to 0$ 
and $\gamma$ is the Euler-Mascheroni constant. We also have
\begin{eqnarray}
C_0(m_1,m_2,m_3;M_1,M_2,M_3) & = & \int \frac{d^4k}{\pi^2} \, 
\frac{1}{(k^2 + m_1^2)\{ (k+p_2)^2 + m_2^2\} \{ (k+p_2+p_3)^2 + m_3^2 \}} 
\nonumber \\
C_{11}p_{2\mu} + C_{12} p_{3\mu} & = & \int \frac{d^4k}{\pi^2} \, 
\frac{k_\mu}{(k^2 + m_1^2)\{ (k+p_2)^2 + m_2^2\} \{ (k+p_2+p_3)^2 + m_3^2 \}} 
\nonumber \\
\end{eqnarray}
where $p_1 = p_2 + p_3$ and $p_i^2 = - M_i^2$ for $i = 1,2,3$ and the 
$C_0, C_{11}$ and $C_{12}$ are naturally finite. In fact, the GIM 
cancellation ensures that all the form factors are finite and hence, we 
keep only the finite parts of the $B$ and $C$ functions. In terms of 
these, we can now compute the $F_1$ form factors
\footnotesize
\begin{eqnarray}
 F_{1i}^{(a)} 
 &=& -\frac{\xi \eta^2 }{16 \pi^2} m_c C^{(a)}_{12}\nonumber\\
 F_{1i}^{(b)} 
 &=& \frac{y_i m_i \eta^2 }{16\pi^2} m_t \big\{ 2 \left(C^{(b)}_{11} - C^{(b)}_{12} \right) +  C^{(b)}_0 \big\}\nonumber\\
 F_{1i}^{(c)} 
 &=& \frac{ y_c \eta^2 m_t}{16 \pi^2 (m_t^2 - m_c^2)} m_t {\tilde B}_1 (m_i,M_\omega;m_t) \nonumber\\
 F_{1i}^{(d)} 
 &=& -\frac{ y_t \eta^2 m_c}{16 \pi^2(m_t^2 - m_c^2)} m_c {\tilde B}_1 (m_i,M_\omega;m_c) 
\end{eqnarray}  
\normalsize
and the $F_2$ form factors
\footnotesize
\begin{eqnarray}
 F_{2i}^{(a)} 
 &=& -\frac{\xi \eta^2 }{16 \pi^2} m_t \left(C^{(a)}_{11} - C^{(a)}_{12}\right) \nonumber\\ 
 F_{2i}^{(b)} 
 &=& \frac{y_i m_i \eta^2 }{16\pi^2} m_c \left( C^{(b)}_0 + 2 C^{(b)}_{12} \right) \nonumber\\ 
 F_{2i}^{(c)} 
 &=& \frac{ y_c \eta^2 m_t}{16 \pi^2 (m_t^2 - m_c^2)} m_c {\tilde B}_1 (m_i,M_\omega;m_t) \nonumber\\ 
 F_{2i}^{(d)} 
 &=& -\frac{ y_t \eta^2 m_c}{16 \pi^2(m_t^2 - m_c^2)} m_t {\tilde B}_1 (m_i,M_\omega;m_c) 
\end{eqnarray}
\normalsize  
where
\begin{eqnarray}
C_X^{(a)} & = & C_X(m_i,M_\omega,M_\omega;m_c,m_t,M_H) \nonumber \\
C_X^{(b)} & = & C_X(M_\omega,m_i,m_i;m_c,m_t,M_H)
\end{eqnarray}
for $X = 0, 11, 12, 22$. These are evaluated using the public domain software
FF \cite{vanOldenborgh:1990yc}.

The Yukawa couplings $y$ are the SM ones, i.e.
\begin{equation}
y_i = \frac{gm_i}{2M_\omega}  \qquad  
y_t = \frac{gm_t}{2M_\omega}  \qquad  
y_c = \frac{gm_c}{2M_\omega} \ .
\label{eqn:SMYukawa}
\end{equation}

The above form factors can be used to evaluate the total form factors 
appearing in Eqn.~(\ref{eqn:formfacS}), which then enables us to compute 
the helicity amplitudes in Eqn.~(\ref{eqn:helampS}). These are then 
convoluted with the $\lambda$ factors in Eqn.~(\ref{eqn:Samplitudes}) and 
used to generate the squared and spin-summed/ averaged matrix element in 
Eqn.~(\ref{eqn:squaredS}). Plugging this into Eqn.~(\ref{eqn:widthS}) 
then produces the desired result.

\subsection{The decay $t \to c + Z$}

We now consider the decay $t(k) \to c(p) + Z(q)$. The kinematics is 
similar to the previous case, with $M_Z$ in place of $M_H$. Accordingly, 
the helicity spinor for the $c$-quark, in the approximation $m_c \ll 
m_t, M_Z$, is
\begin{equation}
u(p,h_c) \simeq \sqrt{\frac{m_t^2 - M_Z^2}{8m_t}}
\left( \begin{array}{cccc} 1 + h_c & 1 - h_c  &  1 + h_c   &  -1 + h_c  
\end{array} 
\right)^T 
\end{equation}
while the helicity spinor for the $t$-quark is identical with that in 
Eqn.~(\ref{eqn:tspinor}). In this case, we also have to consider the 
polarisation vector of the $Z$ boson, which, for the three helicity 
choices $h_Z = 0, \pm1$, has the form
\begin{equation}
\varepsilon(q,h_Z) = \left( \begin{array}{cccc}
-\frac{\left(1 - |h_Z| \right)|\vec{p}|}{M_Z} & \mp\frac{h_Z}{\sqrt{2}}
& -\frac{i|h_Z|}{\sqrt{2}} & \frac{\left(1 - |h_Z|\right)E_Z}{M_Z}
\end{array} \right)
\end{equation}
where, as in Eqn.~(\ref{eqn:fourmom}),
\begin{equation}
|\vec{p}| \simeq E_c \simeq \frac{m_t^2 - M_Z^2}{2m_t}
\qquad\qquad
E_Z \simeq \frac{m_t^2 + M_Z^2}{2m_t}
\end{equation}
The helicity amplitudes ${\cal M}(h_Z;h_c,h_t)$ now have the explicit form
\begin{equation}
{\cal M}(h_Z;h_c,h_t) = \sum_{i=1}^3 \lambda _i \, {\cal A}_i(h_Z;h_c,h_t)
\end{equation}
where $i$ runs over the three $d$-type quarks in the loop, 
$\lambda_i = V_{2i}V_{3i}^\ast$, and we parametrise
\begin{eqnarray}
{\cal A}_i(h_Z;h_c,h_t) & = & \bar{u}(p,h_c) \, i\Gamma^\mu \, u(k,h_t) \, 
\varepsilon_\mu^\ast(q) \nonumber \\
\Gamma^\mu & = & F_{1i}\gamma^\mu P_L + F_{2i}\gamma^\mu P_R 
+ iF_{3i} \sigma^{\mu\nu}q_\nu P_L
+ iF_{4i} \sigma^{\mu\nu}q_\nu P_R
\end{eqnarray}
Of the 12 possible helicity amplitudes, the only nonvanishing ones are 
\begin{eqnarray}
{\cal A}_i(+1;-1,+1) & = & -\sqrt{2(m_t^2 - M_Z^2)} 
\left[ F_{1i} - F_{4i}\left(E_Z +|\vec{p}|\right) 
\right]  \\
{\cal A}_i(-1;+1,-1) & = & -\sqrt{2(m_t^2 - M_Z^2)} 
\left[ F_{2i} -F_{3i}\left(E_Z +|\vec{p}|\right) 
\right] \nonumber \\
{\cal A}_i(0;+1,+1) & = & -\sqrt{m_t^2 - M_Z^2} 
\left[ F_{2i}\sqrt{\frac{E_Z +|\vec{p}|}{E_Z -|\vec{p}|}} - F_{3i} M_Z \right] 
\nonumber \\
{\cal A}_i(0;-1,-1) & = & -\sqrt{m_t^2 - M_Z^2} 
\left[ F_{1i}\sqrt{\frac{E_Z +|\vec{p}|}{E_Z -|\vec{p}|}} - F_{4i}M_Z \right] 
\nonumber 
\label{eqn:helampZ}
\end{eqnarray}
Each of the form factors can be written
\begin{equation}
F_{ni} = F_{ni}^{(a)} + F_{ni}^{(b)} + F_{ni}^{(c)} + F_{ni}^{(d)}
\label{eqn:formfacZ}
\end{equation}
where $n = 1,2,3,4$ and the superscripts refer to the graphs ($a$)--($d$) 
shown in Figure~1 (with $H$ replaced by $Z$). These can be written, as 
before, in terms of the Passarino-'tHooft-Veltman functions. We thus 
obtain the $F_1$ form factors
\footnotesize
\begin{eqnarray}
 F_{1i}^{(a)} &=& \frac{ \xi \eta^2  }{16 \pi^2}  \Big[m_t^2 (C_{11}^{(a)}
 -C_{12}^{(a)}+C_{21}^{(a)}-C_{23}^{(a)}) + m_c m_t (C_{12}^{(a)} + C_{23}^{(a)}) 
 - C_{24}^{(a)} \Big] \nonumber\\
 F_{1i}^{(b)} &=& \frac{\eta^2 }{16\pi^2} \Big[\alpha_i m_i^2 C_0^{(b)} 
+ \beta_i \left( B_0 - M_\omega^2 C_0^{(b)} + m_t^2 (C_{21}^{(b)} - C_{23}^{(b)}) 
- m_c^2 C_{12}^{(b)} - 2 C_{24}^{(b)}\right) \nonumber \\ 
&& \hspace*{2.0in} + \beta_i m_c m_t \left(\frac{3}{2}(C_0^{(b)} + C_{11}^{(b)}) + 
C_{12}^{(b)} + C_{23}^{(b)}\right)\Big] \nonumber\\
 F_{1i}^{(c)} &=& -\frac{ \eta^2}{16 \pi^2 (m_t^2 - m_c^2)}\Big[\alpha m_c^2 B_1 
 (m_i,M_\omega;m_c) + \beta m_c m_t B_1 (m_i,M_\omega;m_c) \Big] \nonumber\\
 F_{1i}^{(d)} &=& \frac{ \eta^2}{16 \pi^2 (m_t^2 - m_c^2)}\Big[\alpha m_t^2 B_1 
 (m_i,M_\omega;m_t) + \beta m_c m_t B_1 (m_i,M_\omega;m_t)\Big]
\end{eqnarray}
\normalsize
the $F_2$ form factors
\footnotesize
\begin{eqnarray}
 F_{2i}^{(a)} &=& -\frac{ \xi \eta^2 }{16 \pi^2} \Big[m_t^2 (C_{11}^{(a)}-C_{12}
 ^{(a)}+C_{21}^{(a)}-C_{23}^{(a)}) -m_c m_t (C_{12}^{(a)} + C_{23}^{(a)}) - C_{24}
 ^{(a)} \Big] \nonumber\\
 F_{2i}^{(b)} &=& -\frac{\eta^2 }{16\pi^2}\Big[\alpha_i m_i^2 C_0^{(b)} + \beta_i 
 \left( B_0 - M_\omega^2 C_0^{(b)} + m_t^2 (C_{21}^{(b)} - C_{23}^{(b)}) - m_c^2 
 C_{12}^{(b)} - 2 C_{24}^{(b)}\right) \nonumber \\ 
&& \hspace*{2.0in} - \beta_i m_c m_t \left(\frac{3}{2}(C_0^{(b)} + C_{11}^{(b)}) + 
C_{12}^{(b)} + C_{23}^{(b)}\right)\Big] \nonumber\\
 F_{2i}^{(c)} &=& \frac{ \eta^2}{16 \pi^2 (m_t^2 - m_c^2)}\Big[\alpha m_c^2 B_1 
 (m_i,M_\omega;m_c) - \beta m_c m_t B_1 (m_i,M_\omega;m_c) \Big] \nonumber\\
 F_{2i}^{(d)} &=& -\frac{ \eta^2}{16 \pi^2 (m_t^2 - m_c^2)}\Big[\alpha m_t^2 B_1 
 (m_i,M_\omega;m_t) - \beta m_c m_t B_1 (m_i,M_\omega;m_t)\Big]
\end{eqnarray}
\normalsize
the nonvanishing $F_3$ form factors
\footnotesize
\begin{eqnarray}
 F_{3i}^{(a)} &=& -\frac{ \xi \eta^2 }{16 \pi^2} \Big[m_t (C_{11}^{(a)}-C_{12}
 ^{(a)}+C_{21}^{(a)}-C_{23}^{(a)}) + m_c (C_{12}^{(a)} + C_{23}^{(a)}) \Big] \\
 F_{3i}^{(b)} &=& -\frac{\eta^2 }{16\pi^2}\beta_i \Big[m_t \left(C_{11}^{(b)} - 
 C_{12}^{(b)} + C_{21}^{(b)} - C_{23}^{(b)}\right) + m_c \left(\frac{1}{2}
 (C_0^{(b)} + C_{11}^{(b)}) + C_{12}^{(b)} + C_{23}^{(b)}\right) \Big] \nonumber
\end{eqnarray}
\normalsize
and the nonvanishing $F_4$ form factors
\footnotesize
\begin{eqnarray}
 F_{4i}^{(a)} &=& -\frac{ \xi \eta^2 }{16 \pi^2} \Big[m_t (C_{11}^{(a)}-C_{12}
 ^{(a)}+C_{21}^{(a)}-C_{23}^{(a)}) - m_c (C_{12}^{(a)} + C_{23}^{(a)})  \Big] \\
 F_{4i}^{(b)} &=& -\frac{\eta^2 }{16\pi^2}\beta_i \Big[m_t \left(C_{11}^{(b)} - 
 C_{12}^{(b)} + C_{21}^{(b)} - C_{23}^{(b)}\right) - m_c \left(\frac{1}{2}(C_0^{(b)} 
 + C_{11}^{(b)}) + C_{12}^{(b)} + C_{23}^{(b)}\right) \Big] \nonumber 
\end{eqnarray}
\normalsize
In the above,
\begin{eqnarray}
 B_0 &=& B_0(m_i,m_i; M_Z) \nonumber\\
 C_X^{(a)} &=& C_X (m_i, M_\omega, M_\omega; m_c, m_t, M_Z) \nonumber \\
 C_X^{(b)} &=& C_X (M_\omega, m_i, m_i; m_c, m_t, M_Z) \nonumber
\end{eqnarray}
where $X = {0,11,12,21,23,24}$. The $Zd_i\bar{d}_i$ couplings are $\alpha 
= - \frac{1}{2} - 2 Q {\rm \sin}^2 \theta_W$ and $\beta = \frac{1}{2}$ 
where $Q = -1/3$ is the charge of the down-type quark.

Once we have these form factors, we sum them up using 
Eqn.~(\ref{eqn:formfacZ}) and use them to calculate the helicity 
amplitudes in Eqn.~(\ref{eqn:helampZ}). These are then convoluted with 
the $\lambda_i$ factors in Eqn.~(\ref{eqn:Zamplitudes}) and used to 
calculate the squared spin-summed/ averaged matrix element in 
Eqn.~(\ref{eqn:squaredZ}). Finally this is used in 
Eqn.~(\ref{eqn:widthZ}) to produce the partial width.

\section{SM and cMSSM amplitudes}

\subsection{The decay $t \to c + H$}

In the Standard Model, as in the toy model, the decay $t \to c + H$ can 
be parametrised in terms of the two nonvanishing helicity amplitudes of 
Eqn.~(\ref{eqn:helampS}). The calculation follows the lines of the toy 
model, except that the diagrams are those of Figure~\ref{fig:FeynSM} 
instead of Figure~\ref{fig:Feyntoy}. Thus, in this Appendix, we only 
require to list the form factors, diagram-wise.

It is convenient, in evaluating these diagrams, to define a set of 
general vertices:

\begin{table} [!htb]
\centering
\begin{tabular}{rcl}
$\overline{u}_i u_i h$ & : & $ig (A^h_{ui} P_L + B^h_{ui} P_R)$  \\ [2mm]
$\overline{d}_i d_i h$ & : & $ig (A^h_{di} P_L + B^h_{di} P_R)$  \\ [2mm]
$h(-q)\phi^+ (p)W_\mu^-$ & : & $ig \alpha^h_\phi (p+q)_\mu$  \\ [2mm]
$h \phi^+ \phi'^-$ & : & $ig M_W \beta_{\phi\phi'}^h$  \\ [2mm]
$h W_\mu^+ W_\nu^-$& : & $ig M_W \omega_h g_{\mu\nu}$  \\ [2mm]
$\overline{u}_i d_j \phi^+$& : & $ig \left(X_{ij}^\phi P_L + Y_{ij}^\phi P_R\right)$  \\
\end{tabular}
\end{table}
in terms of a set of coupling constants $A^h_{ui}$, $B^h_{ui}$, 
$A^h_{di}$, $B^h_{di}$, $\alpha^h_\phi$, $\beta^h_{\phi\phi'}$, 
$\omega_h$, $X^\phi_{ij}$ and $Y^\phi_{ij}$. In order to obtain numerical 
values in the SM, we need to substitute the coupling constants according 
to the table given below.

\begin{table} [!htb]
\centering
\begin{tabular}{rccccccccc}
coupling: & $A^h_{ui}$ & $B^h_{ui}$ & $A^h_{di}$ & $B^h_{di}$ & $\alpha^h_\phi$ & 
$\beta^h_{\phi\phi'}$ & $\omega_h$ & $X^\phi_{ij}$ & $Y^\phi_{ij}$ \\ \hline\hline
SM value : & \Large $\frac{m_i}{2 M_W}$ & \Large $\frac{m_i}{2 M_W}$ 
& \Large $\frac{m_i}{2 M_W}$ & \Large $\frac{m_i}{2 M_W}$ & \Large $-\frac{1}{2}$ 
& \Large $-\frac{m_h^2}{M_W^2}$ & $1$ & \Large $\frac{m_i}{\sqrt{2}M_W}$ 
& $-\frac{m_j}{\sqrt{2}M_W}$ \\ \hline
\end{tabular}
\end{table}

In terms of these, the form factors of type $F_1$ are
\footnotesize
\begin{eqnarray}
 F_{1i}^{(a)} &=& \frac{ig^3  M_W \omega_h} {16 \pi^2}  m_c C_{12}^{(a)} \nonumber\\
 F_{1i}^{(b)} &=& \frac{ig^3 \alpha_{G^+}^h}{16 \sqrt{2} \pi^2} \left[ X_{ci}^G \left( (m_t^2 - 2 M_h^2)(C_{11}^{(b)} - C_{12}^{(b)}) - B_0(2,3) + m_i^2 C_0^{(b)} + 2 m_c^2 C_{11}^{(b)}\right) -m_i m_c Y_{ci}^G (C_{12}^{(b)} + 2 C_0^{(b)})\right] \nonumber\\
 F_{1i}^{(c)} &=& \frac{ig^3 \alpha_{G^+}^h}{16 \sqrt{2} \pi^2} \left[  X_{ti}^G \left( 2 m_t^2 C_{11}^{(c)}  - 2 m_S^2 C_{12}^{(c)}  + m_c^2 C_{12}^{(c)} - B_0(2,3) + m_i^2 C_0^{(c)}\right) - m_i m_t Y_{ti}^G \left(C_{11}^{(c)}-C_{12}^{(c)} + 2 C_0^{(c)}\right)\right] \nonumber\\
 F_{1i}^{(d)} &=& -\frac{ig^3 M_W \beta_{G G}^h}{16 \pi^2} \left[ m_t X_{ci}^G X_{ti}^G (C_{11}^{(d)} - C_{12}^{(d)}) - m_i X_{ci}^G Y_{ti}^G C_0^{(d)} + m_c Y_{ci}^G Y_{ti}^G C_{12}^{(d)}\right] \nonumber\\
 F_{1i}^{(e)} &=& \frac{ig^3  m_i }{16 \pi^2} m_c \left[ (A_{di}^h + B_{di}^h)C_{12}^{(e)} + B_{di}^h C_0^{(e)}\right] \nonumber\\
  F_{1i}^{(f)} &=& -\frac{ig^3}{16 \pi^2} \left[ z_1 \left( B_0 (2,3) - M_W^2 C_0^{(f)} \right) - z_3 C_0^{(f)} - m_t z_5 \frac{X_{ti}^G}{Y_{ti}^G} (C_{11}^{(f)} - C_{12}^{(f)}) - m_c z_2 C_{12}^{(f)} \right] Y_{ti}^G \nonumber\\
  F_{1i}^{(g)} &=& \frac{ig^3}{16 \pi^2 (m_t^2-m_c^2)} m_t^2 A_{uc}^h B_1^{(g)} \nonumber\\
  F_{1i}^{(h)} &=& \frac{ig^3}{16 \pi^2 (m_t^2 - m_c^2)} \left[ m_t\left(m_t A_{uc}^h Y_{ci}^G   + m_c A_{uc}^h \frac{X_{ti}^G}{Y_{ti}^G} X_{ci}^G \right) B_1^{(h)} 
 - m_i \left(m_t A_{uc}^h Y_{ci}^G \frac{X_{ti}^G}{Y_{ti}^G} + m_c A_{uc}^h X_{ci}^G \right) B_0^{(h)} \right] Y_{ti}^G \nonumber\\
  F_{1i}^{(i)} &=& -\frac{ig^3}{16 \pi^2 (m_t^2 - m_c^2)} m_c m_t A_{ut}^h B_1^{(i)} \nonumber\\
  F_{1i}^{(j)} &=& -\frac{ig^3}{16 \pi^2 (m_t^2-m_c^2) } \left[ m_c X_{ti}^G (m_c X_{ci}^G B_1^{(j)} - m_i Y_{ci}^G B_0^{(j)}) + m_t Y_{ti}^G (m_c Y_{ci}^G B_1^{(j)} - m_i X_{ci}^G B_0^{(j)}) \right] A_{ut}^h
\end{eqnarray}
\normalsize
and the form factors of type $F_2$ are
\footnotesize
\begin{eqnarray}
 F_{2i}^{(a)} &=& \frac{ig^3  M_W \omega_h} {16 \pi^2}  m_t (C_{11}^{(a)} - C_{12}^{(a)}) \nonumber\\
 F_{2i}^{(b)} &=& \frac{ig^3 \alpha_{G^+}^h}{16 \sqrt{2} \pi^2} \left[ X_{ci}^G m_c m_t (C_{12}^{(b)} - 2 C_{11}^{(b)}) + Y_{ci}^G m_i m_t (C_0^{(b)} - C_{11}^{(b)} + C_{12}^{(b)})\right] \nonumber\\
 F_{2i}^{(c)} &=& \frac{-ig^3 \alpha_{G^+}^h}{16 \sqrt{2} \pi^2} \left[ X_{ti}^G m_c m_t (C_{12}^{(c)} - 2 C_{11}^{(c)}) + Y_{ti}^G m_i m_t (C_0^{(c)} - C_{11}^{(c)} + C_{12}^{(c)})\right] \nonumber\\
 F_{2i}^{(d)} &=& -\frac{ig^3 M_W \beta_{G G}^h}{16 \pi^2} \left[ m_t Y_{ci}^G Y_{ti}^G (C_{11}^{(d)} - C_{12}^{(d)}) - m_i X_{ti}^G Y_{ci}^G C_0^{(d)} + m_c X_{ci}^G X_{ti}^G C_{12}^{(d)}\right] \nonumber\\
 F_{2i}^{(e)} &=& \frac{ig^3  m_i }{16 \pi^2} m_t \left[ (A_{di}^h + B_{di}^h)(C_{11}^{(e)} -C_{12}^{(e)}) + A_{di}^h C_0^{(e)}\right] \nonumber\\
 F_{2i}^{(f)} &=& -\frac{ig^3}{16 \pi^2} \left[ z_4 \left( B_0 (2,3) - M_W^2 C_0^{(f)} \right) - z_6 C_0^{(f)} - m_t z_2 \frac{Y_{ti}^G}{X_{ti}^G} (C_{11}^{(f)} - C_{12}^{(f)}) - m_c z_5 C_{12}^{(f)} \right] X_{ti}^G  \nonumber
\end{eqnarray}
\begin{eqnarray}
 F_{2i}^{(g)} &=& \frac{ig^3}{16 \pi^2 (m_t^2-m_c^2)} m_c m_t B_{uc}^h B_1^{(g)} \nonumber\\
 F_{2i}^{(h)} &=& \frac{ig^3}{16 \pi^2 (m_t^2 - m_c^2)} \left[ m_t\left(m_t B_{uc}^h X_{ci}^G + m_c B_{uc}^h \frac{Y_{ti}^G}{X_{ti}^G} Y_{ci}^G \right) B_1^{(h)}
 - m_i \left(m_t B_{uc}^h X_{ci}^G \frac{Y_{ti}^G}{X_{ti}^G} + m_c B_{uc}^h Y_{ci}^G \right) B_0^{(h)} \right] X_{ti}^G \nonumber\\
 F_{2i}^{(i)} &=& -\frac{ig^3}{16 \pi^2 (m_t^2 - m_c^2)} m_c^2 B_{ut}^h B_1^{(i)} \nonumber\\
 F_{2i}^{(j)} &=& -\frac{ig^3}{16 \pi^2 (m_t^2-m_c^2) } \left[ m_c Y_{ti}^G (m_c Y_{ci}^G B_1^{(j)} - m_i X_{ci}^G B_0^{(j)}) + m_t X_{ti}^G (m_c X_{ci}^G B_1^{(j)} - m_i Y_{ci}^G B_0^{(j)}) \right] B_{ut}^h
\end{eqnarray}
\normalsize
As in the previous section, the superscripts refer to the diagrams marked 
($a$)--($j$) in Figure~\ref{fig:FeynSM}. 

In the above, we have used the functions
\begin{eqnarray}
  C_X^{(a)} &=& C_X (m_i, M_W, M_W; m_c, m_t, M_h) \qquad\qquad
  B_1^{(g)} = B_1(m_i, M_W; m_t) \nonumber \\
  C_X^{(b)} &=& C_X (m_i, M_W, M_W; m_c, m_t, M_h) \qquad\qquad
  B_1^{(h)} = B_1(m_i, M_W;m_t) \nonumber \\
  C_X^{(c)} &=& C_X (m_i, M_W, M_W; m_c, m_t, M_h) \qquad\qquad
  B_0^{(h)} = B_0(m_i, M_W;m_t) \nonumber \\
  C_X^{(d)} &=& C_X (m_i, M_W, M_W; m_c, m_t, M_h) \qquad\qquad
  B_1^{(i)} = B_1(m_i,M_W; m_c) \nonumber \\
  C_X^{(e)} &=& C_X (M_W, m_i, m_i; m_c, m_t, M_h) \qquad\qquad
\ \,  B_1^{(j)} = B_1(m_i, M_W; m_c) \nonumber \\
  C_X^{(f)} &=& C_X (M_W, m_i, m_i; m_c, m_t, M_h) \qquad\qquad
\ \,  B_0^{(j)} = B_0(m_i, M_W; m_c)
\end{eqnarray}
where $X = {0,11,12,21,23,24}$, as before, and defined a set of effective 
couplings
\begin{eqnarray}
 z_1 &=& X_{ci}^G B_{di}^h \nonumber \\
 z_2 &=& m_t Y_{ci}^G \frac{X_{ti}^G}{Y_{ti}^G} A_{di}^h + m_c X_{ci}^G B_{di}^h + m_i Y_{ci}^G B_{di}^h + m_i Y_{ci}^G A_{di}^h \nonumber \\
 z_3 &=& m_t \frac{X_{ti}^G}{Y_{ti}^G} A_{di}^h (m_i X_{ci}^G + m_c Y_{ci}^G) + m_i^2 A_{di}^h X_{ci}^G + m_i m_c A_{di}^h Y_{ci}^G \nonumber \\
 z_4 &=& Y_{ci}^G A_{di}^h \nonumber \\
 z_5 &=& m_t X_{ci}^G \frac{Y_{ti}^G}{X_{ti}^G} B_{di}^h + m_c Y_{ci}^G A_{di}^h + m_i X_{ci}^G A_{di}^h + m_i X_{ci}^G B_{di}^h \nonumber \\
 z_6 &=& m_t \frac{Y_{ti}^G}{X_{ti}^G} B_{di}^h (m_i Y_{ci}^G + m_c X_{ci}^G) + m_i^2 B_{di}^h Y_{ci}^G + m_i m_c B_{di}^h X_{ci}^G 
\end{eqnarray}
These form factors can now be combined, using $F_{ni} = \sum_{A=a}^j 
F_{ni}^{A}$ for $n = 1,2$ and the results substituted into 
Eqn.~(\ref{eqn:helampS}) as before.

When we come to consider the cMSSM, the SM contributions will not only 
involve modifications of the SM couplings given above, but will also be 
enhanced by contributions from the additional eight diagrams in 
Figure~\ref{fig:FeynSUSY}, which involve superparticles in the loops. 
These involve some additional couplings which are parametrised in a 
general way as
\begin{table} [!htb]
\centering
\begin{tabular}{rcl}
$\chi^+_i \chi^-_j h$ & : & 
$ig \left(A_{ij}^h P_L + B_{ij}^h P_R\right)$ \\ [2mm]
$\chi^+_i \chi^-_j Z^\mu$ & : & 
$ig \gamma^\mu \left(A_{ij}^Z P_L + B_{ij}^Z P_R\right)$ \\ [2mm]
$\tilde{d}^* \tilde{d} h$ & : & 
$ig M_W \beta_{\tilde{d}\tilde{d}}^h$  \\ [2mm]
$\tilde{d}(p) \tilde{d}^*(q) Z^\mu$ & : & 
$ig \alpha^{\tilde{d}}_{\tilde{d}} (p+q)^\mu$  \\ [2mm]
$\tilde{d}_i^*\overline{u}_k \chi^+_j $ & : & 
$ig \left(X_{kj}^i P_L + Y_{kj}^i P_R\right) $ 
\end{tabular}
\end{table}
\vspace{-0.2in}

in terms of an additional set of coupling constants $A_{ij}^h$, 
$B_{ij}^h$, $A_{ij}^Z$, $B_{ij}^Z$, $\beta_{\tilde{d}\tilde{d}}^h$, 
$\alpha^{\tilde{d}}_{\tilde{d}}$, $X_{kj}^i$, $Y_{kj}^i$. For a numerical 
analysis, we require to take the full set of coupling constants as given 
in the table below.

\begin{table} [!htb]
\centering
\begin{tabular}{rcccc}
\hline
coupling : & $A^h_{ui}$ & $B^h_{ui}$ & $A^h_{di}$ & $B^h_{di}$
\\ \hline\hline \\ [-12pt]
cMSSM : & \Large $-\frac{m_i \cos\alpha}{2 M_W \sin\beta}$
& \Large $-\frac{m_i \cos\alpha}{2 M_W \sin\beta}$
& \Large $\frac{m_i \sin\alpha}{2 M_W \cos\beta}$ 
& \Large $\frac{m_i \sin\alpha}{2 M_W \cos\beta}$ 
\\ \hline \\ [-10pt]
coupling: & $\alpha^h_{G^+}$ & $\alpha^h_{h^+}$ & $\beta_{G^+G^-}^h$
& $\beta_{G^+h^+}^h$
\\ \hline\hline \\ [-12pt]
cMSSM : & $-\frac{1}{2} \sin(\beta - \alpha)$ & $-\frac{1}{2} \cos(\beta - \alpha) $
& \Large $\frac{\cos 2\beta \sin(\alpha+\beta)}{2 \cos^2\theta_W}$
& \Large $\frac{\cos(\beta - \alpha) (m_{h^+}^2 - m_{h^0}^2)}{2 M_W^2}$
\\ \hline \\ [-10pt]
coupling: & $\beta_{h^+h^-}^h$ & $\omega_h$ & $X_{ij}^{G^+}$ & $Y_{ij}^{G^+}$
\\ \hline\hline \\ [-12pt]
cMSSM value : & $-\sin(\beta-\alpha)$ & $\sin(\beta-\alpha)$
& \Large $\frac{m_i}{\sqrt{2}M_W}$ & \Large $-\frac{m_j}{\sqrt{2}M_W}$ \\ 
& \Large $-\frac{\cos 2\beta \sin(\alpha+\beta)}{2 \cos^2\theta_W}$ & & & 
\\ \hline \\ [-10pt]
coupling: & $X_{ij}^{h^+}$ & $Y_{ij}^{h^+}$ & &
\\ \hline\hline \\ [-12pt]
cMSSM value : & \Large $\frac{m_i \cot\beta}{\sqrt{2}M_W}$ 
& \Large $\frac{m_j tan \beta}{\sqrt{2}M_W}$ & &
\\ \hline \\ [-10pt]
coupling:
& $A_{ij}^h$ & $B_{ij}^h$ & $A_{ij}^Z$ & $B_{ij}^Z$ 
\\ \hline\hline \\ [-12pt]
cMSSM value : & $Q_{ij}^* \sin \alpha - S_{ij}^* \cos\alpha$ 
& $Q_{ji} \sin \alpha - S_{ji} \cos\alpha$ & $Q_{ij}^V$ & $Q_{ij}^U$ 
\\ \hline \\ [-10pt]
coupling: & $\beta_{\tilde{d}\tilde{d}}^h$ & $\alpha^{\tilde{d}}_{\tilde{d}}$ 
& $X_{kj}^i$ & $Y_{kj}^i$
\\ \hline\hline \\ [-12pt]
cMSSM value : & \Large $-\left(\frac{1}{2}-\frac{\sin^2\theta_W}{3}\right)$ 
& \Large $\frac{1-\frac{2}{3}\sin^2\theta_W}{2\cos\theta_W}$ & $0$  & $U_{j1}$ \\ 
& \Large $+\frac{\sin(\alpha + \beta)}{\cos^2 \theta_W}$ & & & \\
\hline
\end{tabular}
\end{table}
where, in terms of the chargino mixing matrices $U$ and $V$, 
\begin{align*}
Q_{ij} &= \frac{1}{\sqrt{2}} U_{i2} V_{j1};\quad S_{ij} = \frac{1}{\sqrt{2}} U_{i1} V_{j2}\\
Q_{ij}^U &= - U_{i1} U_{j1}^* - \frac{1}{2} U_{i2} U_{j2}^* + \delta_{ij} sin^2 \theta_W ; \quad Q_{ij}^V = - V_{i1} V_{j1}^* - \frac{1}{2} V_{i2} V_{j2}^* + \delta_{ij} \sin^2 \theta_W 
\end{align*}

Evaluating the Feynman diagrams of Figs.~\ref{fig:FeynSM} and 
\ref{fig:FeynSUSY} now leads to the $F_1$ form factors
\footnotesize
\begin{eqnarray}
F_{1i}^{(k)} &=& \frac{ig^3 \alpha_{h^+}^h}{16 \sqrt{2} \pi^2} \left[ X_{ci}^{h^+} \left( (m_t^2 - 2 M_{h^+}^2)(C_{11}^{(k)} - C_{12}^{(k)}) - B_0(2,3) + m_i^2 C_0^{(k)} + 2 m_c^2 C_{11}^{(k)}\right) -m_i m_c Y_{ci}^{h^+} (C_{12}^{(k)} + 2 C_0^{(k)})\right] \nonumber\\
F_{1i}^{(l)} &=& \frac{ig^3 \alpha_{h^+}^h}{16 \sqrt{2} \pi^2} \left[ X_{ti}^{h^+} \left( 2 m_t^2 C_{11}^{(l)}  - 2 M_{h^+}^2 C_{12}^{(l)}  + m_c^2 C_{12}^{(l)} - B_0(2,3) + m_i^2 C_0^{(l)}\right) - m_i m_t Y_{ti}^{h^+} (C_{11}^{(l)}-C_{12}^{(l)} + 2 C_0^{(l)})\right] \nonumber\\
F_{1i}^{(m)} &=& -\frac{ig^3 M_W \beta_{h^+ h^-}^h}{16 \pi^2} \left[ m_t X_{ci}^{h^+} X_{ti}^{h^+} (C_{11}^{(m)} - C_{12}^{(m)}) - m_i X_{ci}^{h^+} Y_{ti}^{h^+} C_0^{(m)} + m_c Y_{ci}^{h^+} Y_{ti}^{h^+} C_{12}^{(m)}\right] \nonumber\\
F_{1i}^{(n)} &=& -\frac{ig^3 M_W \beta_{G^+ h^-}^h}{16 \pi^2} \left[ m_t X_{ci}^{G} X_{ti}^{h^+} (C_{11}^{(n)} - C_{12}^{(n)}) - m_i X_{ci}^{G} Y_{ti}^{h^+} C_0^{(n)} + m_c Y_{ci}^{G} Y_{ti}^{h^+} C_{12}^{(n)}\right] \nonumber\\
F_{1i}^{(o)} &=& -\frac{ig^3 M_W \beta_{G^+ h^-}^S}{16 \pi^2} \left[ m_t X_{ci}^{h^+} X_{ti}^{G} (C_{11}^{(o)} - C_{12}^{(o)}) - m_i X_{ci}^{h^+} Y_{ti}^{G} C_0^{(o)} + m_c Y_{ci}^{h^+} Y_{ti}^{G} C_{12}^{(o)}\right] \nonumber\\
F_{1i}^{(p)} &=& -\frac{ig^3}{16 \pi^2} \left[ z_1 \left( B_0^{(p)} - M_{h^+}^2 C_0^{(p)} \right) - z_3 C_0^{(p)} - m_t z_5 \frac{X_{ti}^{h^+}}{Y_{ti}^{h^+}} (C_{11}^{(p)} - C_{12}^{(p)}) - m_c z_2 C_{12}^{(p)} \right] Y_{ti}^{h^+} \nonumber\\
F_{1i}^{(q)} &=& -\frac{ig^3 M_W \beta_{\tilde{d} \tilde{d}}^h}{16 \pi^2} \left[ m_t X_{cj}^i X_{tj}^i (C_{11}^{(q)} - C_{12}^{(q)}) - m_i X_{cj}^i Y_{tj}^i C_0^{(q)} + m_c Y_{cj}^i Y_{tj}^i C_{12}^{(q)}\right] \nonumber
\end{eqnarray}
\begin{eqnarray}
F_{1i}^{(r)} &=& -\frac{ig^3}{16 \pi^2} \left[ z_1 \left( B_0^{(r)} - M_{\tilde{d}_i}^2 C_0^{(r)} \right) - z_3 C_0^{(r)} - m_t z_5 \frac{X_{tj}^i}{Y_{tj}^i} (C_{11}^{(r)} - C_{12}^{(r)}) - m_c z_2 C_{12}^{(r)} \right] Y_{tj}^i \nonumber\\
F_{1i}^{(s)} &=& \frac{ig^3}{16 \pi^2 (m_t^2 - m_c^2)} \left[m_t\left(m_t A_{uc}^h Y_{ci}^{h^+}   + m_c A_{uc}^h \frac{X_{ti}^{h^+}}{Y_{ti}^{h^+}} X_{ci}^{h^+} \right) B_1^{(s)} \right. \nonumber \\
 && \left. \qquad- m_i \left(m_t A_{uc}^h Y_{ci}^{h^+} \frac{X_{ti}^{h^+}}{Y_{ti}^{h^+}} + m_c A_{uc}^h X_{ci}^{h^+} \right) B_0^{(s)} \right] Y_{ti}^{h^+} \nonumber\\
F_{1i}^{(t)} &=& \frac{ig^3}{16 \pi^2 (m_t^2 - m_c^2)} \left[m_t\left(m_t A_{uc}^h Y_{cj}^i   + m_c A_{uc}^h \frac{X_{ti}^{h^+}}{Y_{ti}^{h^+}} X_{cj}^i \right) B_1^{(t)}
 - m_i \left(m_t A_{uc}^h Y_{cj}^i \frac{X_{ti}^{h^+}}{Y_{ti}^{h^+}} + m_c A_{uc}^h X_{cj}^i \right) B_0^{(t)} \right] Y_{ti}^{h^+} \nonumber\\
F_{1i}^{(u)} &=& -\frac{ig^3}{16 \pi^2 (m_t^2-m_c^2) } \left[m_c X_{ti}^{h^+} \left(m_c X_{ci}^{h^+} B_1^{(u)} - m_i Y_{ci}^{h^+} B_0^{(u)}\right) + m_t Y_{ti}^{h^+} \left(m_c Y_{ci}^{h^+} B_1^{(u)} - m_i X_{ci}^{h^+} B_0^{(u)}\right) \right] A_{ut}^h \nonumber\\
F_{1i}^{(v)} &=& -\frac{ig^3}{16 \pi^2 (m_t^2-m_c^2) } \left[m_c X_{tj}^i \left(m_c X_{cj}^i B_1^{(v)} - m_i Y_{cj}^i B_0^{(v)}\right) + m_t Y_{tj}^i \left(m_c Y_{cj}^i B_1^{(v)} - m_i X_{cj}^i B_0^{(v)}\right) \right] A_{ut}^h 
\end{eqnarray}
\normalsize
and the $F_2$ form factors
\footnotesize
\begin{eqnarray}
F_{2i}^{(k)} &=& \frac{ig^3 \alpha_{h^+}^h}{16 \sqrt{2} \pi^2} \left[X_{ci}^{h^+} m_c m_t (C_{12}^{(k)} - 2 C_{11}^{(k)}) + Y_{ci}^{h^+} m_i m_t (C_0^{(k)} - C_{11}^{(k)} + C_{12}^{(k)})\right] \nonumber\\ 
F_{2i}^{(l)} &=& \frac{-ig^3 \alpha_{h^+}^h}{16 \sqrt{2} \pi^2} \left[ X_{ti}^{h^+} m_c m_t (C_{12}^{(l)} - 2 C_{11}^{(l)}) + Y_{ti}^{h^+} m_i m_t (C_0^{(l)} - C_{11}^{(l)} + C_{12}^{(l)})\right] \nonumber\\ 
F_{2i}^{(m)} &=& -\frac{ig^3 M_W \beta_{h^+ h^-}^h}{16 \pi^2} \left[ m_t Y_{ci}^{h^+} Y_{ti}^{h^+} (C_{11}^{(m)} - C_{12}^{(m)}) - m_i X_{ti}^{h^+} Y_{ci}^{h^+} C_0^{(m)} + m_c X_{ci}^{h^+} X_{ti}^{h^+} C_{12}^{(m)}\right] \nonumber\\
F_{2i}^{(n)} &=& -\frac{ig^3 M_W \beta_{G^+ h^-}^h}{16 \pi^2} \left[ m_t Y_{ci}^{G} Y_{ti}^{h^+} (C_{11}^{(n)} - C_{12}^{(n)}) - m_i X_{ti}^{h^+} Y_{ci}^{G} C_0^{(n)} + m_c X_{ci}^{G} X_{ti}^{h^+} C_{12}^{(n)}\right] \nonumber\\
F_{2i}^{(o)} &=& -\frac{ig^3 M_W \beta_{G^+ h^-}^S}{16 \pi^2} \left[ m_t Y_{ci}^{h^+} Y_{ti}^{G} (C_{11}^{(o)} - C_{12}^{(o)}) - m_i X_{ti}^{G} Y_{ci}^{h^+} C_0^{(o)} + m_c X_{ci}^{h^+} X_{ti}^{G} C_{12}^{(o)}\right] \nonumber\\ 
F_{2i}^{(p)} &=& -\frac{ig^3}{16 \pi^2} \left[ z_4 \left( B_0^{(p)} - M_{h^+}^2 C_0^{(p)} \right) - z_6 C_0^{(p)} - m_t z_2 \frac{Y_{ti}^{h^+}}{X_{ti}^{h^+}} (C_{11}^{(p)} - C_{12}^{(p)}) - m_c z_5 C_{12}^{(p)} \right] X_{ti}^{h^+}  \nonumber\\
F_{2i}^{(q)} &=& -\frac{ig^3 M_W \beta_{\tilde{d} \tilde{d}}^h}{16 \pi^2} \left[ m_t Y_{cj}^i Y_{tj}^i (C_{11}^{(q)} - C_{12}^{(q)}) - m_i X_{tj}^i Y_{cj}^i C_0^{(q)} + m_c X_{cj}^i X_{tj}^i C_{12}^{(q)}\right] \nonumber\\
F_{2i}^{(r)} &=& -\frac{ig^3}{16 \pi^2} \left[ z_4 \left( B_0^{(r)} - M_{\tilde{d}_i}^2 C_0^{(r)} \right) - z_6 C_0^{(r)} - m_t z_2 \frac{Y_{tj}^i}{X_{tj}^i} (C_{11}^{(r)} - C_{12}^{(r)}) - m_c z_5 C_{12}^{(r)} \right] X_{tj}^i  \nonumber\\
\end{eqnarray}
\begin{eqnarray}
F_{2i}^{(s)} &=& \frac{ig^3}{16 \pi^2 (m_t^2 - m_c^2)} \left[m_t\left(m_t B_{uc}^h X_{ci}^{h^+} + m_c B_{uc}^h \frac{Y_{ti}^{h^+}}{X_{ti}^{h^+}} Y_{ci}^{h^+} \right) B_1^{(s)} \right. \nonumber \\
 && \left. \qquad- m_i \left(m_t B_{uc}^h X_{ci}^{h^+} \frac{Y_{ti}^{h^+}}{X_{ti}^{h^+}} + m_c B_{uc}^h Y_{ci}^{h^+} \right) B_0^{(s)} \right] X_{ti}^{h^+} \nonumber \\
F_{2i}^{(t)} &=& \frac{ig^3}{16 \pi^2 (m_t^2 - m_c^2)} \left[m_t\left(m_t B_{uc}^h X_{cj}^i + m_c B_{uc}^h \frac{Y_{ti}^{h^+}}{X_{ti}^{h^+}} Y_{cj}^i \right) B_1^{(t)}
 - m_i \left(m_t B_{uc}^h X_{cj}^i \frac{Y_{ti}^{h^+}}{X_{ti}^{h^+}} + m_c B_{uc}^h Y_{cj}^i \right) B_0^{(t)} \right] X_{ti}^{h^+} \nonumber \\ 
F_{2i}^{(u)} &=& -\frac{ig^3}{16 \pi^2 (m_t^2 - m_c^2)} \left[m_c Y_{ti}^{h^+} \left(m_c Y_{ci}^{h^+} B_1^{(u)} - m_i X_{ci}^{h^+} B_0^{(u)}\right) + m_t X_{ti}^{h^+} \left(m_c X_{ci}^{h^+} B_1^{(u)} - m_i Y_{ci}^{h^+} B_0^{(u)}\right) \right] B_{ut}^h \nonumber\\ 
F_{2i}^{(v)} &=& -\frac{ig^3}{16 \pi^2 (m_t^2 - m_c^2)} \left[m_c Y_{tj}^i \left(m_c Y_{cj}^i B_1^{(v)} - m_i X_{cj}^i B_0^{(v)}\right) + m_t X_{tj}^i \left(m_c X_{cj}^i B_1^{(v)} - m_i Y_{cj}^i B_0^{(v)}\right) \right] B_{ut}^h 
\end{eqnarray}
\normalsize

where
\begin{eqnarray}
  C_X^{(k)} &=& C_X (m_i, M_W, M_{h^+}; m_c, m_t, M_h) \qquad\qquad
  B_X^{(s)} = B_X(m_{\tilde{\chi}^+_i}, M_{\tilde{d}_j}; m_t)  \\
  C_X^{(l)} &=& C_X (m_i, M_{h^+}, M_W; m_c, m_t, M_h) \qquad\qquad
  B_X^{(t)} = B_X(m_i, M_{h^+}; m_t) \nonumber \\
  C_X^{(m)} &=& C_X (m_i, M_{h^+}, M_{h^+}; m_c, m_t, M_h) \qquad\qquad
  B_X^{(u)} = B_X(m_{\tilde{\chi}^+_i}, M_{\tilde{d}_j}; m_c) \nonumber \\
  C_X^{(n)} &=& C_X (m_i, M_{h^+}, M_W; m_c, m_t, M_h) \qquad\qquad
  B_X^{(v)} = B_X(m_i, M_{h^+}; m_c) \nonumber \\
  C_X^{(o)} &=& C_X (m_i, M_W, M_{h^+}; m_c, m_t, M_h) \qquad\qquad
  C_X^{(p)} = C_X (M_{h^+}, m_i, m_i; m_c, m_t, M_h) \nonumber \\
\  C_X^{(q)} &=& C_X (m_{\tilde{\chi}^+_i}, M_{\tilde{d}_j}, M_{\tilde{d}_j}; m_c, m_t, M_h) \qquad\qquad
\!\!  C_X^{(r)} = C_X (M_{\tilde{d}_j}, m_{\tilde{\chi}^+_i}, m_{\tilde{\chi}^+_i}; m_c, m_t, M_h) \nonumber
\end{eqnarray}
where $X = {0,11,12,21,23,24}$, as before, and defined two sets of effective 
couplings
\begin{eqnarray}
 z^{(p)}_1 &=& X_{ci}^{h} B_{di}^h \nonumber \\
 z^{(p)}_2 &=& m_t Y_{ci}^{h} \frac{X_{ti}^{h}}{Y_{ti}^{h}} A_{di}^h + m_c X_{ci}^{h} B_{di}^h + m_i Y_{ci}^{h} B_{di}^h + m_i Y_{ci}^{h} A_{di}^h \nonumber \\
 z^{(p)}_3 &=& m_t \frac{X_{ti}^{h}}{Y_{ti}^{h}} A_{di}^h (m_i X_{ci}^{h} + m_c Y_{ci}^{h}) + m_i^2 A_{di}^h X_{ci}^{h} + m_i m_c A_{di}^h Y_{ci}^{h} \nonumber \\
 z^{(p)}_4 &=& Y_{ci}^{h} A_{di}^h \nonumber \\
 z^{(p)}_5 &=& m_t X_{ci}^{h} \frac{Y_{ti}^{h}}{X_{ti}^{h}} B_{di}^h + m_c Y_{ci}^{h} A_{di}^h + m_i X_{ci}^{h} A_{di}^h + m_i X_{ci}^{h} B_{di}^h \nonumber \\
 z^{(p)}_6 &=& m_t \frac{Y_{ti}^{h}}{X_{ti}^{h}} B_{di}^h (m_i Y_{ci}^{h} + m_c X_{ci}^{h}) + m_i^2 B_{di}^h Y_{ci}^{h} + m_i m_c B_{di}^h X_{ci}^{h} 
\end{eqnarray}
and
\begin{eqnarray}
 z^{(r)}_1 &=& X_{cj}^i B_{ij}^h \nonumber \\
 z^{(r)}_2 &=& m_t Y_{cj}^i \frac{X_{tj}^i}{Y_{tj}^i} A_{ij}^h + m_c X_{cj}^i B_{ij}^h + m_i Y_{cj}^i B_{ij}^h + m_i Y_{cj}^i A_{ij}^h \nonumber \\
 z^{(r)}_3 &=& m_t \frac{X_{tj}^i}{Y_{tj}^i} A_{ij}^h (m_i X_{cj}^i + m_c Y_{cj}^i) + m_i^2 A_{ij}^h X_{cj}^i + m_i m_c A_{ij}^h Y_{cj}^i \nonumber \\
 z^{(r)}_4 &=& Y_{cj}^i A_{ij}^h \nonumber 
 \end{eqnarray}
\begin{eqnarray}
 z^{(r)}_5 &=& m_t X_{cj}^i \frac{Y_{tj}^i}{X_{tj}^i} B_{ij}^h + m_c Y_{cj}^i A_{ij}^h + m_i X_{cj}^i A_{ij}^h + m_i X_{cj}^i B_{ij}^h \nonumber \\
 z^{(r)}_6 &=& m_t \frac{Y_{tj}^i}{X_{tj}^i} B_{ij}^h (m_i Y_{cj}^i + m_c X_{cj}^i) + m_i^2 B_{ij}^h Y_{cj}^i + m_i m_c B_{ij}^h X_{cj}^i
\end{eqnarray}
As before, these form factors can now be combined, using $F_{ni} = \sum_{A=a}^j  
F_{ni}^{A}$ for $n = 1,2$ and the results substituted into Eqn.~(\ref{eqn:helampS})
to get the final amplitude.

\subsection{The decay $t \to c + Z$}

When we turn to the decay process $t \to c + Z$, then, as in the toy 
model, we have to calculate four helicity amplitudes in terms of four 
form factors $F_1$, $F_2$, $F_3$ and $F_4$. For the Standard Model, we 
then evaluate the diagrams of Figure~\ref{fig:FeynSM}, replacing the $H$ 
everywhere by a $Z$. In order to do this, we set up the following general 
vertices.

\begin{table} [!htb]
\centering
\begin{tabular}{rcl}
$\overline{u_i} u_i Z^\mu$ & : & $ig \gamma^\mu (A^Z_{ui} P_L + B^Z_{ui} P_R)$ \\ [2mm]
$\overline{d_i} d_i Z^\mu$ & : & $ig \gamma^\mu (A^Z_{di} P_L + B^Z_{di} P_R)$ \\ [2mm]
$W^{\mu+} Z^\nu \phi^-$ & : &  $ig \omega^\phi_{WZ} g^{\mu\nu}$ \\ [2mm]
$Z^\mu \phi(p)^+ \phi'^-(q)$ & : & $ig \alpha^\phi_{\phi'}(p+q)^\mu$ \\ [2mm]
$\overline{u}_i d_j \phi^+$ & : & $ig \left(X_{ij}^\phi P_L + Y_{ij}^\phi P_R\right)$ 
\end{tabular}
\end{table}
\vspace{-0.1in}
in terms of a set of coupling constants $A^Z_{ui}$, $B^Z_{ui}$, 
$A^Z_{di}$, $B^Z_{di}$, $\omega^\phi_{WZ}$, $\alpha^\phi_{\phi'}$, 
$X_{ij}^\phi$ and $Y_{ij}^\phi$. In the SM, these have values given in 
the table below.

\begin{table} [!htb]
\centering
\begin{tabular}{rcccc}
\hline
coupling : & $A^Z_{ui}$ & $B^Z_{ui}$ & $A^Z_{di}$ & $B^Z_{di}$ \\ \hline\hline
\\ [-12pt]
SM :
& \Large $-\frac{g_L^{u}}{\cos \theta_W}$
& \Large $-\frac{g_R^{u}}{\cos \theta_W}$
& \Large $-\frac{g_L^{u}}{\cos \theta_W}$
& \Large $-\frac{g_R^{u}}{\cos \theta_W}$ \\  \\ [-10pt]
\hline
coupling : & $\omega^\phi_{WZ}$ & $\alpha^\phi_{\phi'}$ & $X_{ij}^\phi$ & $Y_{ij}^\phi$ 
\\\hline\hline \\ [-12pt]
SM :
& $-M_Z \sin^2\theta_W$ 
& \Large $-\frac{\cos2\theta_W}{2 \cos\theta_W}$ 
& \Large $\frac{m_i}{\sqrt{2}M_W} $
& \Large $-\frac{m_j}{\sqrt{2}M_W}$
\\ [8pt] \hline 
\end{tabular}
\end{table}
where 
\begin{eqnarray}
g_L^u & = & \frac{1}{2} - \frac{2}{3}\sin^2\theta_W \qquad  \ \ 
g_R^u = - \frac{2}{3}\sin^2\theta_W \nonumber \\
g_L^d & = & -\frac{1}{2} + \frac{1}{3}\sin^2\theta_W \qquad
g_R^d = \frac{1}{3}\sin^2\theta_W
\end{eqnarray}
As in the previous cases, we can now compute, using the diagrams of 
Figure~\ref{fig:FeynSM} (with $h^0 \to Z$) a set of forms factors. The 
set of $F_1$ form factors are
\footnotesize
\begin{eqnarray}
 F_{1i}^{(a)} &=& \frac{g^3 \cos \theta_W}{16 \pi^2} \left[ 2 m_t^2 (C^{(a)}_{21} - C^{(a)}_{23})- 2 C^{(a)}_{24} + (m_t^2 + m_c^2 - M_Z^2) C^{(a)}_{11} -  m_c^2 C^{(a)}_{12} - (B_0^{(a)} - m_i^2 C^{(a)}_0)\right] \nonumber \\
 F_{1i}^{(b)} &=& - \frac{g^3 \omega_{WZ}^{G^+}}{16 \sqrt{2} \pi^2} \left[ m_t X_{ti}^{G} (C^{(b)}_{11}-C^{(b)}_{12}) - m_i Y_{ti}^{G} C^{(b)}_0 \right] \nonumber \\
 F_{1i}^{(c)} &=& - \frac{g^3 \omega_{WZ}^{G^+}}{16 \sqrt{2} \pi^2} \left[ m_c X_{ci}^{G} C^{(c)}_{12}  - 2 m_i Y_{ci}^{G} C^{(c)}_0 + 2 m_t X_{ci}^{G} (C^{(c)}_{11} -C^{(c)}_{12})\right] \nonumber \\
 F_{1i}^{(d)} &=& - \frac{g^3 \alpha_{G^-}^{G^+}}{16 \sqrt{2} \pi^2} \left[ m_t^2 Y_{ci}^{G}Y_{ti}^{G} (C^{(d)}_{21} - C^{(d)}_{23})  + m_c m_t X_{ci}^{G}X_{ti}^{G} C^{(d)}_{23}  - 2 Y_{ci}^{G}Y_{ti}^{G} C^{(d)}_{24}  - m_i m_t Y_{ti}^{G}X_{ci}^{G} (C^{(d)}_0 + C^{(d)}_{11})  \right]  \nonumber \\
F_{1i}^{(e)} &=& \frac{g^3}{32 \pi^2} \left[ A_{di}^Z \left\lbrace 2 (m_c^2 + m_t^2 - M_Z^2 + m_c m_t) (C^{(e)}_0 + C^{(e)}_{11}) + 2 m_t^2 (C^{(e)}_{11} - C^{(e)}_{12}) + m_c^2 C^{(e)}_{12} + 2 C^{(e)}_{24} \right. \right.\nonumber \\ 
&&\left. \left. + 2 m_i m_t (C^{(e)}_0 + C^{(e)}_{11}) -  B^{(e)}_0 + M_W^2 C^{(e)}_0 \right\rbrace + 2 m_i B_{di}^Z \left\lbrace m_t (C^{(e)}_0 + C^{(e)}_{11}) - m_i C^{(e)}_0\right\rbrace \right] \nonumber 
\end{eqnarray}
\begin{eqnarray}
 F_{1i}^{(f)} &=& -\frac{g^3}{16 \pi^2} \Big[ X_{ci}^{G} \Big\{ \left(m_i + m_t \frac{Y_{ci}^{G}}{X_{ci}^{G}}\right) A_{di}^{Z} \left(m_c X_{ci}^{G}  (C^{(f)}_0 + C^{(f)}_{12})  + m_i Y_{ci}^{G} C^{(f)}_0 \right) \nonumber \\
&& - B_{di}^Z \left( m_c (m_i X_{ci}^{G} + m_c Y_{ci}^{G}) C^{(f)}_{12} - Y_{ci}^{G} ( B^{(f)}_0 - M_W^2 C^{(f)}_0 )
-  m_t Y_{ci}^{G} (m_t C^{(f)}_{21} + m_c C^{(f)}_{23})  - 2 C^{(f)}_{24}\right)\Big\}  \nonumber \\ && - m_t Y_{ci}^{G} A_{di}^Z  (m_c X_{ci}^{G} + m_i Y_{ci}^{G}) C^{(f)}_{11} \Big] \nonumber \\
 F_{1i}^{(g)} &=& - \frac{g^3}{16 \pi^2 (m_t^2 - m_c^2)}  m_c^2 A_{ci}^{Z} B^{(g)}_1 \nonumber \\
 F_{1i}^{(h)} &=& - \frac{g^3}{16 \pi^2 (m_t^2 - m_c^2)} A_{ti}^{Z} \left[ m_t X_{ti}^{G} \left(m_c X_{ci}^{G} B^{(h)}_1 - m_i Y_{ci}^{G} B^{(h)}_0 \right)  + m_c Y_{ti}^{G} \left(Y_{ci}^{G} m_c B^{(h)}_1 - m_i X_{ci}^{G} B^{(h)}_0 \right) \right] \nonumber \\
 F_{1i}^{(i)} &=& \frac{g^3}{32 \pi^2 (m_t^2 - m_c^2)}m_t \left( m_t A_{ti}^{Z} + m_c B{ti}^{Z}\right) B^{(i)}_1  \\
 F_{1i}^{(j)} &=& \frac{g^3}{16 \pi^2 (m_t^2 - m_c^2)}
 A_{ti}^{Z} Y_{ti}^{G} \left[ m_t Y_{ci}^{G} \left(m_t  + m_c \frac{X_{ti}^{G} X_{ci}^{G}}{Y_{ti}^{G} Y_{ci}^{G}} \right) B^{(j)}_1  - m_i X_{ci}^{G} \left(m_t  \frac{Y_{ci}^{G}X_{ti}^{G}}{X_{ci}^{G}Y_{ti}^{G}}+ m_c \right) B^{(j)}_0\right] \nonumber
\end{eqnarray}
\normalsize
The nonvanishing $F_2$ form factors are
\footnotesize
\begin{eqnarray}
 F_{2i}^{(a)} &=& \frac{g^3 \cos \theta_W}{16 \pi^2}  m_c m_t (C^{(a)}_{11}-C^{(a)}_{12}) \nonumber \\
 F_{2i}^{(b)} &=& - \frac{g^3 \omega_{WZ}^{G^+}}{16 \sqrt{2} \pi^2}  (m_t - m_c) X_{ti}^{G} C^{(b)}_{12}  \nonumber \\
 F_{2i}^{(c)} &=&  \frac{g^3 \omega_{WZ}^{G^+}}{16 \sqrt{2} \pi^2} m_t X_{ci}^{G} (C^{(c)}_{11} -C^{(c)}_{12}) \nonumber \\
 F_{2i}^{(d)} &=&  - \frac{g^3 \alpha_{G^-}^{G^+}}{16 \sqrt{2} \pi^2} \left[ m_t^2 X_{ci}^{G}X_{ti}^{G} (C^{(d)}_{21} - C^{(d)}_{23})  + m_c m_t Y_{ci}^{G}Y_{ti}^{G} C^{(d)}_{23}  - 2 X_{ci}^{G}X_{ti}^{G} C^{(d)}_{24}  - m_i m_t X_{ti}^{G}Y_{ci}^{G} (C^{(d)}_0 + C^{(d)}_{11})  \right]  \nonumber \\
 F_{2i}^{(e)} &=& - \frac{g^3}{32 \pi^2} A_{di}^Z \left[ m_t(m_t + m_c)(C^{(e)}_0 + C^{(e)}_{11}) + m_t^2 (C^{(e)}_{11}-C^{(e)}_{12}) + m_c m_t C^{(e)}_{12} - m_t^2 C^{(e)}_{21} - m_t (m_t - m_c) C^{(e)}_{23}\right] \nonumber \\
 F_{2i}^{(f)} &=& -\frac{g^3}{16 \pi^2} \Big[ Y_{ci}^{G} \Big\{ \left(m_i + m_t \frac{X_{ci}^{G}}{Y_{ci}^{G}}\right) B_{di}^{Z} \left(m_c Y_{ci}^{G}  (C^{(f)}_0 + C^{(f)}_{12})  + m_i X_{ci}^{G} C^{(f)}_0 \right) \nonumber \\
&& - A_{di}^Z \left( m_c (m_i Y_{ci}^{G} + m_c X_{ci}^{G}) C^{(f)}_{12} - X_{ci}^{G} ( B^{(f)}_0 - M_W^2 C^{(f)}_0 )
-  m_t Y_{ci}^{G} (m_t C^{(f)}_{21} + m_c C^{(f)}_{23})  - 2 C^{(f)}_{24}\right) \Big\}  \nonumber \\ && - m_t X_{ci}^{G} B_{di}^Z  (m_c Y_{ci}^{G} + m_i X_{ci}^{G}) C^{(f)}_{11} \Big] \nonumber \\
 F_{2i}^{(g)} &=& - \frac{g^3}{16 \pi^2 (m_t^2 - m_c^2)}  m_c m_t B_{ci}^{Z} B^{(g)}_1 
 \nonumber \\
 F_{2i}^{(h)} &=& - \frac{g^3}{16 \pi^2 (m_t^2 - m_c^2)} B_{ti}^{Z} \left[ m_t Y_{ti}^{G} \left(m_c Y_{ci}^{G} B^{(h)}_1 - m_i X_{ci}^{G} B^{(h)}_0 \right)  + m_c X_{ti}^{G} \left(m_c X_{ci}^{G} B^{(h)}_1 - m_i Y_{ci}^{G} B^{(h)}_0 \right) \right] \nonumber \\
 F_{2i}^{(j)} &=& \frac{g^3}{16 \pi^2 (m_t^2 - m_c^2)}
 B_{ti}^{Z} X_{ti}^{G} \left[ m_t X_{ci}^{G} \left(m_t  + m_c \frac{Y_{ti}^{G} Y_{ci}^{G}}{X_{ti}^{G} X_{ci}^{G}} \right) B^{(j)}_1  - m_i Y_{ci}^{G} \left(m_t  \frac{X_{ci}^{G} Y_{ti}^{G}} {Y_{ci}^{G} X_{ti}^{G}}+ m_c \right) B^{(j)}_0\right] 
\end{eqnarray}
\normalsize
The nonvanishing $F_3$ form factors are
\footnotesize
\begin{eqnarray}
 F_{3i}^{(a)} &=& - \frac{g^3 \cos \theta_W}{32 \pi^2} m_c \left[ C^{(a)}_{11} + 2 C^{(a)}_{12}\right] \nonumber \\
 F_{3i}^{(c)} &=& \frac{g^3 \omega_{WZ}^{G^+}}{16 \sqrt{2} \pi^2} X_{ci}^{G} (C^{(c)}_{11} -C^{(c)}_{12}) \nonumber \\
 F_{3i}^{(d)} &=& \frac{g^3 \alpha_{G^-}^{G^+}}{16 \sqrt{2} \pi^2} \left[ m_t X_{ci}^{G}X_{ti}^{G} (C^{(d)}_{21}-C^{(d)}_{23})   - m_c Y_{ci}^{G}Y_{ti}^{G} C^{(d)}_{23}  + m_i Y_{ti}^{G}X_{ci}^{G} (C^{(d)}_0 + C^{(d)}_{11})  \right] \nonumber \\
 F_{3i}^{(e)} &=& \frac{g^3}{32 \pi^2} A_{di}^Z \left[ (m_t + m_c)(C^{(e)}_0 + C^{(e)}_{11}) + m_t (C^{(e)}_{11} - C^{(e)}_{12}) + m_c C^{(e)}_{12} - m_t C^{(e)}_{21} - (m_t - m_c) C^{(e)}_{23} \right] \nonumber \\
 F_{3i}^{(f)} &=& \frac{g^3}{16 \pi^2} X_{ci}^{G} \Big[A_{di}^{V}  \left(m_i + m_t \frac{Y_{ci}^{G}}{X_{ci}^{G}}\right) X_{ci}^{G} (C^{(f)}_{11} - C^{(f)}_{12})  \\
&& - B_{di}^Z \left\lbrace \left(m_i X_{ci}^{G} + m_c Y_{ci}^{G}\right) (C^{(f)}_{11} - C^{(f)}_{12}) + (m_i X_{ci}^{G} + m_c Y_{ci}^{G}) C^{(f)}_{11}  + Y_{ci}^{G}(m_t C^{(f)}_{21} + m_c C^{(f)}_{23})\right\rbrace \Big] \nonumber
\end{eqnarray}
\normalsize
The nonvanishing $F_4$ form factors are
\footnotesize
\begin{eqnarray}
 F_{4i}^{(a)} &=& \frac{g^3 \cos \theta_W}{16 \pi^2} m_t \{2 (C^{(a)}_{11} - C^{(a)}_{12}) - (C^{(a)}_{21} - C^{(a)}_{23}) \} \nonumber \\
 F_{4i}^{(b)} &=& - \frac{g^3 \omega_{WZ}^{G^+}}{16 \sqrt{2} \pi^2}  X_{ti}^{G} C^{(b)}_{12} \nonumber \\
 F_{4i}^{(d)} &=&  \frac{g^3 \alpha_{G^-}^{G^+}}{16 \sqrt{2} \pi^2} \left[ m_t Y_{ci}^{G} Y_{ti}^{G} (C^{(d)}_{21}-C^{(d)}_{23})   - m_c X_{ci}^{G} X_{ti}^{G} C^{(d)}_{23}  + m_i X_{ti}^{G} Y_{ci}^{G} (C^{(d)}_0 + C^{(d)}_{11})  \right] \nonumber \\
 F_{4i}^{(e)} &=& - \frac{g^3}{16 \pi^2} m_i (A_{di}^Z + B_{di}^Z) (C^{(e)}_0 + C^{(e)}_{11}) \nonumber \\
 F_{4i}^{(f)} &=& \frac{g^3}{16 \pi^2} Y_{ci}^{G} \Big[B_{di}^{Z}  \left(m_i + m_t \frac{X_{ci}^{G}}{Y_{ci}^{G}}\right) Y_{ci}^{G} (C^{(f)}_{11} - C^{(f)}_{12})  \\
&& - A_{di}^Z \left\lbrace \left(m_i Y_{ci}^{G} + m_c X_{ci}^{G}\right) (C^{(f)}_{11} - C^{(f)}_{12}) + (m_i Y_{ci}^{G} + m_c X_{ci}^{G}) C^{(f)}_{11}  + X_{ci}^{G}(m_t C^{(f)}_{21} + m_c C^{(f)}_{23})\right\rbrace \Big] \nonumber
\end{eqnarray}
\normalsize
where
\begin{eqnarray}
  C_X^{(a)} &=& C_X (m_i, M_W, M_W; m_c, m_t, M_Z) \qquad\qquad
  B_0^{(e)} = B_0(M_W, m_i; M_Z) \nonumber \\ 
  C_X^{(b)} &=& C_X (m_i, M_W, M_W; m_c, m_t, M_Z) \qquad\qquad
  B_1^{(g)} = B_1(m_i, M_W; m_t) \nonumber \\
  C_X^{(c)} &=& C_X (m_i, M_W, M_W; m_c, m_t, M_Z) \qquad\qquad
  B_1^{(h)} = B_1(m_i, M_W;m_t) \nonumber \\
  C_X^{(d)} &=& C_X (m_i, M_W, M_W; m_c, m_t, M_Z) \qquad\qquad
  B_0^{(h)} = B_0(m_i, M_W;m_t) \nonumber \\
  C_X^{(e)} &=& C_X (M_W, m_i, m_i; m_c, m_t, M_Z) \qquad\qquad
\ \;  B_1^{(i)} = B_1(m_i,M_W; m_c) \nonumber \\
  C_X^{(f)} &=& C_X (M_W, m_i, m_i; m_c, m_t, M_Z) \qquad\qquad
  \ \; B_1^{(j)} = B_1(m_i, M_W; m_c) \nonumber \\
   B_0^{(a)} &=& B_0(M_W, M_W; M_Z) \hspace*{1.3in}
   \ \; B_0^{(j)} = B_0(m_i, M_W; m_c) 
\end{eqnarray}
where $X = {0,11,12,21,23,24}$, as usual. We then calculate the total 
form factors using $F_{ni} = \sum_{A=a}^j F_{ni}^{A}$ for $n = 1,2,3,4$ 
and substitute the results into Eqn.~(\ref{eqn:helampZ}) to get the final 
SM amplitude.

In the cMSSM, we require to evaluate all the diagrams which contribute in 
the SM, i.e. those which are listed in Figure~\ref{fig:FeynSM}. This will 
involve all the vertices we have defined for the SM, but the coupling 
constants will be somewhat different. These are listed in the table 
below.

\begin{table} [!htb]
\centering
\begin{tabular}{rcccc}
\hline
coupling : & $A^Z_{ui}$ & $B^Z_{ui}$ & $A^Z_{di}$ & $B^Z_{di}$ \\ \hline\hline
\\ [-12pt]
SM :
& \Large $-\frac{g_L^{u}}{\cos \theta_W}$
& \Large $-\frac{g_R^{u}}{\cos \theta_W}$
& \Large $-\frac{g_L^{u}}{\cos \theta_W}$
& \Large $-\frac{g_R^{u}}{\cos \theta_W}$ \\  \\ [-10pt]
\hline
coupling : & $\omega^{G^+}_{WZ}$ & $\alpha^{G^+}_{G^-}$ & $X_{ij}^{H^+}$ & $Y_{ij}^{H^+}$ 
\\\hline\hline \\ [-12pt]
SM :
& \Large $- M_Z \sin^2\theta_W$ 
& \Large $-\frac{\cos2\theta_W}{2 \cos\theta_W}$
& \Large $\frac{m_i \cot \beta}{\sqrt{2}M_W}$
& $\frac{m_j \tan \beta}{\sqrt{2}M_W}$ \\ [8pt] \hline 
\end{tabular}
\end{table}
Due to the absence of a $W^\pm H^\mp Z$ vertex (whereas there is a $W^\pm 
H^\mp h^0$ vertex, the list of additional diagrams in the cMSSM can be 
obtained by changing the $H$ lines in Figure~\ref{fig:FeynSUSY} to $Z$ 
lines, provided we discard the diagrams marked ($k$), ($\ell$), ($n$) and 
($o$). Evaluating the remaining ones we get the $F_1$ form factors 

\vspace*{-0.2in}
\footnotesize
\begin{eqnarray}
 F_{1i}^{(m)} &=& - \frac{g^3 \alpha_{h^+}^{h^-}}{16 \sqrt{2} \pi^2} \left[ m_t^2 (C^{(m)}_{21} - C^{(m)}_{23}) Y_{ci}^{h}Y_{ti}^{h} + m_c m_t C^{(m)}_{23} X_{ci}^{h}X_{ti}^{h} - 2 C^{(m)}_{24} Y_{ci}^{h}Y_{ti}^{h} - m_i m_t (C^{(m)}_0 + C^{(m)}_{11}) Y_{ti}^{h}X_{ci}^{h} \right]  \nonumber \\
 F_{1i}^{(p)} &=& -\frac{g^3}{16 \pi^2} \Big[ X_{ci}^{h} \Big\{ \left(m_i + m_t \frac{Y_{ci}^{h}}{X_{ci}^{h}}\right) A_{di}^{Z} \left( X_{ci}^{h} m_c (C_0^{(p)} + C^{(p)}_{12})  + m_i Y_{ci}^{h} C_0^{(p)} \right) \nonumber \\
&& - B_{di}^Z \left( \left(m_i X_{ci}^{h} + m_c Y_{ci}^{h}\right) m_c C^{(p)}_{12} - Y_{ci}^{h} \left( B_0^{(p)} - M_{h^+}^2 C_0^{(p)} \right)
-  m_t Y_{ci}^{h} (m_t C^{(p)}_{21} + m_c C^{(p)}_{23})  - 2 C^{(p)}_{24}\right) \Big\}  \nonumber \\ && - m_t A_{di}^Z Y_{ci}^{h} (m_i Y_{ci}^{h} + m_c X_{ci}^{h}) C^{(p)}_{11} \Big] \nonumber \\
 F_{1i}^{(q)} &=& - \frac{g^3 \alpha_{\tilde{d}}^{\tilde{d}}}{16 \sqrt{2} \pi^2} \left[ m_t^2 Y_{ci}^{j} Y_{ti}^{j} (C^{(q)}_{21} - C^{(q)}_{23})  + m_c m_t X_{ci}^{j}X_{ti}^{j} C^{(q)}_{23}  - 2 Y_{ci}^{j} Y_{ti}^{j} C^{(q)}_{24}  - m_i m_t Y_{ti}^{j} X_{ci}^{j} (C^{(m)}_0 + C^{(q)}_{11})  \right]  \nonumber \\
 F_{1i}^{(r)} &=& -\frac{g^3}{16 \pi^2} \Big[ X_{ci}^{j} \Big\{ \left(m_i + m_t \frac{Y_{ci}^{j}}{X_{ci}^{j}}\right) A_{di}^Z \left( X_{ci}^{j} m_c (C^{(r)}_0 + C^{(r)}_{12})  + m_i Y_{ci}^{j} C^{(r)}_0 \right) \nonumber \\
&& - B_{di}^Z \left( \left(m_i X_{ci}^{j} + m_c Y_{ci}^{j}\right) m_c C^{(r)}_{12} - Y_{ci}^{j} \left( B^{(r)}_0 - M_{\tilde{d}_j}^2 C^{(r)}_0 \right)
- Y_{ci}^{j} (m_t C^{(r)}_{21} + m_c C^{(r)}_{23}) m_t  - 2 C^{(r)}_{24}\right) \Big\}  \nonumber \\ && - m_t A_{di}^Z Y_{ci}^{j} (m_i Y_{ci}^{j} + m_c X_{ci}^{j}) C^{(r)}_{11} \Big] \nonumber \\
F_{1i}^{(s)} &=& \frac{g^3}{16 \pi^2 (m_t^2 - m_c^2)}  A_{ci}^{Z} Y_{ti}^{h} \left[ m_t Y_{ci}^{h}  \left(m_t  + m_c \frac{X_{ti}^{h} X_{ci}^{h}}{Y_{ti}^{h} Y_{ci}^{h}} \right) B^{(s)}_1 - m_i X_{ci}^{h} \left(m_t  \frac{Y_{ci}^{h}X_{ti}^{h}}{X_{ci}^{h}Y_{ti}^{h}}+ m_c \right) B^{(s)}_0 \right] \nonumber \\
  F_{1i}^{(t)} &=& \frac{g^3}{16 \pi^2 (m_t^2 - m_c^2)} A_{ci}^{Z} Y_{ti}^{j} \left[ m_t Y_{ci}^{j}  \left(m_t + m_c \frac{X_{ti}^{j} X_{ci}^{j}}{Y_{ti}^{j} Y_{ci}^{j}} \right) B^{(t)}_1 - m_i X_{ci}^{j} \left(m_t  \frac{Y_{ci}^{j}X_{ti}^{j}}{X_{ci}^{j}Y_{ti}^{j}}+ m_c \right) B^{(t)}_0\right] \nonumber \\
 F_{1i}^{(u)} &=& - \frac{g^3}{16 \pi^2 (m_t^2 - m_c^2)} A_{ti}^Z \left[ m_c Y_{ti}^{h} \left(m_c Y_{ci}^{h} B^{(u)}_1 - m_i X_{ci}^{h} B^{(u)}_0 \right) + m_t X_{ti}^{h} \left(m_c X_{ci}^{h} B^{(u)}_1 - m_i Y_{ci}^{h} B^{(u)}_0 \right)\right] \nonumber \\
 F_{1i}^{(v)} &=& - \frac{g^3}{16 \pi^2 (m_t^2 - m_c^2)} A_{ti}^Z \left[ m_c Y_{ti}^{j} \left(m_c Y_{ci}^{j} B^{(r)}_1 - m_i X_{ci}^{j} B^{(r)}_0 \right) + m_t X_{ti}^{j} \left(m_c X_{ci}^{j} B^{(r)}_1 - m_i Y_{ci}^{j} B^{(r)}_0 \right)\right] \end{eqnarray}
\normalsize
The $F_2$ form factors are
\footnotesize
\begin{eqnarray}
 F_{2i}^{(m)} &=&  - \frac{g^3 \alpha_{h^+}^{h^-}}{16 \sqrt{2} \pi^2} \left[ m_t^2 X_{ci}^{h} X_{ti}^{h} (C^{(m)}_{21} - C^{(m)}_{23})  + m_c m_t Y_{ci}^{h} Y_{ti}^{h} C^{(m)}_{23}  - 2 X_{ci}^{h} X_{ti}^{h} C^{(m)}_{24}  - m_i m_t X_{ti}^{h} Y_{ci}^{h} (C^{(m)}_0 + C^{(m)}_{11})  \right] \nonumber \\
 F_{2i}^{(p)} &=& -\frac{g^3}{16 \pi^2} \Big[ Y_{ci}^{h} \Big\{ \left(m_i + m_t \frac{X_{ci}^{h}}{Y_{ci}^{h}}\right) B_{di}^{Z} \left( Y_{ci}^{h} m_c (C_0^{(p)} + C^{(p)}_{12})  + m_i X_{ci}^{h} C_0^{(p)} \right) \nonumber \\
&& - A_{di}^Z \left( \left(m_i Y_{ci}^{h} + m_c X_{ci}^{h}\right) m_c C^{(p)}_{12} - X_{ci}^{h} \left( B_0^{(p)} - M_{h^+}^2 C_0^{(p)} \right)
-  m_t X_{ci}^{h} (m_t C^{(p)}_{21} + m_c C^{(p)}_{23})  - 2 C^{(p)}_{24}\right) \Big\}  \nonumber \\ && - m_t B_{di}^Z X_{ci}^{h} (m_i X_{ci}^{h} + m_c Y_{ci}^{h}) C^{(p)}_{11} \Big] \nonumber \\
 F_{2i}^{(q)} &=&   - \frac{g^3 \alpha_{\tilde{d}}^{\tilde{d}}}{16 \sqrt{2} \pi^2} \left[ m_t^2 X_{ci}^{j} X_{ti}^{j} (C^{(q)}_{21} - C^{(q)}_{23})  + m_c m_t Y_{ci}^{j} Y_{ti}^{j} C^{(q)}_{23}  - 2 X_{ci}^{j} X_{ti}^{j} C^{(q)}_{24}  - m_i m_t X_{ti}^{j} Y_{ci}^{j} (C^{(m)}_0 + C^{(q)}_{11})  \right]  \nonumber 
\end{eqnarray}
\begin{eqnarray}
 F_{2i}^{(r)} &=& -\frac{g^3}{16 \pi^2} \Big[ Y_{ci}^{j} \Big\{ \left(m_i + m_t \frac{X_{ci}^{j}}{Y_{ci}^{j}}\right) B_{di}^Z \left( Y_{ci}^{j} m_c (C^{(r)}_0 + C^{(r)}_{12})  + m_i X_{ci}^{j} C^{(r)}_0 \right) \nonumber \\
&& - B_{di}^Z \left( \left(m_i Y_{ci}^{j} + m_c X_{ci}^{j}\right) m_c C^{(r)}_{12} - X_{ci}^{j} \left( B^{(r)}_0 - M_{\tilde{d}_j}^2 C^{(r)}_0 \right)
- X_{ci}^{j} (m_t C^{(r)}_{21} + m_c C^{(r)}_{23}) m_t  - 2 C^{(r)}_{24}\right) \Big\}  \nonumber \\ && - m_t B_{di}^Z X_{ci}^{j} (m_i X_{ci}^{j} + m_c Y_{ci}^{j}) C^{(r)}_{11} \Big] \nonumber \\
 F_{2i}^{(s)} &=& \frac{g^3}{16 \pi^2 (m_t^2 - m_c^2)}  B_{ci}^{Z} X_{ti}^{h} \left[ m_t X_{ci}^{h}  \left(m_t  + m_c \frac{Y_{ti}^{h} Y_{ci}^{h}}{X_{ti}^{h} X_{ci}^{h}} \right) B^{(s)}_1 - m_i Y_{ci}^{h} \left(m_t  \frac{X_{ci}^{h}Y_{ti}^{h}}{Y_{ci}^{h}X_{ti}^{h}}+ m_c \right) B^{(s)}_0 \right] \nonumber \\
 F_{2i}^{(t)} &=& \frac{g^3}{16 \pi^2 (m_t^2 - m_c^2)} B_{ci}^{Z} X_{ti}^{j} \left[ m_t X_{ci}^{j}  \left(m_t + m_c \frac{Y_{ti}^{j} Y_{ci}^{j}}{X_{ti}^{j} X_{ci}^{j}} \right) B^{(t)}_1 - m_i Y_{ci}^{j} \left(m_t  \frac{X_{ci}^{j}Y_{ti}^{j}}{Y_{ci}^{j}X_{ti}^{j}}+ m_c \right) B^{(t)}_0\right] \nonumber \\
  F_{2i}^{(u)} &=& - \frac{g^3}{16 \pi^2 (m_t^2 - m_c^2)} B_{ti}^Z \left[ m_c X_{ti}^{h} \left(m_c X_{ci}^{h} B^{(u)}_1 - m_i Y_{ci}^{h} B^{(u)}_0 \right) + m_t Y_{ti}^{h} \left(m_c Y_{ci}^{h} B^{(u)}_1 - m_i X_{ci}^{h} B^{(u)}_0 \right)\right] \nonumber \\
 F_{2i}^{(v)} &=& - \frac{g^3}{16 \pi^2 (m_t^2 - m_c^2)} B_{ti}^Z \left[ m_c X_{ti}^{j} \left(m_c X_{ci}^{j} B^{(r)}_1 - m_i Y_{ci}^{j} B^{(r)}_0 \right) + m_t Y_{ti}^{j} \left(m_c Y_{ci}^{j} B^{(r)}_1 - m_i X_{ci}^{j} B^{(r)}_0 \right)\right] \end{eqnarray}
\normalsize
The nonvanishing $F_3$ form factors are
\footnotesize
\begin{eqnarray}
 F_{3i}^{(m)} &=& \frac{g^3 \alpha_{h^+}^{h^-}}{16 \sqrt{2} \pi^2}\left[ m_t  X_{ci}^{h}X_{ti}^{h} (C^{(m)}_{21}-C^{(m)}_{23})  - m_c Y_{ci}^{h}Y_{ti}^{h} C^{(m)}_{23} + m_i Y_{ti}^{h}X_{ci}^{h} (C^{(m)}_0 + C^{(m)}_{11})  \right] \nonumber \\
 F_{3i}^{(p)} &=& \frac{g^3}{16 \pi^2} X_{ci}^{h} \Big[A_{di}^{Z} \left(m_i + m_t \frac{Y_{ci}^{h}}{X_{ci}^{h}}\right)  X_{ci}^{h} (C^{(p)}_{11} - C^{(p)}_{12}) \nonumber \\
&& - B_{di}^Z \left\lbrace \left(m_i X_{ci}^{h} + m_c Y_{ci}^{h}\right) (C^{(p)}_{11} - C^{(p)}_{12}) + (m_i X_{ci}^{h} + m_c Y_{ci}^{h}) C^{(p)}_{11}  + Y_{ci}^{h}(m_t C^{(p)}_{21} + m_c C^{(p)}_{23})\right\rbrace \Big] \nonumber \\
 F_{3i}^{(q)} &=& \frac{g^3 \alpha_{\tilde{d}}^{\tilde{d}}}{16 \sqrt{2} \pi^2}\left[ m_t X_{ci}^{j}X_{ti}^{j} (C^{(q)}_{21}-C^{(q)}_{23}) - m_c  Y_{ci}^{j}Y_{ti}^{j} C^{(q)}_{23} + m_i Y_{ti}^{j}X_{ci}^{j} (C^{(m)}_0 + C^{(q)}_{11}) \right] \nonumber \\
 F_{3i}^{(r)} &=& \frac{g^3}{16 \pi^2} X_{ci}^{j} \Big[A_{di}^{Z} \left(m_i + m_t \frac{Y_{ci}^{j}}{X_{ci}^{j}}\right)  X_{ci}^{j} (C^{(r)}_{11} - C^{(r)}_{12})  \\
&& - B_{di}^Z \left\lbrace \left(m_i X_{ci}^{j} + m_c Y_{ci}^{j}\right) (C^{(r)}_{11} - C^{(r)}_{12}) + (m_i X_{ci}^{j} + m_c Y_{ci}^{j}) C^{(r)}_{11}  + Y_{ci}^{j}(m_t C^{(r)}_{21} + m_c C^{(r)}_{23})\right\rbrace \Big] \nonumber 
\end{eqnarray}
\normalsize
Finally, the nonvanishing $F_4$ form factors are
\footnotesize
\begin{eqnarray}
 F_{4i}^{(m)} &=& \frac{g^3 \alpha_{h^+}^{h^-}}{16 \sqrt{2} \pi^2}\left[ m_t  Y_{ci}^{h}Y_{ti}^{h} (C^{(m)}_{21}-C^{(m)}_{23})  - m_c X_{ci}^{h}X_{ti}^{h} C^{(m)}_{23}  + m_i X_{ti}^{h}Y_{ci}^{h} (C_0 + C^{(m)}_{11}) \right] \nonumber \\
 F_{4i}^{(p)} &=& \frac{g^3}{16 \pi^2} X_{ci}^{h} \Big[B_{di}^{Z} \left(m_i + m_t \frac{X_{ci}^{h}}{Y_{ci}^{h}}\right)  Y_{ci}^{h} (C^{(p)}_{11} - C^{(p)}_{12}) \nonumber \\
&& - A_{di}^Z \left\lbrace \left(m_i Y_{ci}^{h} + m_c X_{ci}^{h}\right) (C^{(p)}_{11} - C^{(p)}_{12}) + (m_i Y_{ci}^{h} + m_c X_{ci}^{h}) C^{(p)}_{11}  + X_{ci}^{h}(m_t C^{(p)}_{21} + m_c C^{(p)}_{23})\right\rbrace \Big] \nonumber \\
 F_{4i}^{(q)} &=&  \frac{g^3 \alpha_{\tilde{d}}^{\tilde{d}}}{16 \sqrt{2} \pi^2}\left[ m_t Y_{ci}^{j}Y_{ti}^{j} (C^{(q)}_{21}-C^{(q)}_{23}) - m_c  X_{ci}^{j}X_{ti}^{j} C^{(q)}_{23} + m_i X_{ti}^{j}Y_{ci}^{j} (C_0 + C^{(q)}_{11}) \right] \nonumber \\
 F_{4i}^{(r)} &=& \frac{g^3}{16 \pi^2} X_{ci}^{j} \Big[B_{di}^{Z} \left(m_i + m_t \frac{X_{ci}^{j}}{Y_{ci}^{j}}\right)  Y_{ci}^{j} (C^{(r)}_{11} - C^{(r)}_{12}) \\
&& - A_{di}^Z \left\lbrace \left(m_i Y_{ci}^{j} + m_c X_{ci}^{j}\right) (C^{(r)}_{11} - C^{(r)}_{12}) + (m_i Y_{ci}^{j} + m_c X_{ci}^{j}) C^{(r)}_{11}  + X_{ci}^{j}(m_t C^{(r)}_{21} + m_c C^{(r)}_{23})\right\rbrace \Big] \nonumber
\end{eqnarray}
\normalsize
where we have used
\begin{eqnarray}
  C_X^{(m)} &=& C_X (m_i,M_{h^+}, M_{h^+}; m_c, m_t, M_Z) \qquad\qquad
  B_0^{(m)} = B_0(M_{h^+}, M_{h^+};M_Z) \nonumber \\
  C_X^{(p)} &=& C_X (M_{h^+},m_i, m_i; m_c, m_t, M_Z) \qquad\qquad
\ \ \, B_0^{(q)} = B_X(M_{\tilde{d}_j}, M_{\tilde{d}_j}; M_Z) \nonumber \\
  C_X^{(q)} &=& C_X (m_{\tilde{\chi}^+_i},M_{\tilde{d}_j}, M_{\tilde{d}_j}; m_c, m_t, M_Z) \qquad\qquad
\!\!  B_X^{(s)} = B_X(m_i, M_{h^+}; m_t) \nonumber \\
  C_X^{(r)} &=& C_X (M_{\tilde{d}_j}, m_{\tilde{\chi}^+_i},m_{\tilde{\chi}^+_i}; m_c, m_t, M_Z) \qquad\qquad
\!\!  B_X^{(t)} = B_X(m_{\tilde{\chi}^+_i}, M_{\tilde{d}_j};m_t) \nonumber \\
  B_X^{(u)} &=& B_X(m_i, M_{h^+}; m_t) \hspace*{1.55in}
  B_X^{(v)} = B_X(m_{\tilde{\chi}^+_i}, M_{\tilde{d}_j};m_c) 
\end{eqnarray}
for $X = {0,11,12,21,23,24}$, as usual. It is now a simple matter to calculate the 
total form factors
using $F_{ni} = \sum_{A=a}^j  F_{ni}^{A}$ for $n = 1,2,3,4$ and substitute the results 
into Eqn.~(\ref{eqn:helampZ}) to get the final cMSSM amplitude.

\section{RPV-MSSM amplitudes}

\subsection{The decay $t \to c + H$}

Since the RPV-MSSM is merely an extension of the MSSM, it will contain 
all the diagrams of Figures~\ref{fig:FeynSM} and \ref{fig:FeynSUSY}. 
However, as we have seen in the text, these contributions are small, and 
the $R$-parity violating contributions can be much larger. It is 
sensible, therefore, to calculate these alone. To have a unified picture, 
we include both $\lambda'_{ijk}$ and $\lambda''_{ijk}$ couplings when 
listing the diagrams in Figure~\ref{fig:FeynRPV}, though only one set at 
a time can contribute. In terms of these, the $F_1$ form factors are
\vspace*{-0.2in}

\footnotesize
\begin{eqnarray}
F^{1a}_{1ik} &=& g M_W \beta_{\tilde{e}_i \tilde{e}_i}^h\frac{ \lambda^{'}_{i2k}\lambda^{'}_{i3k}}{16 \pi^2} m_c C^{(a)}_{12} \nonumber \\
F^{1b}_{1ik} &=& \frac{y_{d_k} \lambda^{'}_{i2k}\lambda^{'}_{i3k} }{16\pi^2} m_c M_{\tilde{d}_k}  \left[C^{(b)}_0 + 2 C^{(b)}_{12}\right] \nonumber \\ 
F^{1c}_{1ik} &=& g M_W \beta_{\tilde{d}_k \tilde{d}_k}^h\frac{ \lambda^{'}_{i2k}\lambda^{'}_{i3k}}{16 \pi^2} m_c C^{(c)}_{12} \nonumber \\
F^{1d}_{1ik} &=& \frac{y_{l_i} \lambda^{'}_{i2k}\lambda^{'}_{i3k} }{16\pi^2} m_c m_{l_i}  \left[C^{(d)}_0 + 2 C^{(d)}_{12}\right] \nonumber \\  
F^{1e}_{1ik} &=& - \frac{y_{t} \lambda^{'}_{i2k}\lambda^{'}_{i3k} }{16\pi^2 (m_t^2 - m_c^2)} m_c m_t B^{(e)}_1 \nonumber \\ 
F^{1f}_{1ik} &=& \frac{y_{c} \lambda^{'}_{i2k}\lambda^{'}_{i3k} }{16\pi^2 (m_t^2 - m_c^2)} m_t (m_t + m_c) B^{(f)}_1 \nonumber \\ 
F^{1g}_{1jk} &=& g M_W \beta_{\tilde{d}_k \tilde{d}_k}^h\frac{ \lambda^{''}_{2jk}\lambda^{''}_{3jk}}{16 \pi^2} m_t \left[ C^{(g)}_{11}-C^{(g)}_{12}\right] \nonumber \\
F^{1h}_{1jk} &=& \frac{y_{d_k} \lambda^{''}_{2jk}\lambda^{''}_{3jk} }{16\pi^2} m_t m_{d_i}  \left[ C^{(h)}_0 + 2 \left( C^{(h)}_{11}-C^{(h)}_{12}\right)\right] \nonumber \\
F^{1i}_{1jk} &=& \frac{y_{t} \lambda^{''}_{2jk}\lambda^{''}_{3jk} }{16\pi^2 (m_t^2 - m_c^2)} m_c m_t B^{(i)}_1 \nonumber \\
F^{1j}_{1jk} &=&  - \frac{y_{c} \lambda^{''}_{2jk}\lambda^{''}_{3jk} }{16\pi^2 (m_t^2 - m_c^2)} m_t(m_t + m_c) B^{(j)}_1
\end{eqnarray}
\normalsize

and the $F_2$ form factors are
\footnotesize
\begin{eqnarray}
F^{2a}_{1ik} &=& g M_W \beta_{\tilde{e}_i \tilde{e}_i}^h \frac{ \lambda^{'}_{i2k}\lambda^{'}_{i3k}}{16 \pi^2} m_t \left[C^{(a)}_{11}-C^{(a)}_{12}\right] \nonumber \\  
 F^{2b}_{1ik} &=& \frac{y_{d_k} \lambda^{'}_{i2k}\lambda^{'}_{i3k} }{16\pi^2} m_t m_{d_k}  \left[ C^{(b)}_0 + 2 (C^{(b)}_{11}-C^{(b)}_{12})\right] \nonumber \\
F^{2c}_{1ik} &=& g M_W \beta_{\tilde{d}_k \tilde{d}_k}^h\frac{ \lambda^{'}_{i2k}\lambda^{'}_{i3k}}{16 \pi^2} m_t \left[ C^{(c)}_{11}-C^{(c)}_{12}\right] \nonumber \\ 
 F^{2d}_{1ik} &=& \frac{y_{l_i} \lambda^{'}_{i2k}\lambda^{'}_{i3k} }{16\pi^2} m_t m_{l_i}  \left[C^{(d)}_0 + 2 \left(C^{(d)}_{11}-C^{(d)}_{12}\right)\right] \nonumber \\
F^{2e}_{1ik} &=& - \frac{y_{t} \lambda^{'}_{i2k}\lambda^{'}_{i3k} }{16\pi^2 (m_t^2 - m_c^2)} m_c^2 B^{(e)}_1 \nonumber \\ 
 F^{2g}_{1jk} &=& g M_W \beta_{\tilde{d}_k \tilde{d}_k}^h \frac{ \lambda^{''}_{2jk}\lambda^{''}_{3jk}}{16 \pi^2} m_c C^{(g)}_{12} \nonumber \\ 
F^{2h}_{1jk} &=& \frac{y_{d_k} \lambda^{''}_{2jk}\lambda^{''}_{3jk} }{16\pi^2} m_c M_{\tilde{d}_k}  \left[C^{(h)}_0 + 2 C^{(h)}_{12}\right] \nonumber \\ 
F^{2i}_{1jk} &=& \frac{y_{t} \lambda^{''}_{2jk}\lambda^{''}_{3jk}} {16\pi^2 (m_t^2 - m_c^2)} m_c^2 B^{(i)}_1 
\end{eqnarray}
\normalsize
in terms of
\begin{eqnarray}
C_X^{(a)} &=& C_X (m_k, M_{\tilde{e}_i}, M_{\tilde{e}_i}; m_c, m_t, M_h) \quad\qquad
  B_1^{(e)} = B_1(m_k, M_{\tilde{e}_i}; m_c)  \nonumber \\
C_X^{(b)} &=& C_X (M_{\tilde{e}_i}, m_k, m_k; m_c, m_t, M_h) \qquad\qquad
\! \! \!    B_1^{(f)} = B_1(m_k, M_{\tilde{e}_i};m_c) \nonumber \\
C_X^{(c)} &=& C_X (m_i, M_{\tilde{d}_k}, M_{\tilde{d}_k}; m_c, m_t, M_h) \quad\qquad
 B_1^{(i)} = B_1(m_j,M_{\tilde{d}_k}; m_t) \nonumber \\
  C_X^{(d)} &=& C_X (M_{\tilde{d}_k}, m_{e_i}, m_{e_i}; m_c, m_t, M_h) \quad\qquad
\!    B_1^{(j)} = B_1(m_j, M_{\tilde{d}_k}; m_t)  \\
  C_X^{(g)} &=& C_X (m_j,M_{\tilde{d}_k}, M_{\tilde{d}_k}; m_c, m_t, M_h) \qquad\quad
  C_X^{(h)} = C_X (M_{\tilde{d}_k}, m_j, m_j; m_c, m_t, M_h) \nonumber 
\end{eqnarray}
where, as usual, $X = {0,11,12,21,23,24}$. As before, we go on to compute 
total form factors using $F_{ni} = \sum_{A=a}^j F_{ni}^{A}$ for $n = 1,2$ 
and substitute the results into Eqn.~(\ref{eqn:helampS}) to get the 
amplitude in the RPV-MSSM.

\subsection{The decay $t \to c + Z$}

The Feynman diagrams for the decay $t \to c + Z$ are the same as those in 
Figure~\ref{fig:FeynRPV}, with $h^0 \to Z$, as we have seen before. As 
before, we present the amplitudes for the $\lambda'$ and $\lambda''$ 
couplings together, though either one or the other must be zero.

The $F_1$ form factors are
\footnotesize
\begin{eqnarray}
 F^{1a}_{1ik} &=& \frac{ g_{Ze} \lambda^{'}_{i2k}\lambda^{'}_{i3k}}{16 \pi^2} \left[ m_t^2 \left(C^{(a)}_{11}-C^{(a)}_{12} + C^{(a)}_{21} - C^{(a)}_{23}\right) - 2 C^{(a)}_{24} \right] \nonumber \\
 F^{1b}_{1ik} &=& \frac{ \lambda^{'}_{i2k}\lambda^{'}_{i3k} }{16\pi^2} \left[ g_{dR} \left( m_t^2 (C^{(b)}_{21} - C^{(b)}_{23}) - 2 C^{(b)}_{24} + B^{(b)}_0 - M_{\tilde{e}_i}^2 C^{(b)}_0\right) + g_{dL} m_{k}^2 C^{(b)}_0 \right] \nonumber \\
 F^{1c}_{1ik} &=& \frac{ g_{Zd} \lambda^{'}_{i2k}\lambda^{'}_{i3k}}{16 \pi^2} \left[ m_t^2 \left(C^{(c)}_{11}-C^{(c)}_{12} + C^{(c)}_{21} - C^{(c)}_{23}\right) - 2 C^{(c)}_{24} \right] \nonumber \\
 F^{1d}_{1ik} &=& \frac{ \lambda^{'}_{i2k}\lambda^{'}_{i3k} }{16\pi^2} \left[ g_{eR} \left( m_t^2 (C^{(d)}_{21} - C^{(d)}_{23}) - 2 C^{(d)}_{24} + B^{(d)}_0 - M_{\tilde{d}_k}^2 C^{(d)}_0\right) + g_{eL} m_i^2 C^{(d)}_0 \right] \nonumber \\
 F^{1e}_{1ik} &=& - \frac{ \lambda^{'}_{i2k}\lambda^{'}_{i3k} }{16\pi^2 (m_t^2 - m_c^2)} g_{uL} m_t^2 B^{(e)}_1 \nonumber \\
 F^{1f}_{1ik} &=&  \frac{\lambda^{'}_{i2k}\lambda^{'}_{i3k} }{16\pi^2 (m_t^2 - m_c^2)} g_{uL} m_c^2 B^{(f)}_1 \nonumber \\
 F^{1g}_{1jk} &=& \frac{ g_{Zd} \lambda^{''}_{2jk}\lambda^{''}_{3jk}}{16 \pi^2} m_t m_c \left(C^{(g)}_{12} + C^{(g)}_{23}\right) \nonumber \\
 F^{1h}_{1jk} &=& \frac{ \lambda^{''}_{2jk}\lambda^{''}_{3jk} }{16\pi^2} g_{dL} \left[ m_c m_t C^{(h)}_{23} +  m_t \left( m_t (C^{(h)}_{11} - C^{(h)}_{12}) + m_c C^{(h)}_{12}\right) + m_c m_t C^{(h)}_{11} \right] \nonumber \\
 F^{1i}_{1jk} &=& - \frac{ \lambda^{''}_{2jk}\lambda^{''}_{3jk} }{16\pi^2 (m_t^2 - m_c^2)} g_{uL} m_c m_t B^{(i)}_1 \nonumber \\
  F^{1j}_{1jk} &=& \frac{\lambda^{''}_{2jk}\lambda^{''}_{3jk} }{16\pi^2 (m_t^2 - m_c^2)}  g_{uL} m_c m_t B^{(j)}_1 \nonumber 
\end{eqnarray}
\begin{eqnarray}
\end{eqnarray}
\normalsize

The $F_2$ form factors are
\footnotesize
\begin{eqnarray}
F^{2a}_{1ik} &=& \frac{ g_{Ze} \lambda^{'}_{i2k}\lambda^{'}_{i3k}}{16 \pi^2} m_t m_c \left(C^{(a)}_{12} + C^{(a)}_{23}\right) \nonumber \\ 
 F^{2b}_{1ik} &=& \frac{ \lambda^{'}_{i2k}\lambda^{'}_{i3k} }{16\pi^2} g_{dR} \left( m_t^2 (C^{(b)}_{11} - C^{(b)}_{12})  + m_c m_t (C^{(b)}_{11} + C^{(b)}_{12} + C^{(b)}_{23})\right) \nonumber \\ 
 F^{2c}_{1ik} &=& \frac{ g_{Zd} \lambda^{'}_{i2k}\lambda^{'}_{i3k}}{16 \pi^2} m_t m_c \left(C^{(c)}_{12} + C^{(c)}_{23}\right) \nonumber \\ 
 F^{2d}_{1ik} &=& \frac{ \lambda^{'}_{i2k}\lambda^{'}_{i3k} }{16\pi^2} g_{eR} \left[ m_c m_t C^{(d)}_{23} +  m_t \left( m_t (C^{(d)}_{11} - C^{(d)}_{12}) + m_c C^{(d)}_{12}\right) + m_c m_t C^{(d)}_{11} \right] \nonumber \\ 
 F^{2e}_{1ik} &=& - \frac{ \lambda^{'}_{i2k}\lambda^{'}_{i3k} }{16\pi^2 (m_t^2 - m_c^2)} g_{uR} m_c m_t B^{(e)}_1 \nonumber \\
  F^{2f}_{1ik} &=&  \frac{\lambda^{'}_{i2k}\lambda^{'}_{i3k} }{16\pi^2 (m_t^2 - m_c^2)} g_{uR} m_c m_t B^{(f)}_1 \nonumber \\ 
 F^{2g}_{1jk} &=& \frac{ g_{Zd} \lambda^{''}_{2jk}\lambda^{''}_{3jk}}{16 \pi^2} \left[m_t^2 \left(C^{(g)}_{11}-C^{(g)}_{12} + C^{(g)}_{21} - C^{(g)}_{23}\right) - 2 C^{(g)}_{24}\right] \nonumber \\ 
 F^{2h}_{1jk} &=& \frac{ \lambda^{''}_{2jk}\lambda^{''}_{3jk} }{16\pi^2} \left[ g_{dL} \left( m_t^2 (C^{(h)}_{21} - C^{(h)}_{23}) - 2 C^{(h)}_{24} + B^{(h)}_0 - M_{\tilde{d}_k}^2 C^{(h)}_0\right) + g_{dR} m_j^2 C^{(h)}_0 \right] \nonumber \\ 
 F^{2i}_{1jk} &=& - \frac{ \lambda^{''}_{2jk}\lambda^{''}_{3jk} }{16\pi^2 (m_t^2 - m_c^2)} g_{uR} m_t^2 B^{(i)}_1 \nonumber \\ 
 F^{2j}_{1jk} &=& \frac{\lambda^{''}_{2jk}\lambda^{''}_{3jk}}{16\pi^2 (m_t^2 - m_c^2)} g_{uR} m_c^2 B^{(j)}_1 
\end{eqnarray}
\normalsize

The $F_3$ form factors are
\footnotesize
\begin{eqnarray}
F^{3a}_{1ik} &=& - \frac{ g_{Ze} \lambda^{'}_{i2k}\lambda^{'}_{i3k}}{16 \pi^2}  m_c \left(C^{(a)}_{12} + C^{(a)}_{23}\right) \nonumber \\ 
 F^{3b}_{1ik} &=& - \frac{ \lambda^{'}_{i2k}\lambda^{'}_{i3k} }{16\pi^2} g_{dL} m_c \left( C^{(b)}_{11}  +  C^{(b)}_{23}\right) \nonumber \\ 
 F^{3c}_{1ik} &=& - \frac{ g_{Zd} \lambda^{'}_{i2k}\lambda^{'}_{i3k}}{16 \pi^2}  m_c \left(C^{(c)}_{12} + C^{(c)}_{23}\right) \nonumber \\ 
 F^{3d}_{1ik} &=& - \frac{ \lambda^{'}_{i2k}\lambda^{'}_{i3k} }{16\pi^2} g_{eL} m_c \left( C^{(d)}_{11} + C^{(d)}_{23} \right) \nonumber \\
 F^{3g}_{1jk} &=& - \frac{ g_{Zd} \lambda^{''}_{2jk}\lambda^{''}_{3jk}}{16 \pi^2} m_t \left(C^{(g)}_{11}-C^{(g)}_{12} + C^{(g)}_{21} - C^{(g)}_{23}\right) \nonumber \\ 
 F^{3h}_{1jk} &=& - \frac{ \lambda^{''}_{2jk}\lambda^{''}_{3jk} }{16\pi^2} g_{dL} m_t \left( C^{(h)}_{21} - C^{(h)}_{23}\right) \nonumber \\ 
\end{eqnarray}
\normalsize
and, finally the $F_4$ form factors are
\footnotesize
\begin{eqnarray}
F^{4a}_{1ik} &=& - \frac{ g_{Ze} \lambda^{'}_{i2k}\lambda^{'}_{i3k}}{16 \pi^2} m_t \left(C^{(a)}_{11}-C^{(a)}_{12} + C^{(a)}_{21} - C^{(a)}_{23}\right) \nonumber \\ 
 F^{4b}_{1ik} &=& \frac{ \lambda^{'}_{i2k}\lambda^{'}_{i3k} }{16\pi^2} g_{dR} m_c \left( C^{(b)}_{21} - C^{(b)}_{23}\right) \nonumber \\ 
 F^{4c}_{1ik} &=& - \frac{ g_{Zd} \lambda^{'}_{i2k}\lambda^{'}_{i3k}}{16 \pi^2} m_t \left(C^{(c)}_{11}-C^{(c)}_{12} + C^{(c)}_{21} - C^{(c)}_{23}\right) \nonumber \\ 
 F^{4d}_{1ik} &=& - \frac{ \lambda^{'}_{i2k}\lambda^{'}_{i3k} }{16\pi^2} g_{eR} m_t \left( C^{(d)}_{21}  - C^{(d)}_{23}\right) \nonumber \\
 F^{4g}_{1jk} &=& - \frac{ g_{Zd} \lambda^{''}_{2jk}\lambda^{''}_{3jk}}{16 \pi^2} m_c \left(C^{(g)}_{12} + C^{(g)}_{23}\right) \nonumber \\ 
 F^{4h}_{1jk} &=& - \frac{ \lambda^{''}_{2jk}\lambda^{''}_{3jk} }{16\pi^2} g_{dR} m_c \left(C^{(h)}_{11}  + C^{(h)}_{23}\right) \nonumber \\ 
\end{eqnarray}
\normalsize
where
\begin{eqnarray}
C_X^{(a)} &=& C_X (m_k, M_{\tilde{e}_i}, M_{\tilde{e}_i}; m_c, m_t, M_Z) 
\quad \qquad
B_0^{(b)} = B_0(m_k, m_k; M_Z) \nonumber\\
C_X^{(b)} &=& C_X (M_{\tilde{e}_i}, m_k, m_k; m_c, m_t, M_Z) 
\qquad\qquad
\!\!\!\!   B_0^{(d)} = B_0(m_i,m_i; M_Z) 
\nonumber \\
C_X^{(c)} &=& C_X (m_i,M_{\tilde{d}_k}, M_{\tilde{d}_k}; m_c, m_t, M_Z)
\quad \qquad
\!  B_1^{(e)} = B_1(m_k, M_{\tilde{e}_i}; m_c) 
\nonumber \\
C_X^{(d)} &=& C_X (M_{\tilde{d}_k}, m_i, m_i; m_c, m_t, M_Z) 
\qquad\qquad
\!\!\!   B_1^{(f)} = B_1(m_k,M_{\tilde{e}_i}; m_t) 
\nonumber \\
C_X^{(g)} &=& C_X (m_i, M_{\tilde{d}_k}, M_{\tilde{d}_k}; m_c, m_t, M_Z) 
\qquad
\ \ \ \! B_0^{(h)} = B_0(m_j, m_j; M_Z) 
\nonumber\\
C_X^{(h)} &=& C_X (M_{\tilde{d}_k}, m_j, m_j; m_c, m_t, M_Z)  
\qquad\qquad
\! \! \! B_1^{(i)} = B_1(m_j, M_{\tilde{d}_k};m_c) 
\nonumber \\
B_1^{(j)} &=& B_1(m_j, M_{\tilde{d}_k}; m_t) 
\end{eqnarray}
and we have defined effective couplings
\begin{eqnarray}
 g_{Zd} &=& -\frac{\sin^2 \theta_W}{6 \cos \theta_W} \qquad\qquad\qquad \ \ \
 g_{Ze} = \frac{1- 2 \sin^2\theta_W}{2 \cos \theta_W} \nonumber\\
 g_{uL} &=& -\frac{1-2 q_u \sin^2 \theta_W}{2 \cos \theta_W} \qquad\qquad 
 g_{uR} = \frac{q_u \sin^2 \theta_W}{\cos \theta_W} \nonumber \\
 g_{dL} &=& \frac{1+2 q_d \sin^2 \theta_W}{2 \cos \theta_W} \qquad\qquad\quad
 g_{dR} = \frac{q_d \sin^2 \theta_W}{\cos \theta_W} \nonumber \\
 g_{eL} &=& \frac{1-2 \sin^2 \theta_W}{2 \cos \theta_W}\qquad\qquad\qquad 
 g_{eR} = - \frac{\sin^2 \theta_W}{\cos \theta_W}
\end{eqnarray}
It is now a straightforward matter to calculate the total form factors
using $F_{ni} = \sum_{A=a}^j  F_{ni}^{A}$ for $n = 1,2,3,4$ and substitute the results 
into Eqn.~(\ref{eqn:helampZ}) to get the final RPV-MSSM amplitude.
\end{appendices}

\newpage

\end{document}